\documentclass[11pt,a4paper]{article}
\usepackage{jcappub}

\usepackage{aas_macros}

\usepackage[T1]{fontenc}
\usepackage[utf8]{inputenc} 
\usepackage[UKenglish]{babel} 

\usepackage{soul}

\usepackage{siunitx}
\DeclareSIUnit\year{yr}
\DeclareSIUnit\parsec{pc}
\DeclareSIUnit\erg{erg}

\newcommand{\BURST}{Burst-like}
\newcommand{\burst}{burst-like}
\newcommand{\CREDIT}{Energy-dependent escape}
\newcommand{\credit}{energy-dependent escape}
\newcommand{\TDD}{Time-dependent diffusion}
\newcommand{\tdd}{time-dependent diffusion}

\title{Stochastic modelling of cosmic-ray sources for Galactic diffuse emissions}

\author[a]{Anton Stall}
\author[a]{and Philipp Mertsch}

\affiliation[a]{Institute for Theoretical Particle Physics and Cosmology (TTK), RWTH Aachen University, 52056 Aachen, Germany}

\emailAdd{stall@physik.rwth-aachen.de}
\emailAdd{pmertsch@physik.rwth-aachen.de}

\abstract{
Galactic diffuse emissions in gamma rays and neutrinos arise from interactions of cosmic rays with the interstellar medium and probe the cosmic-ray intensity away from the Solar system. 
Model predictions for those are influenced by the properties of cosmic-ray sources, and understanding the impact of cosmic-ray sources on Galactic diffuse emissions is key for interpreting measurements by LHAASO, Tibet AS-gamma, IceCube, and the upcoming SWGO. 
We consider supernova remnants as prototypical cosmic-ray sources and study the impact of their discreteness on the Galactic diffuse emissions in different source injection and near-source transport models in a stochastic Monte Carlo study.

Three lessons exemplify the results of our simulations:
First, the distributions of Galactic diffuse emission intensities can be described by a mixture model of stable laws and Gaussian distributions.
Second, the maximal deviations caused by discrete sources across the sky depend on energy, reaching typically tens of percent in \burst{} and \credit{} scenarios but order unity or larger in a \tdd{} scenario.
Third, the additional model uncertainty from source stochasticity is subdominant in \burst{} and \credit{} scenarios, but becomes sizeable above some tens of \si{\tera\electronvolt} in the \tdd{} scenario, where it can help reconcile model predictions with LHAASO measurements.
With increased spatial resolution, especially at energies beyond tens of \si{\tera\electronvolt}, measurements of Galactic diffuse emissions can be expected to constrain source models and locate cosmic ray sources. 
}

\begin{document}
\hfill{TTK-25-27}

\maketitle
\flushbottom

\section{Introduction\label{sec:introduction}}

The discovery of cosmic rays (CRs) dates back to the early 20th century.
More than a hundred years later, many fundamental questions still need to be answered.
Importantly, we still do not know what the sources of CRs in our Galaxy are and how exactly they are distributed.
At the same time, we are entering an era of high-precision measurements of the CR spectrum at Earth with experiments like, e.g., AMS-02~\cite{2021AguilarAliCavasonzaPhR} and DAMPE~\cite{2024AlemannoAltomarePhRvD} with the prospect of even more precise measurements in the future by instruments like the proposed AMS-100~\cite{2019SchaelAtanasyanNIMPA}.
Those measurements might reveal features in the CR spectrum that could inform us about the nature of the sources~\cite{2025StallLooApJL}.
However, those measurements only capture the influence of CR sources at the position of the Solar system.

Another way to learn about the sources of CRs is to study Galactic diffuse emissions (GDEs) produced by the interaction of CRs with the interstellar medium (ISM)~\cite{2021TibaldoGaggeroUniv}.
Diffuse gamma rays have been measured with great precision by \textit{Fermi}-LAT in the gigaelectronvolt energy range~\cite{2012AckermannAjelloApJ}.
Recently, also measurements up to hundreds of teraelectronvolts have been published by LHAASO for gamma rays~\cite{2023CaoAharonianPhRvL, 2025CaoAharonianPhRvL} and IceCube for neutrinos from the Galactic plane~\cite{2023IcecubeCollaborationAbbasiSci}.
Due to inelasticity, those diffuse emissions are produced by CRs of a factor of $\mathcal{O}\left(10\right)$ larger energies (see, e.g.,~\cite{2006KelnerAharonianPhRvD,2021KoldobskiyKachelriessPhRvD}).
Thus, they are sensitive to the high-energy end of the Galactic CR component.
Predictions on the GDE intensity are obtained by solving the CR transport equation with a source term that describes the injection of CRs into the ISM, a model for the ISM, and the interaction cross sections of the CRs with their target material.
The influence of those different ingredients on the predictions of GDEs and their uncertainties have been discussed in Refs.~\cite{2023SchweferMertschApJ,2025VecchiottiPeronJCAP}.
This work focuses specifically on the influence of the source modelling and extends previous work by the consideration of the discrete nature of the CR sources on the predictions of GDEs.

Among the various CR source candidates, supernova remnants (SNRs) are commonly considered as prime candidates for the acceleration of Galactic CRs~\cite{2019GabiciEvoliIJMPD}.
Through the process of diffusive shock acceleration, they are able to accelerate particles to very high energies and inject them into the ISM with a power-law spectrum ($\propto \mathcal{R}^{-\alpha}$) in particle rigidity $\mathcal{R} = \frac{p c}{Z e}$ where $\alpha \approx 2$~\cite{1977KrymskiiSPhD}.
All CRs of rigidities below the so-called CR knee at around \SI{3}{\peta\volt} are believed to be of Galactic origin.
While there is a lot of debate whether SNRs can in fact accelerate particles up to petaelectronvolt energies, it is widely accepted that they are responsible for the bulk of the CR flux below the CR knee~\cite{2019GabiciEvoliIJMPD}.
Other contenders for the acceleration of Galactic CRs are star clusters~\cite{2021MorlinoBlasiMNRAS} which can potentially accelerate particles to higher energies at their termination shock.
This is possible as up- and downstream of the shock are effectively swapped compared to the SNR shock case, giving this source class a geometric advantage for confining CRs close to the shock for further acceleration.
The resulting injection spectrum is also similar to the one of SNRs~\cite{2021MorlinoBlasiMNRAS}.
For simplicity, we will assume that SNRs are the main source of CRs in our Galaxy.
We refer to this as the \textit{supernova paradigm}.

SNRs have spatial dimensions and injection time scales that are much smaller than the corresponding Galactic propagation lengths and times of the CRs they inject.
Thus, SNRs are often modelled well as point sources~\cite{2017GenoliniSalatiA&A}.
However, the exact positions and ages of the CR sources are not known~\cite{2011MertschJCAP,2017GenoliniSalatiA&A, 2021EvoliAmatoPhRvDa, 2012BlasiAmatoJCAP, 2021PhanSchulzePhRvL}.
So instead of modelling the sources with their limited extent in space and time, they are usually treated as a continuous source distribution~\cite{2019GabiciEvoliIJMPD}.
It is conceivable that the consideration of the point-like nature of CR sources can lead to predictions that deviate clearly from the ones obtained with a smooth source distribution~\cite{2015KachelriessNeronovPhRvL,2025StallLooApJL}.
In particular, this is the case for high-energy CRs, where only a few sources contribute effectively to the flux at Earth due to shorter escape timescales at higher energies.

Our limited knowledge of the exact source coordinates only allows us to study the effects of modelling individual sources in a stochastic way.
We will refer to this approach as \textit{stochastic modelling of sources}.
The CR flux at any given position in the Galaxy is dominated by the contributions of young and nearby sources.
However, it is not immediately clear which among the close-by young sources are the most important ones.
This depends, e.g., on the rigidity of the CRs and their injection mechanism.
In our stochastic model, we will investigate the probability distribution of the CR intensity for randomly drawn realisations of the source distribution by conducting a Monte Carlo simulation.
We will refer to the influence of the discrete nature of sources on the CR intensity and GDEs as \textit{source stochasticity}.

Observations of CRs locally and gamma rays produced in CR interactions in other parts of the Galaxy will most likely not be influenced by the same sources, which inherently limits the predictability of GDEs from local CR measurements alone.
Observations of GDEs rather open up an opportunity to study the accumulated effect of the stochastic distribution of CRs throughout the Galaxy and thus give us another potential handle on testing CR source and propagation models.

\begin{figure}
    \centering
    \includegraphics[width=\textwidth]{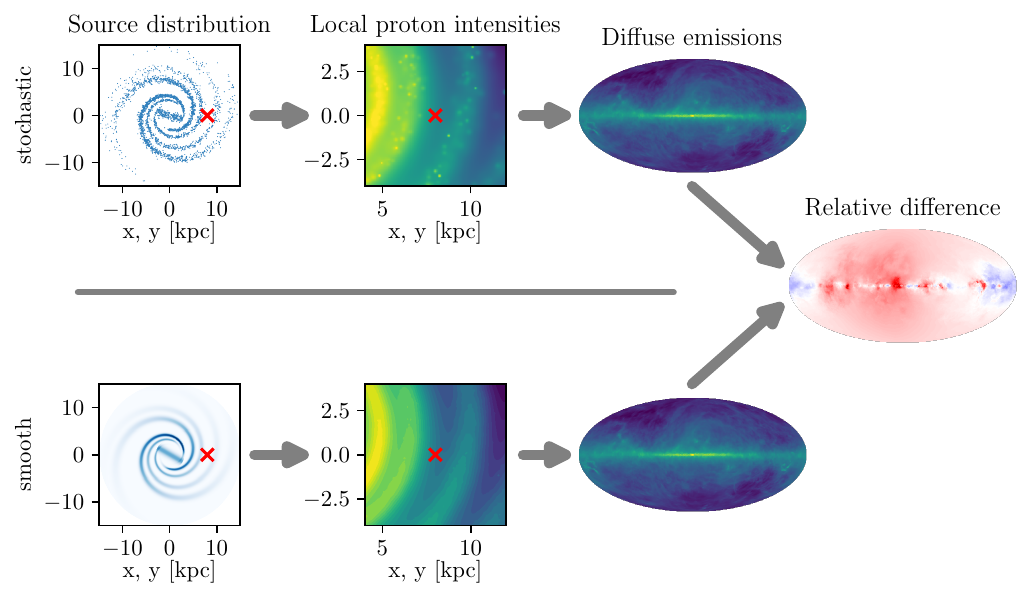}
    \caption{This schematic overview illustrates the stochastic modelling of CR sources (upper half) and how to obtain GDE predictions and contrasts it to the smooth source model (lower half).
    In the left panels, the source distribution is illustrated.
    While point sources can be seen in the stochastic case, the source distribution is a smooth continuum in the smooth case.
    The position of the Sun is marked with a red cross.
    In the next step to the right, the CR proton intensities a few kiloparsecs around the Sun are shown as an illustration for the different imprints of the source distributions.
    The intensity distribution is patchier and imprints of individual sources can be seen by eye.
    From the CR proton intensities throughout the Galaxy, hadronic GDEs can be predicted for both models.
    The relative difference between them reveals imprints of individual sources and allows us to study the influence of the stochastic nature of sources quantitatively.}
    \label{fig:overview}
\end{figure}

It has been pointed out that the CR intensity distribution at any given position follows a so-called stable law with divergent variance manifested in long tails~\cite{2017GenoliniSalatiA&A}.
We will extend this work by studying the influence of the discrete nature of sources on the GDEs.
There has been some effort in the past to investigate the influence of discrete sources on the gamma-ray and neutrino intensities in the Galaxy~\cite{2023ThalerKissmannAPh,2023MarinosRowellMNRAS,2025MarinosPorterApJ,2025KaciGiacintiJCAP}.
By considering large numbers of realisations for different source injection and transport models, we want to adequately cover the space of possible source realisations and give a comprehensive overview of the influence of the discrete nature of sources on the GDEs.
Our working principle is illustrated in Fig.~\ref{fig:overview}. In the stochastic model, we draw a source realisation, calculate the CR proton intensities in the Galaxy, and obtain hadronic diffuse emission predictions specific for this source realisation.
By comparison to the GDE predictions for the corresponding smooth source distribution, where CRs are injected by a source continuum, we can quantify the influence of modelling discrete sources explicitly.

In this work, we are only considering the hadronic contribution to the GDEs.
For the gamma-ray observations, also leptonic contributions may need to be considered.
Depending on the energies of interest, however, their contribution might be subdominant.
It has been estimated that diffuse emissions caused by the interactions of leptons with the ISM make up less than \SI{10}{\percent} for small latitudes ($|b| < 10^\circ$)~\cite{2018LipariVernettoPhRvD}.
Beyond some tens of teraelectronvolts, it is also not clear if there even are leptonic sources in our Galaxy and whether electrons and positrons can escape the source regions due to their short cooling times~\cite{2018MertschJCAP}.
There have been studies of GDEs with discrete CR sources that also take into account a leptonic component~\cite{2023ThalerKissmannAPh,2025MarinosPorterApJ}.
Due to their subdominant contribution, we focus on the hadronic GDEs for our stochastic study in this work (see also~\cite{2025KaciGiacintiJCAP}).
\section{Method}\label{sec: method}
We built a stochastic model for the calculation of hadronic GDEs.
We model the injection of protons, the main component of Galactic CRs, by stochastic sources.
In this section, we will explain in detail how we solved the CR transport equation, modelled the injection by the sources, and obtained the GDEs.
Particularly, we want to highlight the importance of numerically efficient methods to perform the calculations.
Without them, only the analysis of a handful of CR source realisations would be feasibly, limiting the statistical power of this Monte Carlo approach.
We start with a description of the CR transport equation and its solution for discrete sources.
\subsection{Cosmic ray transport model}\label{sec: Transport model}
To study the influence of source stochasticity on GDEs, we have to build a model that allows us to calculate the CR proton intensities for a given source realisation (see also~\cite{2025StallLooApJL}).
For that, we consider a simplified model of the Galactic transport volume.
We assume that all CRs are injected by sources that are distributed close to the Galactic disk ($z\approx0$) and that CRs diffuse through the Galactic halo that extends to $z=\pm H$, where we use a fiducial value of $H = \SI{4}{\kilo\parsec}$~\cite{2025StallLooApJL,2018MertschJCAP}.
We adopt a Galactic centre distance of $R_{\odot} = \SI{8}{\kilo\parsec}$ as inferred in the Galactic gas reconstruction used in this work~\cite{2025SodingEdenhoferA&A} (see Sec.~\ref{sec: diffuse emissions}).

Source ages are drawn from a uniform distribution adopting the fiducial supernova rate $\nu$ of \SI{0.03}{\per\year}~\cite{1991vandenBerghTammannARA&A,1994TammannLoefflerApJS}.
For the distribution of source positions, we consider a spiral model with a Galactic bar inspired by Refs.~\cite{2021EvoliAmatoPhRvDb,2021EvoliAmatoPhRvDa}, which in turn is based on the spiral model in Ref.~\cite{2010Steiman-CameronWolfireApJ}.
Specifically, the Galactic bar extends to about \SI{2.8}{\kilo\parsec} and the centre of the closest spiral arm towards the Galactic centre is \SI{1.5}{\kilo\parsec} away from the Sun's position in our implementation.
We use the distribution of sources in galactocentric radius described in Ref.~\cite{2001FerriereRvMP}, rescaled for our $R_{\odot}$.
Further, we assume that no sources lie beyond a Galactic radius of \SI{15}{\kilo\parsec}.
For the distribution in the $z$-direction out of the Galactic plane, we use a normal distribution with a scale height of $z_0 = \SI{70}{\parsec}$~\cite{2010Steiman-CameronWolfireApJ}.

We will only consider CR protons with rigidities above \SI{10}{\giga\volt} which simplifies the general CR transport equation significantly~\cite{1990BerezinskiiBulanovacr} as inelastic collision and advection are subdominant and can be ignored.
The transport equation then simplifies to
\begin{equation}\label{eq:transport_equation}
    \frac{\partial \psi_{\mathcal{R}}\left(\mathcal{R}, t, \mathbf{x}\right)}{\partial t} - \kappa\left(\mathcal{R}\right) \nabla^2 \psi_{\mathcal{R}} \left(\mathcal{R}, t, \mathbf{x}\right) = Q\left(\mathcal{R}, t, \mathbf{x}\right) \, ,
\end{equation}
where $\psi_{\mathcal{R}}\left(\mathcal{R}, t, \mathbf{x}\right)= {\mathrm{d} n}/{\mathrm{d} \mathcal{R}}$ denotes the isotropic differential CR density ($n$ is the number density).
It is related to the differential intensity $\Phi_{\mathcal{R}} = (\mathrm{d}^4N)/(\mathrm{d}\mathcal{R} \, \mathrm{d}A \, \mathrm{d}t \, \mathrm{d}\Omega) = v/(4 \pi) \ \psi_{\mathcal{R}}$ and to the phase-space density $f=f\left(p, t, \mathbf{x}\right) $ through $\psi_{\mathcal{R}} = (4 \pi p^2 Z e)/c \ f$.

We use the diffusion coefficient determined by fitting to locally measured CR data in Ref.~\cite{2023SchweferMertschApJ}.
The diffusion coefficient is assumed to be isotropic and to follow a broken power law in rigidity.
It is given by
\begin{equation}\label{eq: diffusion_coefficient}
    \kappa\left(\mathcal{R}\right)=\kappa_0 \ \left(\frac{H}{\SI{6}{\kilo\parsec}}\right) \ \beta\left(\frac{\mathcal{R}}{\mathcal{R}_{12}}\right)^{\delta_{1}} \prod_{i=1}^{4} \left(1+\left(\frac{\mathcal{R}}{\mathcal{R}_{i(i+1)}}\right)^{1/s_{i(i+1)}}\right)^{s_{i(i+1)}(\delta_{i+1}-\delta_i)} \, ,
\end{equation}
where $\beta$ is the ratio of the particle's speed to the speed of light.
The parameters are given in Tab.~1 in Ref.~\cite{2023SchweferMertschApJ}.
The normalisation of the diffusion coefficient is not computed from first principles but determined by fitting to secondary-to-primary ratios like B/C.
As the diffusion coefficient, $\kappa$, and the halo height, $H$, approximately enter into the predicted ratio in the combination ($\kappa/H$), the data really only constrain this ratio.
One can therefore rescale the diffusion coefficient fitted under the assumption of one halo height to another halo height.

For primary CRs above some gigavolts, the propagated CR spectrum is approximately proportional to the ratio of the source spectrum and the diffusion coefficient.
This degeneracy is deliberately used to accommodate transport and injection effects alike~\cite{2023SchweferMertschApJ}.
While secondary-to-primary ratios indicate that the break at $\mathcal{R}\simeq\SI{300}{\giga\volt}$ is due to transport effects, likely related to a transition in the turbulence power spectrum~\cite{2012BlasiAmatoPhRvL,2018EvoliBlasiPhRvL}, the break associated with the CR knee at $\mathcal{R}\simeq\SI{3}{\peta\volt}$ is usually attributed to a cut-off of Galactic CR sources.
Although this seems statistically unlikely~\cite{2017GenoliniSalatiA&A}, if some of the spectral breaks in the CR spectrum are caused by the strong influence of a local source~\cite{2021MalkovMoskalenkoApJ,2021FornieriGaggeroPhRvD}, the respective breaks would need to be removed from the parametrisation.
The fit in Ref.~\cite{2023SchweferMertschApJ} remains agnostic to the origin of the breaks.

The source term on the right-hand side of Eq.~\eqref{eq:transport_equation} for a single source located at $\left(t_i, \mathbf{x}_i\right)$ injecting CRs of rigidity $\mathcal{R}$ at time $t_{\text{esc},i}\left(\mathcal{R}\right)$ is given by
\begin{equation}
    Q_i\left(\mathcal{R}, t, \mathbf{x}\right) = Q_{\mathcal{R}} \left(\mathcal{R}\right) \delta\left(\mathbf{x} - \mathbf{x}_i\right) \delta\left(t - t_{\text{esc},i}\left(\mathcal{R}\right)\right) \, , 
\end{equation}
We assume that a single population of SNRs injects all CR protons up to the knee rigidity $\mathcal{R}_{\text{knee}} \approx \SI{3}{\peta\volt}$
(corresponding to a proton energy of $E_{\text{knee}} \approx \SI{3}{\peta\electronvolt}$).
$Q_{\mathcal{R}}(\mathcal{R})$ is the common source spectrum for all of these sources.
Motivated by diffusive shock acceleration~\cite{2001MalkovDruryRPPh}, it has a power-law dependence on rigidity given by
\begin{equation}\label{eq: source spectrum}
    Q_{\mathcal{R}}(\mathcal{R}) = Q_0 \left(\frac{\mathcal{R}}{\SI{1}{\giga\volt}}\right)^{-\alpha} \, ,
\end{equation}
where $Q_0$ is the normalization factor and $\alpha$ is the spectral index, for which we assume a value of $\alpha=2.383$ as fitted in Ref.~\cite{2023SchweferMertschApJ}.
Note that $\alpha$ deviates from the canonical value of $2$ predicted by diffusive shock acceleration.
Possible reasons for that are the relative speed of the scattering centres~\cite{1978BellMNRAS,2020CaprioliHaggertyApJ} or shock obliquity~\cite{2011BellSchureMNRAS}.

The source spectrum normalisation $Q_0$ can be determined as described in Ref.~\cite{2021PhanSchulzePhRvL}.
They assume that a typical SNR has a total kinetic energy of $10^{51}\,\si{\erg}$ and an acceleration efficiency of \SI{8.7}{\percent}.
Further, we assume that CR protons are injected from \SI{100}{\mega\electronvolt} onwards and attribute the first break of the CR spectrum to the source spectrum (see discussion above and in~\cite{2023SchweferMertschApJ}).
This way, we find a value of $Q_0 = {2.828 \cdot 10^{52} \, \si{\per\giga\volt}}$, which we fix for all sources.

The transport equation \eqref{eq:transport_equation} can be solved analytically under the assumption of two free-escape boundary conditions, i.e., $\psi\left(z = \pm H \right) = 0$, using the method of mirror charges.
We do not consider an explicit radial boundary condition as its influence on the GDEs can be expected to be negligible.
Most of the sources in the Galaxy are closer to the free-escape boundary in the $z$-direction than to the radial boundary.
The influence of CR sources far away from the Galactic centre on the GDEs is also strongly suppressed due to small gas densities in the outer parts of the Galaxy.

The Green's function of Eq.~\eqref{eq:transport_equation}, i.e., the solution for a single source, is given by~\cite{1959SyrovatskiiSvA} 
\begin{equation}\label{eq: analytic Greens function}
    G\left(t, \mathbf{x}; t_i, \mathbf{x}_i\right) =  \frac{Q_{\mathcal{R}}\left(\mathcal{R}\right)}{\left(2 \pi \sigma^2\right)^{\frac{3}{2}}} e^{-\frac{\left(\mathbf{x}_i-\mathbf{x}\right)^2}{2 \sigma^2}} \vartheta\left(z, z_i, \sigma^2, H\right) \, .
\end{equation}
Besides the source spectrum described in Eq.~\eqref{eq: source spectrum}, the Green's function contains a Gaussian whose variance
\begin{equation}\label{eq: variance Green's function}
    \sigma^2\left(\mathcal{R}, t; t_i \right) = 2 \ \int_{t_{\text{esc},i}\left(\mathcal{R}, t_i\right)}^t \mathrm{d} t' \kappa\left(\mathcal{R}, t'\right)    
\end{equation}
depends on the diffusion coefficient~\eqref{eq: diffusion_coefficient} and the time since injection.
We will discuss the possibility of a time-dependent diffusion coefficient and a rigidity-dependent injection time in Sec.~\ref{sec: source injection and near-source transport}.
There is also a correction function $\vartheta$, which accounts for the free-escape boundary condition (see App.~\ref{app: Jacobi}).
This function is an infinite sum related to the Jacobi theta function (see, e.g.,~\cite{2011MertschJCAP}) that suppresses the Green's function's value for $z \approx \pm H$ or if the diffusion length ($\approx \sigma$) is comparable with $H$.

There are issues with the diffusion equation for times shortly after the injection.
For once, the approximation of an infinitesimally small injection region might not be justified for positions that are close to a source.
To remedy this, we introduce a finite source extent by adding a constant $\sigma_0^2 = \left(\SI{10}{\parsec}\right)^2$ to the Green's functions' variance parameter.\footnote{We found that \SI{10}{\parsec} should be a reasonable guess for the typical source extent (see, e.g.,~\cite{2022RecchiaGalliA&A}).}
Another concern is that the diffusion equation is not covariant, which means that it allows a small fraction of CRs to propagate faster than the speed of light.
We implement a light-cone condition that excludes contributions from sources that lie outside of a point-of-interest's past light-cone~\cite{2017GenoliniSalatiA&A,2025StallLooApJL}.
There have been suggestions to implement further exclusions of sources which are young and close to the point-of-interest (see, e.g.,~\cite{2012BlasiAmatoJCAP}), but we refrain from that.
A source configuration with high CR intensities might not be realised in our position in the Galaxy, but their occurrence is an inherent prediction of the assumed source and transport models.
Implementing cuts biases the underlying probability distribution of CR intensities in the Galaxy, and it is not at all clear how they should be chosen (see also discussions in~\cite{2018MertschJCAP,2025StallLooApJL}).

As the discrete source distribution is a sum of $N$ point sources, the total isotropic differential CR density is 
\begin{equation}\label{eq: total CR density}
    \psi_{\mathcal{R}}\left(\mathcal{R}, t, \mathbf{x}\right) = \sum_{i=1}^N G\left(\mathcal{R}, \mathbf{x}, t; \mathbf{x}_i, t_i\right) \, .
\end{equation}
We will need the CR intensity $\Phi$ differential in kinetic energy $E_{\mathrm{kin}}$ in the calculation of the GDEs later.
We obtain it from the isotropic CR density by
\begin{equation}
    \Phi = \frac{v}{4 \pi} \ \frac{\mathrm{d \mathcal{R}}}{\mathrm{d} E_{\mathrm{kin}}} \ \psi_{\mathcal{R}} = \frac{c}{4 \pi} \ \psi_{\mathcal{R}} \, .
\end{equation}
\subsection{Source injection and near-source transport}\label{sec: source injection and near-source transport}
The Green's function in Eq.~\eqref{eq: analytic Greens function} describes the contribution of a single CR source to the total CR proton intensity.
Thanks to the linearity of the diffusion transport Eq.~\eqref{eq:transport_equation}, physically motivated scenarios beyond an instantaneous \textit{burst-like} injection of the source spectrum by the source can be modelled with ease.

One option is to consider a rigidity-dependence of the escape time from a source $t_{\text{esc}}\left(\mathcal{R}\right)$ in Eq.~\eqref{eq: analytic Greens function}.
An earlier escape of CRs of higher rigidities from their sources compared to their lower-rigidity counterparts has been considered in the context of CR acceleration and escape models~\cite{2012BlasiAmatoJCAP,2012ThoudamHorandelMNRAS,2009CaprioliBlasiMNRAS} and the gamma-ray emissions in the near-source region~\cite{2009GabiciAharonianMNRAS,2019CelliMorlinoMNRAS}.
We use the \textit{CREDIT} model described in Ref.~\cite{2025StallLooApJL}, where further motivations for this model are discussed in detail.
Here, the escape time is given as $t_{\text{esc}}(\mathcal{R}) = t_0 + \Delta t_{\text{esc}}(\mathcal{R}) $ and depends on the time of the supernova explosion $t_0$ as well as the time it takes a proton of a rigidity $\mathcal{R}$ to escape the source, $\Delta t_{\text{esc}}(\mathcal{R})$.
We assume that the latter follows a power law at high rigidities:
\begin{equation}\label{eq:CREDIT energy dependent escape}
    \Delta t_{\text{esc}}\left(\mathcal{R}\right) = t_{\text{Sed}} \!\left(\frac{\mathcal{R}}{\mathcal{R}_{\text{knee}}}\right)^a \, , \; a = \frac{\ln\left(t_{\text{life}}/t_{\text{Sed}}\right)}{\ln\left(\mathcal{R}_{\text{b}}/\mathcal{R}_{\text{knee}}\right)} \, ,
\end{equation}
for $\mathcal{R}>\mathcal{R}_{\text{b}}$ and $\Delta t_{\text{esc}}\left(\mathcal{R}\right) = t_{\text{life}}$ otherwise.
$t_{\text{Sed}}$ marks the start of the Sedov-Taylor phase when the first particles escape the source.
For this study, we choose the fiducial parameters from Ref.~\cite{2025StallLooApJL}, i.e., $t_{\text{life}} = \SI{100}{\kilo\year}$, $t_{\text{Sed}} = \SI{1}{\kilo\year}$, and $\mathcal{R}_{\text{b}} = \SI{10}{\tera\volt}$.

It is also possible to study the impact of inhomogeneous diffusion around CR sources (see, e.g.,~\cite{2022JacobsMertschJCAP}) in this context.
We implement a model of \textit{time-dependent diffusion} (TDD) introduced in Ref.~\cite{2025KaciGiacintiJCAP} that is designed to capture the effect of suppressed diffusion in a spatial region around CR sources.
In this model, the diffusion is suppressed to a rigidity-independent constant diffusion coefficient $\kappa_{\text{start}} = 10^{28} \, \si{\centi\meter\squared\per\second}$ for a time $t_{\text{change}} = \SI{10}{\kilo\year}$.
This is in line with models of self-generated turbulence around some CR sources, which find even smaller approximately constant values for the diffusion coefficient at rigidities below some tens of teravolts~\cite{2022MukhopadhyayLindenPhRvD}.
With this choice of $\kappa_{\text{start}}$, the spatial extent of the low-diffusion zone is given by the diffusion length after a time $t_{\text{change}}$, i.e., $r_{\text{change}} \simeq \SI{50}{\parsec}$.
This is a typical size determined by models of low-diffusion zones~\cite{2018FangBiApJ}.
After the time $t_{\text{change}}$, the standard rigidity-dependent diffusion coefficient (see Eq.~\eqref{eq: diffusion_coefficient}) sets in.

We implement this by assuming a step function for the diffusion coefficient in this scenario:
\begin{equation}
    \kappa_{\mathrm{TDD}} \left(\mathcal{R}, t'\right) = \begin{cases}
        \kappa_{\text{start}} & \text{if } t' < t_{\text{change}} \\
        \kappa\left(\mathcal{R}\right) & \text{if } t' \geq t_{\text{change}}
    \end{cases} \, ,
\end{equation}
where $\kappa\left(\mathcal{R}\right)$ is defined as in Eq.~\eqref{eq: diffusion_coefficient} and $t' = t-t_{\text{esc}}$ is the time since the escape of the respective particle of rigidity $\mathcal{R}$.
The variance, $\sigma^2\left(\mathcal{R}, t; t_i \right)$, entering the Green's function (Eq.~\eqref{eq: analytic Greens function}), can easily be calculated using Eq.~\eqref{eq: variance Green's function}.

In this study, we consider the following three different scenarios:
\begin{enumerate}
    \item \textit{\BURST} scenario: Protons of all rigidities escape the source at the same time and experience a time-independent diffusion coefficient.
    \item \textit{\CREDIT} scenario: There is a time-dependent injection of protons by the source, where high-rigidity protons get injected earlier, but they experience a time-independent diffusion coefficient.
    \item \textit{\TDD} scenario: Protons of all rigidities escape the source at the same time, but experience a time-dependent diffusion coefficient.
\end{enumerate}
\subsection{Diffuse emissions}\label{sec: diffuse emissions}
Using this Green's function method, we are able to calculate the CR proton intensity at arbitrary positions in the Galaxy which allows us to obtain predictions for hadronic Galactic diffuse emissions (GDEs).
Those are produced by proton-proton ($pp$) interactions creating neutral ($\pi^{0}$) and charged pions ($\pi^{\pm}$).
The neutral pions decay into gamma rays, while the charged pions produce neutrinos~\cite{2016GaisserEngelcrpp}.
In our simulations, we will consider diffuse gamma-ray emissions only.
The results can be related to diffuse neutrino emissions~\cite{2014AhlersMurasePhRvD} as we will discuss at the end of this section.

The intensity $J$ of hadronic diffuse gamma-ray emissions in a certain direction $(l,b)$ at an energy $E$ is calculated by a line-of-sight integration from the position of the Sun at a distance of $R_{\odot} = \SI{8}{\kilo\parsec}$ away from the Galactic centre
\begin{equation}\label{eq: diffuse emissions general}
    J(l, b, E) = \sum_{m,n} \int_0^{\infty} \mathrm{d} s \ e^{- \tau\left(E, \boldsymbol{x}\right)} \ n_{\text{gas},n}(\boldsymbol{x}) \int_E^{\infty} \mathrm{d} E' \ \frac{\mathrm{d} \sigma_{m,n}}{\mathrm{d} E}(E', E) \ \Phi_{m}(\boldsymbol{x}, E') \Big|_{\boldsymbol{x} = \boldsymbol{x}(l, b, s)} \, ,
\end{equation}
where $\Phi_m$ is the CR intensity of species $m$ (in our case just protons), $(\mathrm{d} \sigma_{m,n} / \mathrm{d} E)(E', E)$ is the differential cross section for the production of gamma rays of energy $E$ by inelastic collisions of CRs of kinetic energy $E'$ with gas of species $n$, $n_{\text{gas},n}(\boldsymbol{x})$ describes the 3D distribution of Galactic gas of species $n$~\cite{2023SchweferMertschApJ,2023KaiserMaster}.
Finally, $\tau\left(E, \boldsymbol{x}\right)$ is the optical depth which captures the attenuation of gamma rays due to electron-positron pair-production.
We will discuss this in more detail in Sec.~\ref{sec : absorption}.

The most relevant target material for the production of hadronic GDEs is electrically neutral hydrogen in atomic (HI) or molecular form ($\text{H}_2$).
Other components can be included with scaling factors, assuming that heavier components of the ISM are spatially correlated to those.
We will discuss this in the next paragraph.
For the distribution of the hydrogen components, we use the reconstructions of HI and CO in the Galaxy of~\cite{2025SodingEdenhoferA&A}.
The distribution of $\text{H}_2$ can then be inferred from the CO distribution by assuming a linear relation between the $\text{H}_2$ column density and the velocity-integrated CO brightness temperature. Details can be found in Ref.~\cite{2025SodingEdenhoferA&A} where the final $\text{H}_2$ densities $n_{\mathrm{H}_2}$ are already provided.
The total hydrogen gas number density is then given by
\begin{equation}
    n_{\mathrm{H}} = n_{\mathrm{HI}} + 2 \cdot n_{\mathrm{H}_2}.
\end{equation}
The results of this reconstruction are provided on a grid which is centered on the Sun's position and employs logarithmically spaced radial positions in the angular directions given by a HEALPix~\cite{2005GorskiHivonApJ} pixelation of the different lines of sight.
The provided resolution corresponds to an $N_{\text{side}} = 64$, which equals $49\,152$ grid points for each of the in total $388$ radial distances.
We adopt this grid for the calculation of the diffuse emissions as it is especially suited for the calculation of the line-of-sight integral and captures the angular resolution of the structures in the Galactic gas best.
In total, eight different samples of the Galactic gas are provided in Ref.~\cite{2025SodingEdenhoferA&A} to represent the uncertainty of the reconstruction.
We chose to present the results obtained with the first realisation.

As mentioned above, gamma-ray emissions are not only produced through the interaction of CR protons with Galactic hydrogen gas, but do also include both heavier components of the CR intensity and the ISM.
We chose to include these components in an approximate way by assuming that the heavier components of the ISM follow the spatial distribution of the hydrogen component and that heavier CRs, which are a subdominant component of the total CR intensity, follow the CR proton intensities with respective scaling factors.
For the scaling factors, we follow the procedure in Refs.~\cite{2015CasandjianApJ} and~\cite{2004HondaKajitaPhRvD}, where the proton-proton cross section $\sigma_{pp}$ is rescaled to not only include the $pp$-interactions, but also those of CR protons with the heavier components of the ISM and heavier CRs with the complete ISM (including hydrogen, helium, and heavier components).
We can rewrite Eq.~\eqref{eq: diffuse emissions general} to
\begin{equation}\label{eq: diffuse emissions emissivity}
    J\left(l, b, E\right) = \frac{1}{4 \pi} \int_0^{\infty} \mathrm{d} s \ e^{- \tau\left(E, \boldsymbol{x}\left(l, b, s\right)\right)} \ \epsilon\left(\boldsymbol{x}\left(l, b, s\right), E\right) \, .
\end{equation}
The emissivity $\epsilon$ is then given by
\begin{equation}\label{eq: emissivity}
    \epsilon\left(\boldsymbol{x}(l, b, s), E\right) = 4 \pi \ n_{\mathrm{H}}(\boldsymbol{x}) \int_E^{\infty} \mathrm{d} E' \ \frac{\mathrm{d} \sigma}{\mathrm{d} E}\left(E', E\right) \ \Phi\left(\boldsymbol{x}, E'\right) \, ,
\end{equation}
where $\Phi$ is the CR proton intensity as described in Sec.~\ref{sec: Transport model} and the cross section depends on the p-p and the p-He cross sections as
\begin{equation}\label{eq: cross sections}
    \frac{\mathrm{d} \sigma}{\mathrm{d} E}\left(E', E\right) = \left(1+\xi\right) \frac{\mathrm{d} \sigma_{\mathrm{pp}}}{\mathrm{d} E}\left(E', E\right) + \frac{n_{\mathrm{He}}}{n_{\mathrm{H}}} \frac{\mathrm{d} \sigma_{\mathrm{p},\mathrm{He}}}{\mathrm{d} E}\left(E', E\right) \, .
\end{equation}
We assume that $\frac{n_{\mathrm{He}}}{n_{\mathrm{H}}} = 0.096$~\cite{2015CasandjianApJ}. The scaling factor $\xi = \xi_{\mathrm{p},\mathrm{heavy}} + \xi_{\mathrm{He+heavier},\mathrm{ISM}} = 0.418$ accounts for interactions of CR protons with components of the ISM heavier than helium, and for interactions of heavier CR species with the complete ISM.
According to page 10 in~\cite{2015CasandjianApJ}, $\xi_{\mathrm{p},\mathrm{heavy}} = 0.021$ and $\xi_{\mathrm{He+heavier},\mathrm{ISM}} = 0.397$, where we assumed that the CR helium intensity follows the CR proton one with a ratio of $0.055$.
For the $pp$ and $pHe$ cross sections, we used the \textsc{AAfrag} cross sections~\cite{2019KachelriessMoskalenkoCoPhC} which are provided by a python package \textsc{aafragpy}~\cite{2021KoldobskiyKachelriessPhRvD}.

To obtain the intensity of hadronic GDEs, $J$, according to Eqs.~\eqref{eq: emissivity} and~\eqref{eq: diffuse emissions emissivity}, we need to calculate the CR proton intensity $\Phi$ on the spatial grid described above for a list of kinetic energies $E'$ that contribute to the production of gamma rays of energy $E$.
For the full-sky GDE simulations, we chose to calculate diffuse emissions at $12$ energies around \SI{10}{\giga\electronvolt} (in the range $\left[5, 16\right]$~\si{\giga\electronvolt}) and at $12$ energies around \SI{100}{\tera\electronvolt} (in the range $\left[40, 110\right]$~\si{\tera\electronvolt}).
For this, we evaluated the CR proton intensity at $48$ kinetic energies for both ranges, respectively.
Both the kinetic energy integral in Eq.~\eqref{eq: emissivity} and the line-of-sight integral in~\eqref{eq: diffuse emissions emissivity} can be performed in a series of matrix multiplications.

The calculation of the GDEs includes two main steps: the calculation of the CR proton intensity on the grid and the calculation of the line-of-sight integral.
Depending on the energy range, the CR intensity at a certain position in the Galaxy can have contributions from tens of millions of sources.
All of those have to be added up in an efficient way.
Also the matrix multiplications involved in the line-of-sight integrations require the efficient manipulation of large arrays.
This is computationally an \textit{embarrassingly} parallel problem which we approach by using the computational power of suitable accelerators.
Specifically, we use graphical processing units (GPUs) with the help of the \textsc{python} package \textsc{jax}~\cite{jax2018github} which links the power of array programming~\cite{2020HarrisMillmanNatur} with GPU acceleration.
This helps us speed up the computations significantly by more than an order of magnitude compared to using classical CPUs\footnote{We compare parallelisation on $24$ CPUs to the use of a single H100 NVIDIA GPU on the RWTH HPC cluster CLAIX.} and allows us to study the effects of different source realisations in a Monte Carlo simulation.

To further increase the statistical power of our analysis, we have limited the age range of sources that have to be recalculated in the Monte Carlo simulation and split the remaining sources in a young and an old contribution that we evaluate on the fine spatial grid described above and a coarser spatial grid, respectively.
CRs that have been injected a sufficiently long time ago will contribute to a background CR intensity that does almost not vary across the ensemble of different source realisations as the exact source positions do not matter as much after their injected CRs diffused through the Galaxy for a sufficiently long time.
Depending on the rigidity, we set a maximum threshold on ages that have to be considered for the foreground component: $t_{\mathrm{bkg}}\left(\mathcal{R}\right) = \min \left[ \SI{20}{\mega\year},  5 \ \kappa\left(\mathcal{\SI{1}{\tera\volt}}\right) / \kappa\left(\mathcal{R}\right) \right]$.
The maximum time considered for the background contribution can be chosen as $t_{\mathrm{max}}\left(\mathcal{R}\right) = 5 \ \frac{H^2}{\kappa\left(\mathcal{R}\right)}$, i.e., about $5$ times the time it takes for the diffusion length to be comparable to the halo height.
We have checked that this split does not lead to artefacts in our simulations.

As is well-known, the gamma-ray emissions from $pp$-interactions as discussed in this section can be related to the respective neutrino emissions.
In $pp$-interactions, the ratio of charged to neutral pions is about $K=N_{\pi^{\pm}}/N_{\pi^{0}} \simeq 2$.
Neutral pions decay directly into gamma rays, $\pi^0 \rightarrow \gamma\gamma$, where each gamma ray takes half of the pion energy.
However, the three neutrinos produced in the decay chain $\pi^+ \rightarrow \mu^+\nu_{\mu} $ and $ \mu^+ \rightarrow e^+\nu_e\bar{\nu}_{\mu} $ take only a quarter of the pion energy~\cite{2016GaisserEngelcrpp,2014AhlersMurasePhRvD}.
With these considerations, the GDE intensities for gamma rays and neutrinos can be related.
As absorption only influences gamma rays and not neutrinos, we consider the GDE intensities for gamma rays without absorption, $J_{\gamma}^{\text{no abs}}$, and the GDE intensities for gamma rays and neutrinos, $J_{\nu}$:
\begin{equation}\label{eq:neutrino intensity connection}
    \left.E_{\gamma} J_{\gamma}^{\text{no abs}}\left(E_{\gamma}\right) \approx \frac{1}{3} \sum_{\nu_{\alpha}} E_{\nu} J_{\nu_{\alpha}}\left(E_{\nu}\right) \right|_{E_{\gamma} \approx 2 E_{\nu}} \, ,
\end{equation}
where the sum runs over the neutrino flavours~\cite{2014AhlersMurasePhRvD}.
This allows us to relate the results in this work to diffuse neutrino emissions.
\subsection{Absorption}\label{sec : absorption}
Up to energies of some tens of teraelectronvolts, gamma rays can travel through the Galaxy almost without any obstruction.
However, at higher energies, absorption due to interactions with interstellar radiation fields becomes relevant and limits the range of gamma rays to scales relevant for Galactic transport.
The mechanism at play is the creation of electron-positron pairs in photon-photon interactions $\gamma \gamma \rightarrow e^+ e^-$~\cite{2021DundovicEvoliA&A}.

The relevant radiation fields are the cosmic microwave background (CMB) and infrared emissions from dust heated by starlight.
To include the absorption effects into our calculations of the GDEs according to Eq.~\eqref{eq: diffuse emissions general}, we need the optical depth $\tau\left(E,\mathbf{x}\left(l,b,s\right)\right)$.
In App.~\ref{app:Absorption coefficients}, we discuss in more detail, how those can be obtained from the radiation fields.
For an isotropic homogeneous target radiation field as the CMB, the optical depth is just a function of the gamma-ray energy and the distance along the line of sight.

However, the dust infrared radiation field has a non-trivial distribution in energy, space, and angle.
For our work, we use the precomputed absorption coefficients provided in \textsc{GALPROP v.57}~\cite{2022PorterJohannessonApJS} whose calculation is described in Ref.~\cite{2018PorterRowellPhRvD}.
Two models for the dust component are presented in that work.
We chose the model based on Ref.~\cite{2012RobitailleChurchwellA&A} (R12 in Ref.~\cite{2018PorterRowellPhRvD}) as our fiducial absorption model.
As these absorption coefficients were obtained under the assumption of a Galactic centre distance of \SI{8.5}{\kilo\parsec}, while we are assuming $R_{\odot} = \SI{8}{\kilo\parsec}$, we rescale the results from Ref.~\cite{2018PorterRowellPhRvD} accordingly, assuming the same total energy in infrared radiation in the Galaxy.
The spectral number density of the radiation field produced by the dust emissions is thus scaling by a factor of $\left(\frac{8.5}{8}\right)^3$ while the path length element in Eq.~\eqref{eq: absorption coefficient} scales as $\left(\frac{8}{8.5}\right)$.
We interpolate the results to the spatial and energy grid described in Sec.~\ref{sec: diffuse emissions} and use it to calculate the hadronic GDEs after absorption.

Other components of the interstellar radiation field like the extragalactic background light and radiation emitted by stars directly are subdominant and can be neglected in this work~\cite{2016VernettoLipariPhRvD}.
At sub-teraelectronvolt gamma-ray energies, attenuation due to pair-production can be neglected altogether.
\subsection{Stochastic vs. smooth models}\label{sec:stochastic vs smooth models}
In this section, we demonstrate how we can obtain the GDE intensity predictions of a smooth source model.
By comparing the GDE predictions in a discrete source model to those, it is possible to pin down the influence of discrete source models (see overview in Fig.~\ref{fig:overview}).
We will also point out how imprints of discrete sources on the GDE sky maps can be understood qualitatively.

We can compare the source term in the CR transport equation~\eqref{eq:transport_equation} in a discrete and in a smooth source model:
\begin{equation}
    Q\left(\mathcal{R}, t, \mathbf{x}\right) = \begin{cases}
        \sum_{i=1}^N Q_{\mathcal{R}}(\mathcal{R}) \ \delta\left(t - t_{\text{esc},i}\left(\mathcal{R}\right)\right) \delta\left(\mathbf{x} - \mathbf{x}_i\right) &\text{(discrete)}\\
        Q_{\mathcal{R}}(\mathcal{R}) \ \rho \left(t, \mathbf{x}\right) &\text{(smooth)}
    \end{cases} \, .
\end{equation}
In the discrete source model, we add $N$ sources whose ages $t_i$ and positions $\mathbf{x}_i$ are drawn from a source distribution function with density $\rho \left(t, \mathbf{x}\right)$.
In our case, this describes a uniform distribution of source ages and a spatial distribution of the spiral model described in Sec.~\ref{sec: Transport model}.

We already described how to solve the transport equation in the discrete case with the Green's function method.
The transport equation with a smooth source term can be solved by convolving the Green's function~\eqref{eq: analytic Greens function} with the source density.
This is exactly the same as the mean of the ensemble of all possible source realisations in the discrete source model:
\begin{equation}\label{eq:smooth convolution ensemble mean}
    \psi_{\mathcal{R}}^{\mathrm{smooth}}\left(\mathcal{R}, t, \mathbf{x}\right) = \iint \mathrm{d} t' \mathrm{d} \mathbf{x'} \ G\left(t, \mathbf{x}; t', \mathbf{x'}\right) \rho \left(t', \mathbf{x'}\right) = \bigl\langle \psi_{\mathcal{R}}\left(\mathcal{R}, t, \mathbf{x}\right) \bigl\rangle \, ,
\end{equation}
where $\langle \cdot \rangle$ denotes the ensemble mean over all possible source realisations.
The ensemble mean can be approximated by calculating the average of some samples of source realisations.
This approach is comparable to using Monte Carlo integration for performing the integral in Eq.~\eqref{eq:smooth convolution ensemble mean}.

We use this to obtain the smooth model predictions for the GDEs from our Monte Carlo simulations.
The GDEs depend linearly on the CR intensities as described in Sec.~\ref{sec: diffuse emissions}.
Thus, the GDE intensities for the smooth source model can be approximated by the sample mean of the discrete model predictions:
\begin{equation}\label{eq:smooth approx intensity}
    J^{\mathrm{smooth}} = \langle J \rangle \approx \bar{J} \, ,
\end{equation}
where the bar denotes the sample mean.
The goodness of the approximation of the ensemble mean by the sample average can be quantified by the relative standard error of the sample mean.
We find that this is much lower in the \burst{} and \credit{} scenario than in the \tdd{} scenario.
For the \burst{} scenario, the relative standard error of the sample mean when averaging $1\,000$ realisations of the sky maps is below \SI{0.1}{\percent} for \SI{99}{\percent} of lines of sight at a GDE energy of \SI{10}{\giga\electronvolt} and below \SI{0.22}{\percent} at \SI{100}{\tera\electronvolt}.
Thus, in the rest of this work, we will use the sample mean in the \burst{} scenario as an approximation for the smooth model predictions.

\begin{figure}  
     \centering
    \includegraphics[width=\textwidth]{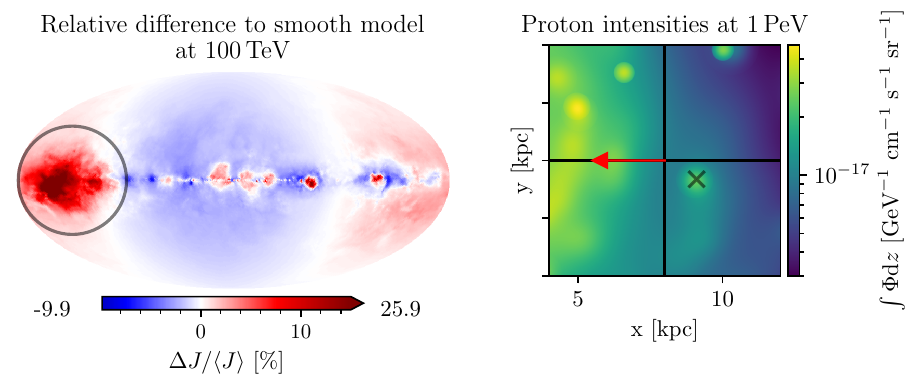}
    \caption{For one specific realisation in the \burst{} scenario, we show the relative difference of GDEs from the smooth model predictions (left) and the proton intensities at \SI{1}{\peta\electronvolt} around the position of the Sun integrated along the $z$-direction (right).
    The red arrow points towards the Galactic centre, which is in the centre of the sky map on the left.
    The region marked in the sky map by a grey circle can be explained by the excess proton intensities marked by the crosses in the figure on the right.
    }
    \label{fig:2}
\end{figure}
Effects of the discrete source models on the GDEs can be understood best by comparison to the smooth source model GDE predictions.
The production mechanism of GDEs described in Sec.~\ref{sec: diffuse emissions} shows that variations of GDEs can be caused by variations in the CR intensity and are modulated by the Galactic gas density in the respective region of the Galactic volume.
Additionally, contributions from multiple source regions along a specific line of sight accumulate.
In Fig.~\ref{fig:2}, we show how regions of enhanced GDE emission can be matched to regions of an increased CR proton intensity in the same direction.
To demonstrate this, we show the sky map of the relative difference to the smooth model predictions at \SI{100}{\tera\electronvolt} for one realisation in the \burst{} scenario on the left.
Multiple regions with enhanced emissions can be seen along the Galactic plane.

As an example, we focus on the most extended region marked with a grey circle and try to understand the cause of this excess emission by looking at a projection of the CR proton intensities at \SI{1}{\peta\electronvolt} on the right side of Fig.~\ref{fig:2}.
Given the typical inelasticity of pion production, the CR proton intensity at this energy is a main contributor to the GDEs at \SI{100}{\tera\electronvolt}.
We integrate the intensities along the vertical direction to also capture enhanced CR proton intensities at $z \neq 0$ in this 2D projection.
The increased GDEs in the marked region can be explained by the stronger proton intensities around the position indicated by the grey cross.
The morphology of the enhanced CR proton intensities identifies this as the imprint of a relatively young source situated in this region.

It should be noted that the excess GDEs show much more structure than the excess in the CR proton intensities, which can be attributed to the structure in the Galactic gas distribution.
This hints to complications with classifying excess regions with sources in general.
Both the extent and the maximum relative difference of an excess region depend on the age and the distance of the contributing SNRs, as well as the Galactic gas density in the near-source region, in a degenerate way.
Drawing conclusions about source parameters from specific excess regions would require more sophisticated morphological studies and will not be discussed in this work.
\subsection{Cosmic rays and stable law distributions}\label{sec:Cosmic rays and stable law distributions}
Discrete sources influence the GDEs through their modification of the CR intensity in their vicinity.
Considering the source positions and ages as random variables with given probability distribution functions renders the CR intensities into stochastic quantities as well.
We are interested in their probability distribution functions and how they influence the GDE distribution functions.
The CR intensity distribution in a \burst{} model has been studied in Ref.~\cite{2017GenoliniSalatiA&A}.
We summarise these results and discuss the correspondence between the CR and GDE intensity distributions.
The GDE intensity distributions will be shown to either follow so-called stable laws or, depending on the line of sight, a mixture of normal distributions and stable laws.
\paragraph{Stable laws for CR intensity distributions}
Stable distributions, also known as stable laws, are an important tool to describe the distributions of CR intensities.
First discussed by Paul Lévy~\cite{1925Levy}, stable laws are useful in many fields from finance to astrophysics~\cite{1999UchaikinZolotarev}.
We will use the notation and results about univariate stable distributions from Ref.~\cite{2020Nolan}.
A stable distribution $S\left(\alpha, \beta, \gamma, \delta\right)$ (in the notation $1$ according to Ref.~\cite{2020Nolan}) is described by $4$ parameters.
$\alpha \in \left(0,2\right]$ is the index of stability that determines the rate at which tail probabilities decay.\footnote{By that, we mean that if $X$'s probability distribution follows a stable law with a certain $\alpha$, the cumulative distribution function $C$ fulfils $\lim_{X\rightarrow\infty} X^{\alpha} C(X) = c > 0$ with a constant $c$.}
$\beta \in \left[-1,1\right]$ describes the skewness of the tails, where $\beta = 1$ describes a tail solely in the positive direction.
$\gamma$ is the scale parameter and $\delta$ the location parameter.\footnote{The mean directly corresponds to the location parameter $\delta$ for $\alpha > 1$, but it is not defined otherwise.
The standard deviation $\sigma$ is only defined for $\alpha=2$, where its relation to the scale parameter is given by $\sigma = \sqrt{2}\gamma$.}
This family of distributions includes, among many others, the normal distribution ($\alpha = 2$).
For $\alpha < 2$ however, power-law tails emerge in the distributions and the variance of the distributions diverges.

In Ref.~\cite{2017GenoliniSalatiA&A}, distributions of CR intensities in a \burst{} discrete source model have been studied.
It has been shown that they are well-described by stable law distributions that exhibit such power-law tails.
In this derivation, it is assumed that CR intensity contributions from individual sources diverge if the source gets arbitrarily close and young (see Eq.~\eqref{eq: analytic Greens function}).
The distribution of the total CR intensity is the sum of contributions from many individual sources which can be studied with the central limit theorem.
But because of the diverging variance of the distributions of CR intensity contributions from individual sources, the standard central limit theorem is not applicable.
Instead, a generalised version of central limit theorem (see, e.g.,~\cite{2020Nolan}) for stable laws has to be used.
Two idealised cases of a uniform distribution of CR sources in a disk around the observer (2D) or in a thin box (3D) are contrasted in Ref.~\cite{2017GenoliniSalatiA&A}.
The resulting probability density function $P(\Phi)$ of the CR intensities $\Phi$ are described by stable laws
\begin{equation}
    P(\Phi) \sim S\left(\alpha, \beta = 1, \gamma = \sigma_N, \delta = \bar{\Phi}\right) \, ,
\end{equation}
where $\alpha = 4/3$ in the 2D case and $\alpha = 5/3$ in the 3D case.
The location parameter is given by the expectation value of the CR intensity $\bar{\Phi}$ and the scale parameter $\sigma_N$ depends on the source distribution and transport parameters~\cite{2017GenoliniSalatiA&A}.

The inclusion of causality or a finite source extent as discussed in Sec.~\ref{sec: Transport model} effectively limits the contributions from individual CR sources.
Without the variance of the single source CR intensity distribution actually diverging, the standard central limit theorem should be applicable for the total CR intensity.
However, the distributions of CR intensities in the Galactic disk are often still better described by stable laws, even if intensity limitations like the causality criterion (see Sec.~\ref{sec: Transport model}) are used~\cite{2017GenoliniSalatiA&A}.
A normal distribution would only be approached in the limit of an infinite source rate.
Still, the implementation of a causality criterion like discussed in Sec.~\ref{sec: Transport model} alters the high-intensity part of the distributions.
For high CR energies, this can lead to deviations of the CR intensity distributions obtained in simulations from the stable law predictions.

The choice of the distribution best suited for the description of the CR intensities at a specific point in the Galaxy can be made based on considerations of the sources surrounding it and the transport parameters.
In Ref.~\cite{2017GenoliniSalatiA&A}, a position in the centre of the Galactic disk has been studied.
It has been argued that the 2D case with $\alpha = 4/3$ generally provides a better description for intensities close to the mean intensity, while the 3D case with $\alpha = 5/3$ is more suitable for more extreme high intensities.
This is because the high intensities are dominated by the contribution from a single close-by source, for which the 3D distribution of source positions can be expected to have a clear imprint.
Less extreme CR intensity realisations receive their dominating intensity contributions from sources farther away, for which the flatness of the Galactic disk and thus the 2D distribution of source positions yields a better description.
For a more detailed discussion of the intensity value when one description is better than the other, we refer to Ref.~\cite{2017GenoliniSalatiA&A}.

For the diffuse emissions, we are not solely interested in the CR intensities at positions where there are many sources, but also in CR intensities at positions outside of the Galactic disk.
CR intensities there are much less likely to receive high contributions from sources in their vicinity.
Thus, the standard central limit theorem predicts that the total CR intensity, being the sum of many contributions from individual sources, should be distributed according to a normal distribution.
In the transition from positions in the Galactic disk towards positions farther away from it, we expect a transition from intensity distributions better described by stable laws to those that are better described by normal distributions.
\paragraph{Stable laws for GDE intensity distributions}
Now, we want to use this understanding of CR intensity distributions at different positions in the Galaxy to make informed predictions about the distributions of GDEs.
To understand the distribution of the GDEs, we approximate the diffuse intensity $J\left(l,b,E\right)$ described in Eq.~\eqref{eq: diffuse emissions general}.
We only consider contributions to the intensity from one CR energy $E'$ and approximate the line-of-sight integral as a sum of contributions of $M$ voxels~(i.e., volume pixels):
\begin{equation}\label{eq: LOS approx voxel}
    J\left(l,b,E\right) \simeq \sum_{i=1}^M n_{\text{gas}}\left(\boldsymbol{x_i}\right) \ \frac{\mathrm{d} \sigma}{\mathrm{d} E}(E', E) \ \Phi\left(\boldsymbol{x_i}, E'\right) = \sum_{i=1}^M n_{\text{gas}, i} \ \sigma \ \Phi_i \, ,
\end{equation}
with $\sigma$, $n_{\text{gas}, i}$, and $\Phi_i$ as shorthand notations for the differential cross section as well as the gas densities and the CR intensity in voxel $i$, respectively.

We can use two properties of random variables that are distributed according to stable laws.
First, if $X \sim S\left(\alpha, \beta, \gamma, \delta\right)$, then $aX \sim S\left(\alpha, \mathrm{sign}(a)\beta, |a| \gamma, a\delta\right)$ for $a \neq 0$, and second, if $X_i \sim S\left(\alpha, \beta, \gamma_i, \delta_i\right)$ and are independent, then $\sum_i X_i \sim S\left(\alpha, \beta, \left( \sum_i \gamma_i^{\alpha}\right)^{1/\alpha}, \sum_i \delta_i\right)$.
Mind, that the second property only holds if the index of stability is the same for the distributions of all random variables.

We can use this to make a prediction about the distribution of the GDEs on different lines of sight (LOS).
First, we consider LOS which pass mainly through the Galactic disk, where we can expect that all underlying CR intensity distributions follow stable laws.
We assume that the voxels in the approximation (Eq.~\ref{eq: LOS approx voxel}) are spaced sufficiently far apart such that the CR intensities are statistically independent from each other and that the CR intensity distributions in all voxels can be described by stable laws with the same $\alpha$ of either $4/3$ or $5/3$.
From the properties of stable laws, we conclude that the intensity for such LOS should be distributed as
\begin{equation}\label{eq: approx fit plane}
    P\left(J\left(l,b,E\right)\right) \simeq S\left(\alpha, 1, \gamma_{\text{tot}}, \bar{J}\right) \quad \alpha \in \{4/3, 5/3\} \, ,
\end{equation}
where $\gamma_{\text{tot}}$ is the scale parameter of the distribution and $\bar{J}$ is the mean GDE intensity at this energy that is the sum of the contributions from all voxels.

The scale parameter $\gamma_{\text{tot}}$ receives contributions from different voxels along the LOS as well as from CRs of different energies.\footnote{Mind, that the property about the sums of random variables distributed following a stable law cannot be applied here as intensities of CRs at different energies are not independent.
However, it still holds that the scale parameter increases with contributions from different energies.}
We do not attempt to determine this parameter a priori.
Instead, we are going to fit this parameter by fitting a stable law distribution to our simulation data.
For that, we use a quantile estimation method~\cite{1986McCulloch} that is implemented in \textsc{SciPy}~\cite{2020VirtanenGommersNatMe} while fixing the other parameters of the stable distribution according to Eq.~\eqref{eq: approx fit plane}.

Now, we make a prediction for the GDE distribution along a LOS out of the Galactic plane.
According to the approximation in Eq.~\eqref{eq: LOS approx voxel}, the GDE intensity gets contributions from multiple voxels.
Only few of them contain nearby CR sources and can be expected to add GDE intensity contributions that are distributed following a stable law.
There will also be voxels farther out of the disk, where there are rarely any CR sources and CR intensity contributions are better described by normal distributions.
This will result in a mixture of contributions of a stable law and a normal distribution to the total GDE distribution.
To capture this effect, we predict that a fraction $f$ of the contributions to the GDEs along this LOS are determined by CR intensities that follow a stable law and $\left(1-f\right)$ by those that follow a normal distribution $\mathcal{N}\left(\mu,\sigma\right)$.\footnote{As usual, this is has the probability density function $\mathcal{N}\left(\mu,\sigma\right) = \frac{1}{\sqrt{2\pi\sigma^2}}e^{-\frac{(x-\mu)^2}{2\sigma^2}}$.}
The distribution of the total GDEs is thus the sum of two random variables.
As those have different indices of stability, we have to use the general convolution formula to obtain the distribution of their sum
\begin{equation}\label{eq: approx fit UP}
     P\left(J\left(l,b,E\right)\right) \simeq S\left(\alpha, 1, f^{1/\alpha} \cdot \gamma_{\text{tot}}, f \cdot \bar{J}\right) \ast \mathcal{N}\left(\left(1-f\right) \cdot \bar{J}, \left(1-f\right)^{1/2} \cdot \sigma_{\text{sim}}\right) \, ,
\end{equation}
where $\gamma_{\text{tot}}$ is as in the Galactic centre LOS and $\sigma_{\text{sim}}$ is the sample standard deviation.
While weighing the mean with the factors $f$ and $\left(1-f\right)$, respectively, is straightforward, the weighing of the scale parameters is not.
We chose the weighing presented in Eq.~\eqref{eq: approx fit UP} based on the properties presented for sums of random variables with stable-law distributions.

We will compare our predictions with the Monte Carlo simulation results in Sec.~\ref{sec:High-statistics assessment of diffuse emission distributions}.
Deviation of the actual distributions from a perfect stable law will be expected due to the implementation of a causality criterion and our simplified assumptions.
It can also be expected that a description of GDE distributions by stable laws or mixtures of stable laws and normal distributions does not work equally well in the different source injection and near-source transport scenarios.
\section{Results and Discussion}
To study the effects of source stochasticity on the Galactic diffuse emissions (GDEs), we ran extensive Monte Carlo simulations where we sampled the CR source coordinates from a source distribution.
These samples include $1\,000$ realisations of the GDE predictions for the whole sky at various different gamma-ray energies.
This allows us to understand the morphology of the GDEs and link fluctuations in the simulated GDEs to individual sources as pointed out in Sec.~\ref{sec:stochastic vs smooth models}.
Furthermore, we ran $10^6$ realisations along selected lines of sight to study the distribution of the GDE intensities and test predictions for their distributions as described in Sec.~\ref{sec:Cosmic rays and stable law distributions}.
We further investigated the GDEs in selected sky windows used by experimental collaborations over a broader energy range to quantify the systematic modelling uncertainties coming from discrete sources by simulating $1\,000$ realisations for each sky window.
We also studied longitude profiles at selected gamma-ray energies.

All of these analyses have been done for each of the three source injection and near-source transport scenarios introduced in Sec.~\ref{sec: source injection and near-source transport}: the \burst{}, the \credit{}, and the \tdd{} scenario.
In the following sections, we will present the results of these analyses and discuss the influence of discrete CR sources on GDEs and what can be learnt from that.
\subsection{First glimpse into the diffuse sky}\label{sec: A glimpse into the diffuse sky}
\begin{figure}  
    \centering
    \includegraphics[width=\textwidth]{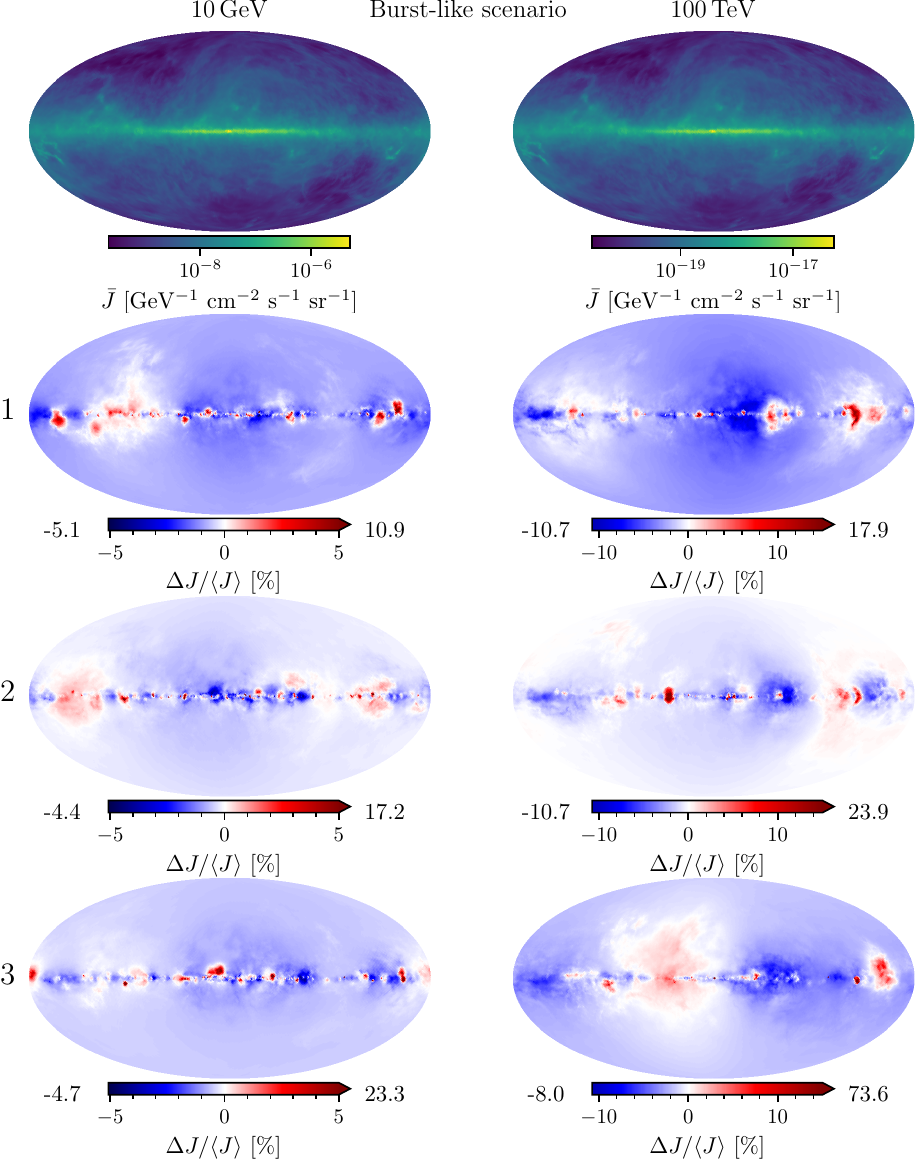}
    \caption{\BURST{} scenario.
    Sky maps at two different energies are shown (\SI{10}{\giga\electronvolt} left and \SI{100}{\tera\electronvolt} right).
    The first row shows the sample means over the $1\,000$ realisations.
    The remaining rows show the relative deviations of three of those source realisations from this mean, used as an approximation for the smooth model prediction (Eq.~\eqref{eq:smooth approx intensity}).
    The colour bars are symmetric and saturate at $ \pm \SI{5}{\percent}$ (\SI{10}{\giga\electronvolt}) and $ \pm \SI{15}{\percent}$ (\SI{100}{\tera\electronvolt}).
    The numbers on either end of the colour bars indicate the maximum relative deviations for each sky map.
    }
    \label{fig:1a}
\end{figure}

\begin{figure}  
    \centering
    \includegraphics[width=\textwidth]{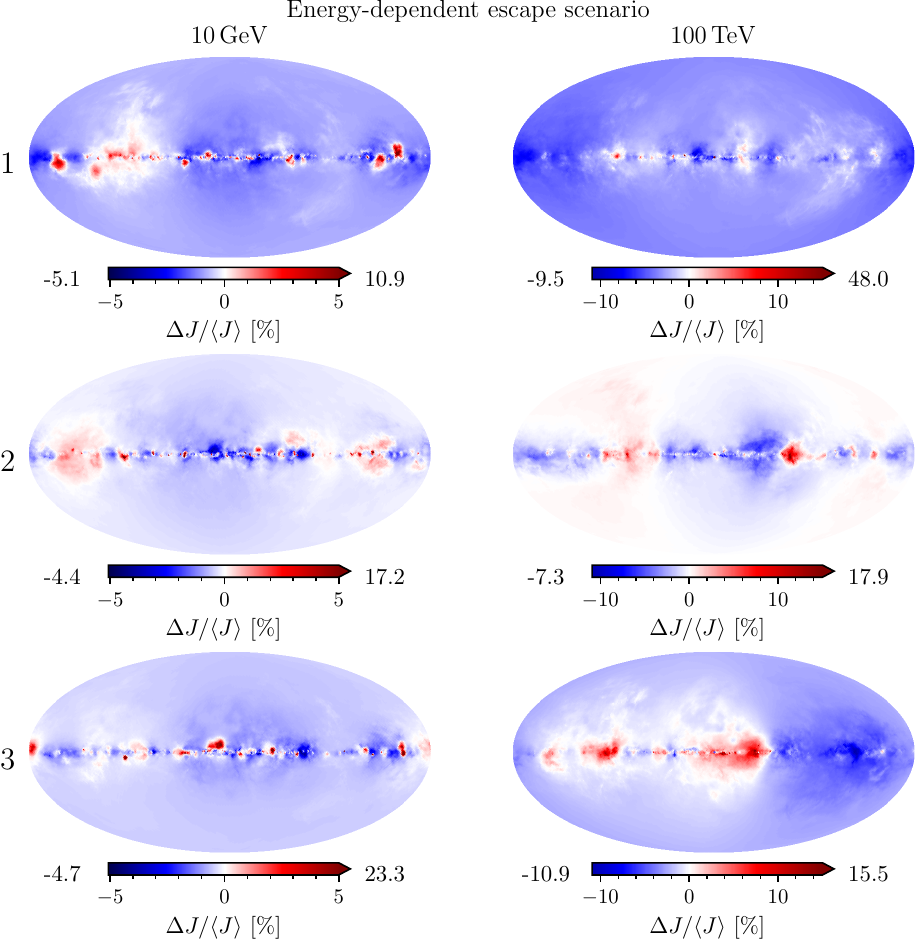}
    \caption{\CREDIT{} scenario.
    This panel shows the relative differences in the \credit{} scenario for the same three source configurations as in Fig.~\ref{fig:1a}.
    }
    \label{fig:1b}
\end{figure}

\begin{figure}[t]  
    \centering
    \includegraphics[width=\textwidth]{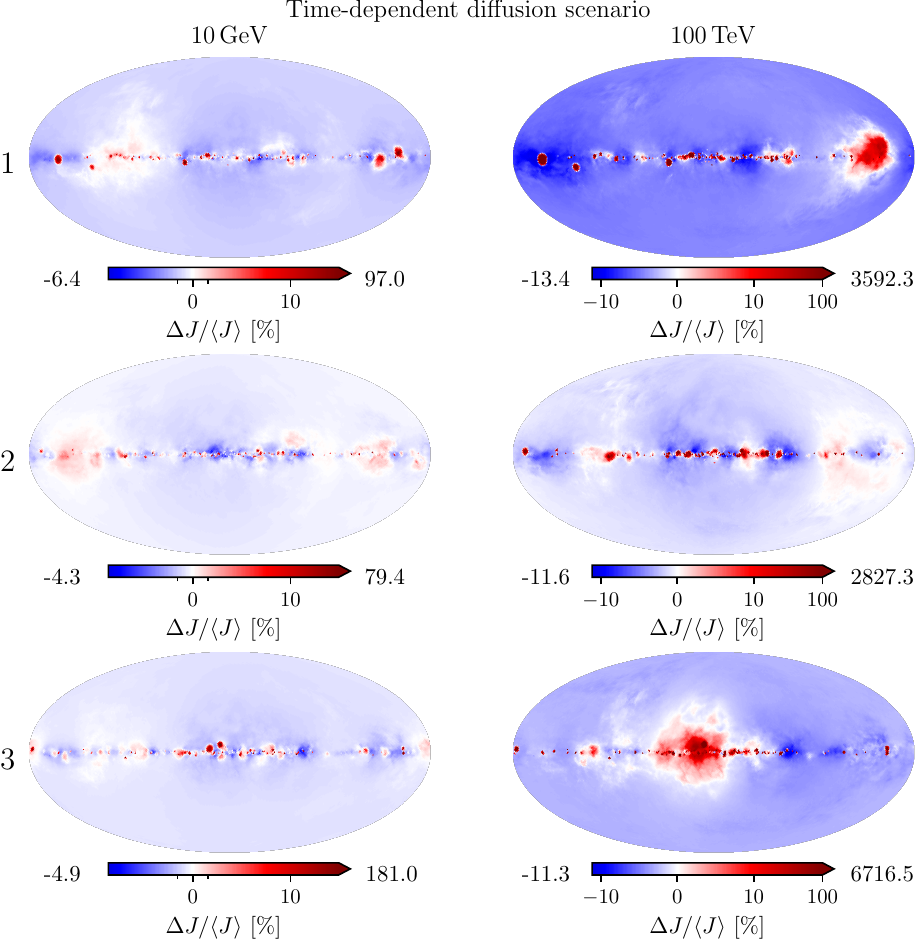}
    \caption{\TDD{} scenario.
    This panel shows the relative differences in the \tdd{} scenario for the same three source configurations as in Fig.~\ref{fig:1a}
    The confinement around the sources leads to more extreme deviations.
    The colour bars are symmetric and saturate at $ \pm \SI{50}{\percent}$ (\SI{10}{\giga\electronvolt}) and $ \pm \SI{100}{\percent}$ (\SI{100}{\tera\electronvolt}).
    They are linear up to $ \pm \SI{5}{\percent}$ (\SI{10}{\giga\electronvolt}) and $ \pm \SI{10}{\percent}$ (\SI{100}{\tera\electronvolt}) and logarithmic beyond that.
    }
    \label{fig:1c}
\end{figure}

With the procedures described in Sec.~\ref{sec: method}, we can calculate the GDEs at a range of energies for different source configurations in varying injection and near-source transport scenarios.
The results can be displayed as sky maps at fixed energies.
We used the HEALPix pixelation scheme~\cite{2005GorskiHivonApJ} in our calculations and to visualise the GDE predictions.
The panels in Figs.~\ref{fig:1a}, \ref{fig:1b}, and~\ref{fig:1c} show such HEALPix sky maps for the various scenarios at two different energies, \SI{10}{\giga\electronvolt} (left) and \SI{100}{\tera\electronvolt} (right).
\paragraph{\BURST{} scenario}
The first row in Fig.~\ref{fig:1a} shows the sample mean, $\bar{J}$, of the GDEs over the $1\,000$ realisations at the two energies.
In section~\ref{sec:stochastic vs smooth models}, we have explained that this represents the GDE predictions of a smooth source model.
The colour bar is chosen to be logarithmic to highlight the large dynamical range of the intensities from the Galactic centre in the centre of the sky map to the intensities in the lines of sight towards outer regions of the Galaxy and out of the Galactic disk.
The emissions mostly follow the underlying gas structure in the Galaxy (see~\cite{2025SodingEdenhoferA&A} for a gas sky map).
The main difference is the total intensity normalisation, as the CR proton intensity is much larger at lower energies.

To visualise the influence of discrete sources on the GDE predictions, the remaining sky maps in Fig.~\ref{fig:1a} show the relative deviations of three of the $1\,000$ realisations from the smooth model prediction, $\Delta J/\langle J \rangle$, where $\Delta J = J - \langle J \rangle$.
Close to the Galactic plane, multiple small regions of increased intensities can be seen.
These are caused by individual sources that are close to the respective line of sight and have a strong individual contribution to the GDEs as we have already discussed in Sec.~\ref{sec:stochastic vs smooth models}.
The colour bars are chosen to be symmetric and saturate at $\pm \SI{5}{\percent}$ for the lower energy and $\pm \SI{15}{\percent}$ for the higher energy.
The maximum relative deviation for each individual sky map is given at the endpoints of the colour bars.
Although the GDEs are produced by CRs injected by the same sources at the same time for both energies, the relative difference sky maps for the same realisations show different morphologies at different energies.
This can be understood by the much larger diffusion coefficient at \SI{100}{\tera\electronvolt} compared to at \SI{10}{\giga\electronvolt}.
While morphological similarities between regions of higher intensities can be found between the sky maps at the two energies, they are usually more extended in the higher energy case (e.g., regions on the right side of realisation $1$).
\paragraph{\CREDIT{} scenario}
The morphological correspondence of the sky maps at different energies can be completely abandoned in the \credit{} scenario shown in Fig.~\ref{fig:1b}.
Here, the time-dependent injection of the CRs means that the CR sources causing excess GDEs at \SI{10}{\giga\electronvolt} are not the same as those causing excess GDEs at \SI{100}{\tera\electronvolt}.
In Fig.~\ref{fig:1b}, we show the relative differences for the same three source configurations as in Fig.~\ref{fig:1a}.
Due to our choice of the parameter $\mathcal{R}_{\text{b}} = \SI{10}{\tera\volt}$, the GDEs at \SI{10}{\giga\electronvolt} are identical to the ones in the \burst{} scenario shown in Fig.~\ref{fig:1a}.
At \SI{100}{\tera\electronvolt}, however, the GDEs in the \credit{} scenario are clearly different from the \burst{} scenario.
It can also be seen that the deviations at high energies are slightly less extreme compared to the \burst{} scenario seen in Fig.~\ref{fig:1a}.
Due to the time-dependent injection, the influence of a source is limited to a narrower energy range (see~\cite{2025StallLooApJL}).
Although this can lead to extreme CR intensity excesses, this limitation to a narrow energy range leads to a less extreme influence of CR sources on the GDEs in this scenario.
We discuss this effect in more detail in App.~\ref{app:Burst vs. CREDIT}.
\paragraph{\TDD{} scenario}
The \tdd{} scenario (Fig.~\ref{fig:1c}) shows the most extreme relative deviations from the smooth model predictions.
We use different colour bars because of this.
They are symmetric and saturate at $ \pm \SI{50}{\percent}$ (\SI{10}{\giga\electronvolt}) and $ \pm \SI{100}{\percent}$ (\SI{100}{\tera\electronvolt}).
Further, they are linear up to $ \pm \SI{5}{\percent}$ (\SI{10}{\giga\electronvolt}) and $ \pm \SI{10}{\percent}$ (\SI{100}{\tera\electronvolt}) and logarithmic beyond that.
For the lower energy, relative deviations of more than a hundred percent and for the higher energy even several thousand percent can be seen.
A morphological correspondence between the sky maps at different energies is observable.
We can also see morphological similarities between the sky maps in this scenario and the \burst{} scenario, where the same source positions and ages were used.

The confinement of the CRs around the sources leads to a very inhomogeneous distribution of the GDEs~\cite{2025KaciGiacintiJCAP}.
Diffuse emission enhancements in such small regions around CR sources blur the line between diffuse and discrete gamma-ray emissions.
It has been pointed out that the number of observed sources with LHAASO could constrain source rates and near-source transport models.
The observation of less than a certain number of enhancement regions could either point towards the \tdd{} scenario being false or the PeVatron source rate being lower.
The \tdd{} scenario with a PeVatron rate of \SI{100}{\percent} as implemented here clearly shows imprints of the near-source transport model.
A lower source rate would lead to a lower number of excess regions, but we do not extend the discussion of these scenarios here.
For a discussion of these aspects, we refer to Ref.~\cite{2025KaciGiacintiJCAP}.
\paragraph{Summary}
We presented sky maps for three realisations of GDEs at \SI{10}{\giga\electronvolt} and \SI{100}{\tera\electronvolt} in each of the three source injection and near-source transport scenarios.
We used the same source positions and ages across the three scenarios.
The morphology of the sky maps can be understood as imprints of individual CR sources as described in Sec.~\ref{sec:stochastic vs smooth models}.
Differences and morphological similarities between the source injection and near-source transport scenarios as well as different energies can be observed and understood qualitatively.
The extreme source imprints in the \tdd{} scenario can also be used to constrain this model (see~\cite{2025KaciGiacintiJCAP}).
We also briefly touched upon the maximum relative deviation that can be expected in a sky map and how they can be understood in the different source models.
We will discuss this in more detail in the following section.
\subsection{Most extreme deviations}\label{sec:The most extreme deviations}
The sky maps shown in Sec.~\ref{sec: A glimpse into the diffuse sky} indicate a wide range of relative deviations depending on the injection and near-source transport model and the specific source realisation.
In this section, we want to investigate the dynamic range of our simulation samples by determining the maximum relative deviation across all pixels of a sky map for all samples and presenting their frequency in histograms.
This allows us to understand which maximum GDE enhancements can be expected due to discrete sources.
Again, we show the results for GDE energies of \SI{10}{\giga\electronvolt} and \SI{100}{\tera\electronvolt}.
The histograms are shown on a logarithmic $x$-axis to reflect the wide range of maximum values in our simulations.
We show the data up to the $99$-percentile, thus not displaying the $10$ most extreme realisations, for visual clarity.
We also explicitly mark the mean ($\bar{X}$) and median ($\tilde{X}$) relative difference of the data in the histograms in percent.
\paragraph{\BURST{} scenario}
\begin{figure}  
     \centering
    \includegraphics[width=\textwidth]{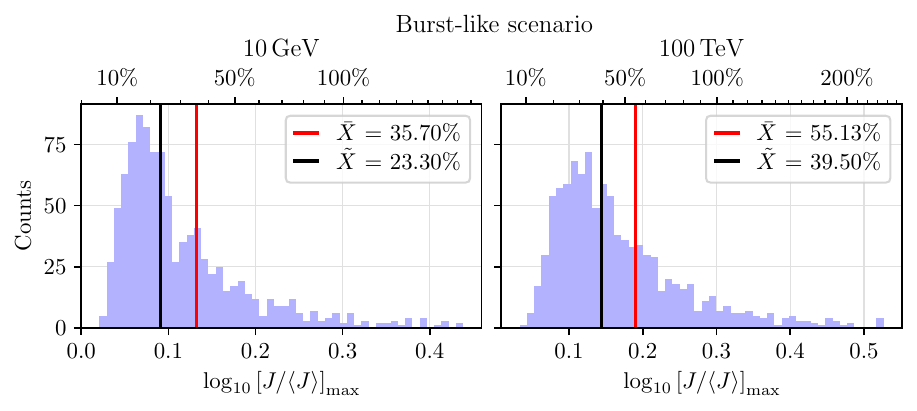}
    \caption{\BURST{} scenario.
    For each of the $1\,000$ realisations, we determine the line of sight with the most extreme relative deviation from the ensemble mean corresponding to the smooth model predictions (Sec.~\ref{sec:stochastic vs smooth models}).
    This plot shows the histograms of this analysis for \SI{10}{\giga\electronvolt} (left) and \SI{100}{\tera\electronvolt} (right) on a logarithmic $x$-axis up to the $99$-percentile.
    We also give the sample mean $\bar{X}$ and the sample median $\tilde{X}$ relative deviation in percent.
    For reference, a secondary axis is added to the top that shows the maximum relative difference $\left[\Delta J / \langle J \rangle \right]_{\text{max}}$ in percent.
    }
    \label{fig:3a}
\end{figure}
Fig.~\ref{fig:3a} shows the histograms for the \burst{} scenario.
It can be seen that the distributions are skewed, with long tails extending towards higher values for both energies.
These distributions share those features with the underlying stable law distributions of CR intensities, which we introduced in Sec.~\ref{sec:Cosmic rays and stable law distributions}.
As the deviations displayed here represent the most extreme GDE enhancements across a sky map, we can expect those histograms to point towards the influence of the CR sources that dominate most strongly in their position of the Galaxy.
Those are especially young sources.
From those histograms, we learn how large their influence can be expected to be.

Further, we observe that the relative deviations are higher for the higher GDE energy.
This can be attributed to the smaller number of CR sources effectively contributing to the CR proton intensities at higher energies where the escape time-scale is smaller than at lower energies.
Hence, a single source has a stronger relative influence on the CR proton intensities, also leading to a higher relative difference in the GDE sky map.
\paragraph{\CREDIT{} scenario}
\begin{figure}  
     \centering
    \includegraphics[width=\textwidth]{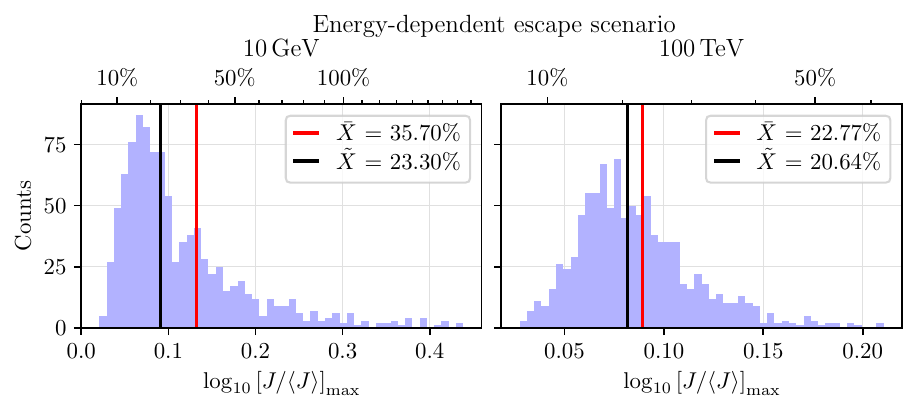}
    \caption{Same as Fig.~\ref{fig:3a}, but for the \credit{} scenario.
    }
    \label{fig:3b}
\end{figure}
We show the histograms for the \credit{} scenario in Fig~\ref{fig:3b}.
As we used the same source realisations for the \burst{}  and the \credit{} scenario, the histogram for \SI{10}{\giga\electronvolt} is identical to the one shown in the \burst{} scenario in Fig.~\ref{fig:3a}.
For a GDE energy of \SI{100}{\tera\electronvolt}, we observe a similarly skewed shape of the distribution as in the \burst{} scenario.
However, we find the deviations tend to be smaller.
As discussed in App.~\ref{app:Burst vs. CREDIT}, this is due to the limitation of the influence of individual sources to a smaller energy range.
As the diffuse emissions are produced by CR protons over a broad energy range, the \credit{} scenario effectively limits the impact of an individual source on the GDEs.
Remarkably, the deviations at \SI{100}{\tera\electronvolt} tend to be less extreme than the ones at \SI{10}{\giga\electronvolt}.
\paragraph{\TDD{} scenario}
\begin{figure}  
     \centering
    \includegraphics[width=\textwidth]{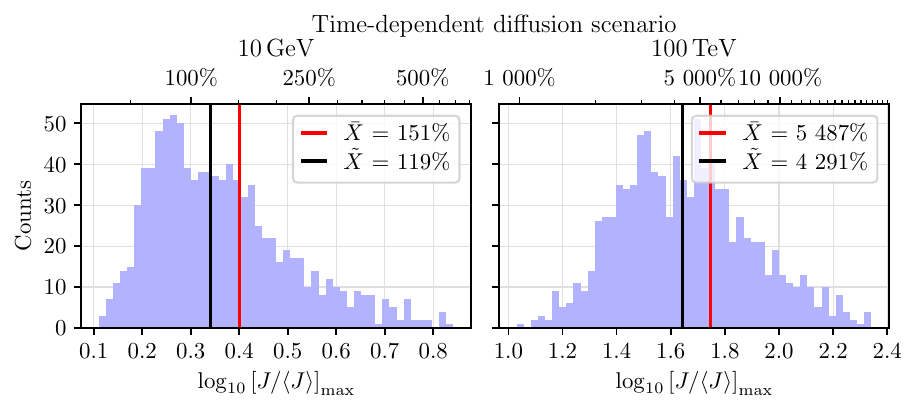}
    \caption{Same as Fig.~\ref{fig:3a}, but for the \tdd{} scenario.
    }
    \label{fig:3c}
\end{figure}
The histograms for the \tdd{} scenario displayed in Fig.~\ref{fig:3c} also show skewed distributions with long tails, but the deviations tend to be much more extreme than in the other scenarios.
For the high-energy case, we can even see that we can find relative differences of over \SI{900}{\percent} in all realisations.
The difference in the extent of the maximum relative deviations at \SI{10}{\giga\electronvolt} and \SI{100}{\tera\electronvolt} can be explained in the same way as in the \burst{} scenario:
The lower number of sources effectively contributing to the CR intensity at higher energies increases the influence a single source can have on the GDE intensities.
The much increased relative deviations at \SI{100}{\tera\electronvolt}, however, are due to the very strong confinement of CRs close to their source at this energy.
The reduced diffusion coefficient $\kappa_{\mathrm{start}}$ is orders of magnitude smaller than in the \burst{} and \credit{} scenarios (see Sec.~\ref{sec: source injection and near-source transport}).
\paragraph{Summary}
These histograms show the scales over which source imprints on the GDEs can vary in the different scenarios.
We find that relative deviations are more extreme at higher GDE energies in the \burst{} and \tdd{} scenario due to a lower number of CR sources effectively contributing to the CR intensity, resulting in a higher relative influence of individual sources.
In the \credit{} scenario, we see the limiting effect that the energy-dependent escape has on source imprints.
For more details, see App.~\ref{app:Burst vs. CREDIT}.
In a next step, we want to understand the distribution of GDE intensities along different specific lines of sight and test our predictions made in Sec.~\ref{sec:Cosmic rays and stable law distributions}.
\subsection{High-statistics assessment of diffuse emission distributions}\label{sec:High-statistics assessment of diffuse emission distributions}
So far, we have investigated the strongest GDE deviations per sky map of the discrete source model from the the smooth model.
Those deviations are caused by the influence of single young sources on the CR intensity in their vicinity.
In this section, we want to investigate this connection further by using our knowledge about CR intensity distributions and predictions based on those that we derived in Sec.~\ref{sec:Cosmic rays and stable law distributions}.
To scrutinise the GDE intensity distributions, we need much higher sample sizes than the $1\,000$ realisations used in the previous sections.
Instead, we restrict ourselves to four selected lines of sight (LOS) where we increase our sample size to $10^6$ realisations each.
Here, we chose to present the results for four different LOS: $(l,b)=(0^{\circ}, 0^{\circ})$ towards the Galactic centre, $(180^{\circ}, 0^{\circ})$ towards the Galactic anticentre, $(0^{\circ}, 90^{\circ})$ directly upwards out of the Galactic plane, and $(-65^{\circ}, 0^{\circ})$ right into the neighbouring Sagittarius-Carina spiral arm.

We will describe the simulation results for the different source injection and near-source transport models, and discuss how an understanding of the underlying distribution of CR intensities as described in Sec.~\ref{sec:Cosmic rays and stable law distributions} can be used to describe the distribution of GDEs.
Again, we show results for GDE energies of \SI{10}{\giga\electronvolt} and \SI{100}{\tera\electronvolt} as exemplary low and high energies.
\paragraph{\BURST{} scenario}
\begin{figure}  
     \centering
    \includegraphics[width=\textwidth]{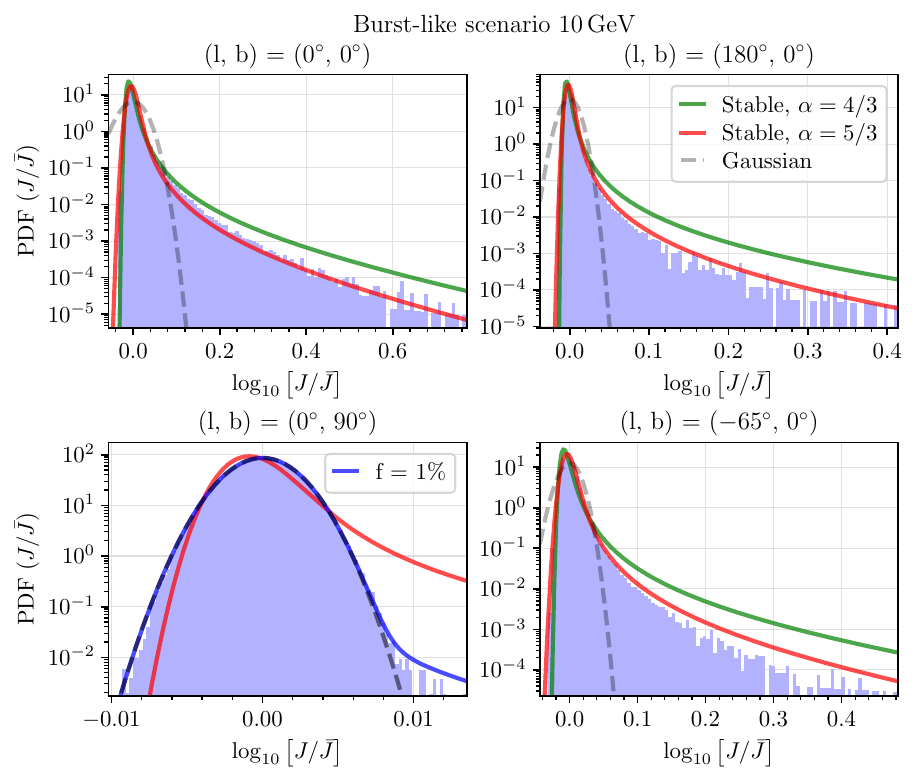}
    \caption{\BURST{} scenario at \SI{10}{\giga\electronvolt}.
    We simulated the GDE intensity along different LOS for $10^6$ realisations of source positions and ages.
    The panels show the different LOS indicated by their respective Galactic longitude $l$ and latitude $b$.
    The histograms are shown on a double-logarithmic plot, excluding the $10$ highest values.
    They are normalised such that the probability density function (PDF) for the random variable $J/\bar{J}$ is shown.
    The mean intensity is calculated as the sample mean of the $10^6$ realisations.
    We also plot stable law distributions with $\alpha=4/3$ (green), $\alpha=5/3$ (red), and normal distributions (gray dashed) with fitted scale parameters.
    For the $(0^{\circ}, 90^{\circ})$ LOS, we show a mixture model as described in Sec.~\ref{sec:Cosmic rays and stable law distributions} in blue, with the parameter $f$ given in the legend.
    }
    \label{fig:LOS_deviation_log_Fiducial_low}
\end{figure}
\begin{figure}  
     \centering
    \includegraphics[width=\textwidth]{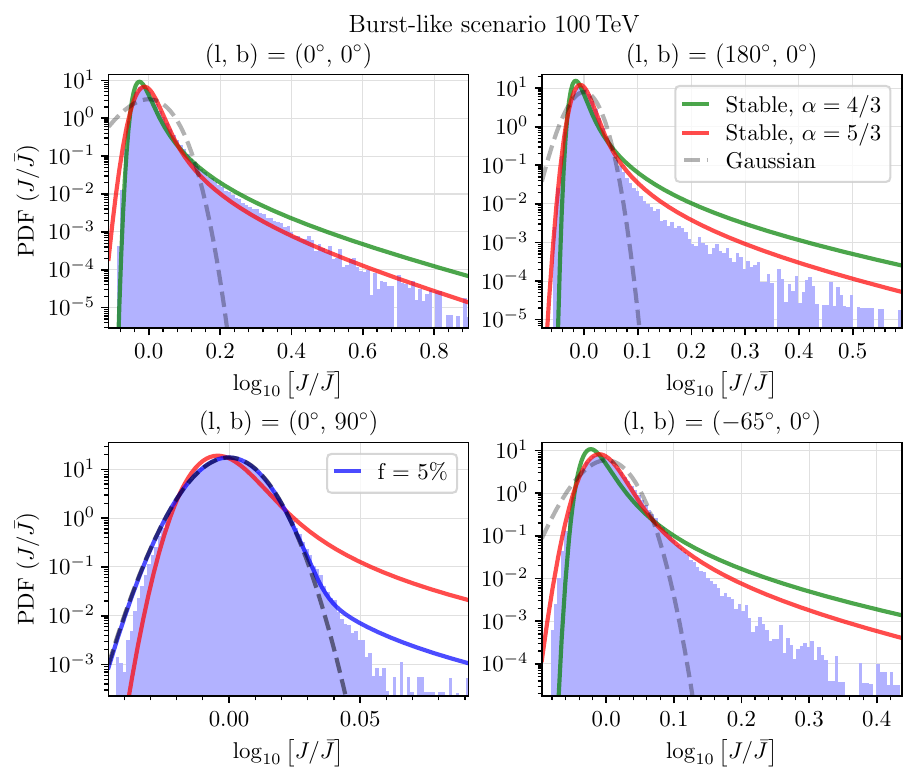}
    \caption{Same as Fig.~\ref{fig:LOS_deviation_log_Fiducial_low}, but for \SI{100}{\tera\electronvolt}.
    }
    \label{fig:LOS_deviation_log_Burst_high}
\end{figure}
The results of the \burst{} scenario are presented in Fig.~\ref{fig:LOS_deviation_log_Fiducial_low} for a gamma-ray energy of \SI{10}{\giga\electronvolt} and in Fig.~\ref{fig:LOS_deviation_log_Burst_high} for \SI{100}{\tera\electronvolt}.
The histograms are presented in a double-logarithmic plot.
The high sample size helps us to resolve the probability density functions (PDFs) far into their tails, which requires to resolve many orders of magnitude.

In addition to the histograms, we also present the predictions from Sec.~\ref{sec:Cosmic rays and stable law distributions}.
For the three LOS that trace the Galactic plane ($b = 0^{\circ}$), we have displayed the stable laws with an index of stability of $\alpha = 4/3$ and $\alpha = 5/3$, respectively (see Eq.~\eqref{eq: approx fit plane}).
The scale parameter has been fitted to the data.
We also show a normal distribution with the sample mean and sample standard deviation.
For the LOS out of the plane with $(l,b)=(0^{\circ}, 90^{\circ})$, we show the mixture model of a stable law with $\alpha = 5/3$ and a normal distribution with a suitable mixing fraction $f$.

In Fig.~\ref{fig:LOS_deviation_log_Fiducial_low}, the results for the lower energy are shown.
The LOS towards the Galactic centre $(0^{\circ}, 0^{\circ})$ shows the highest GDE intensity values.
The shape of the histogram is in general captured by the stable laws.
For low values of $J$, the stable law with $\alpha = 4/3$ describes the data better, while for high $J$-values, $\alpha = 5/3$ is preferred.
We can follow the interpretation of these indices of stability to indicate the dominant influence of sources with a 2D disk-like distribution ($\alpha = 4/3$) or a 3D box-like distribution ($\alpha = 5/3$)~\cite{2017GenoliniSalatiA&A}.
For high values of $J$, the most contributing sources have to be close to the LOS.
At any point close to the Galactic disk, the source distribution is roughly 3D uniform on scales not larger than the vertical scale height, for which we assume \SI{70}{\parsec} (see Sec.~\ref{sec: Transport model}).
For low values of $J$ however, the most influential sources are farther away from the LOS and the flatness of the Galactic disk becomes dominant (see also Sec.~\ref{sec:Cosmic rays and stable law distributions}).
Thus, we can expect the data to be fitted better by the stable law with $\alpha = 5/3$ for high $J$, similar to the description of the CR intensity distributions discussed in Ref.~\cite{2017GenoliniSalatiA&A}.
These observations also hold for the other two LOS with $b = 0^{\circ}$.
There, the stable law with $\alpha = 5/3$ describes the data best for high $J$.
For low $J$, the histograms lies in-between the two stable laws.
The normal distribution (Gaussian) does not describe the histograms well in any of the three LOS along the Galactic plane.

For the LOS $(0^{\circ}, 90^{\circ})$, the intensity values are much less extreme because it traces through regions with much less CR sources.
The normal distribution describes the bulk of the data well in this case, but it does not capture the tail of the distribution.
The mixture model between a stable law with $\alpha = 5/3$ and the normal distribution works well to describe the bulk of the data as well as the tail of the distribution if we fix $f=\SI{1}{\percent}$.
We see this as a confirmation of our prediction that the GDEs along LOS out of the Galactic plane receive contributions from voxels with CR intensities following stable law distributions and ones where they are normally distributed.

In Fig.~\ref{fig:LOS_deviation_log_Burst_high}, we show the results in the \burst{} scenario at \SI{100}{\tera\electronvolt}.
The deviations are in general more extreme than in the low-energy case due to a stronger relative influence of individual CR sources as discussed before (e.g., Sec.~\ref{sec:High-statistics assessment of diffuse emission distributions}).
We can argue for the same model predictions to fit the data best as in the \SI{10}{\giga\electronvolt} case.
There is just one caveat: The distributions seem to describe the histograms worse at higher values for $J$.
We can attribute this to the influence of the causality condition we implemented to remedy the missing causality in the transport equation~\eqref{eq:transport_equation}.
This does only influence high intensity values in the distributions at high energies as has been discussed in Ref.~\cite{2017GenoliniSalatiA&A}.

Additionally, the fraction $f$ in the mixture model for the LOS $(0^{\circ}, 90^{\circ})$ has to be chosen higher to describe the data at this energy.
This can be interpreted as a higher influence of close-by sources at this higher energy.
\paragraph{\CREDIT{} scenario}
\begin{figure}  
     \centering
    \includegraphics[width=\textwidth]{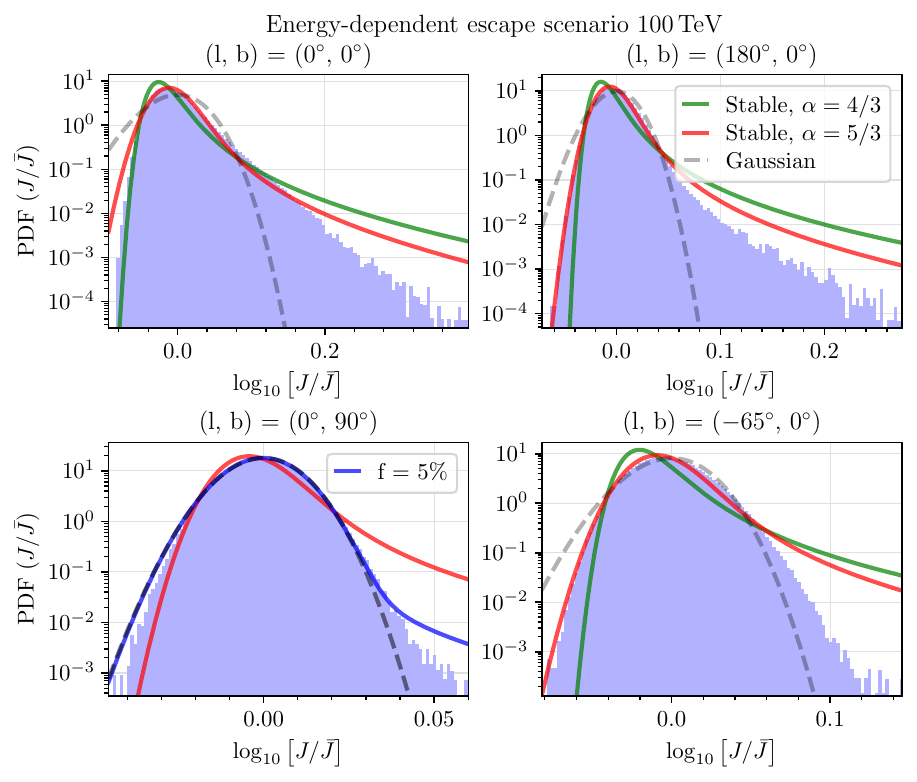}
    \caption{Same as Fig.~\ref{fig:LOS_deviation_log_Burst_high}, but for the \credit{} scenario.
    }
    \label{fig:LOS_deviation_log_CREDIT_high}
\end{figure}
For this scenario, we only show the results at \SI{100}{\tera\electronvolt} as the results at \SI{10}{\giga\electronvolt} are identical to the ones in the \burst{} scenario.
Comparing the results in Fig.~\ref{fig:LOS_deviation_log_CREDIT_high} to the GDEs at the same energy in the \burst{} scenario in Fig.~\ref{fig:LOS_deviation_log_Burst_high}, we see that the probability of high intensities is smaller in the \credit{} scenario.
Similarly to our evaluation in the previous Sec.~\ref{sec:The most extreme deviations}, this can be attributed to the energy-dependent injection times (see App.~\ref{app:Burst vs. CREDIT}).

The shape of the distributions is slightly different compared to their \burst{} counterparts, but the data is still described well by the stable laws.
As in the \burst{} scenario at this energy, the discrepancy between the histograms and the stable laws at high $J$-values can be attributed to our implementation of causality.
The mixture model for the LOS $(0^{\circ}, 90^{\circ})$ also describes the features of the distribution reasonably well.
\paragraph{\TDD{} scenario}
\begin{figure}  
     \centering
    \includegraphics[width=\textwidth]{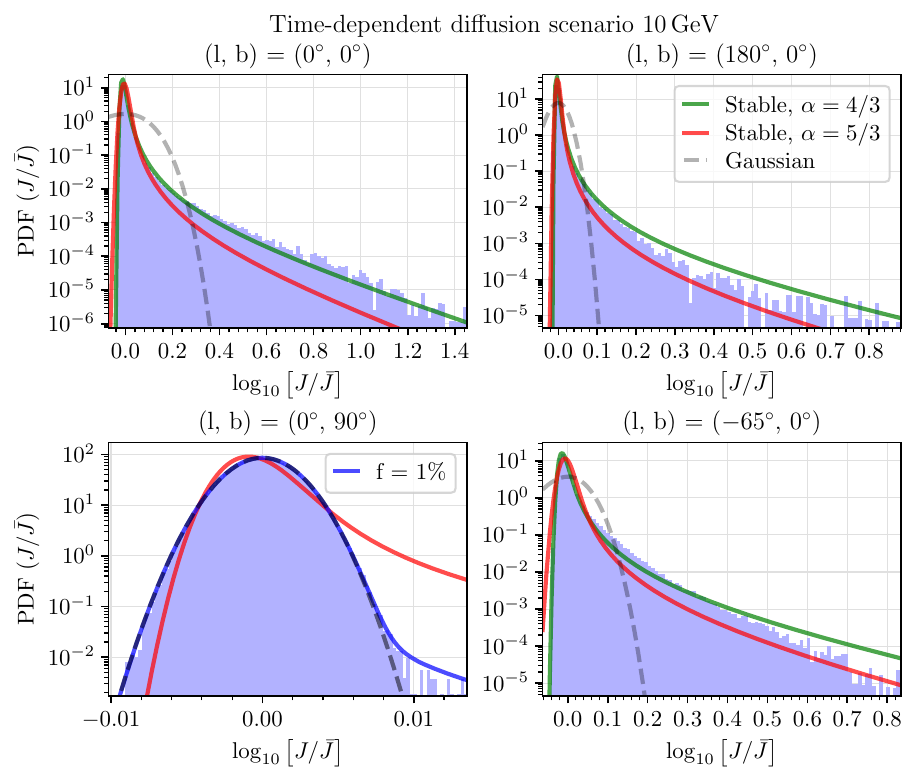}
    \caption{Same as Fig.~\ref{fig:LOS_deviation_log_Fiducial_low}, but for the \tdd{} scenario.
    }
    \label{fig:LOS_deviation_log_TDD_low}
\end{figure}
\begin{figure}  
     \centering
    \includegraphics[width=\textwidth]{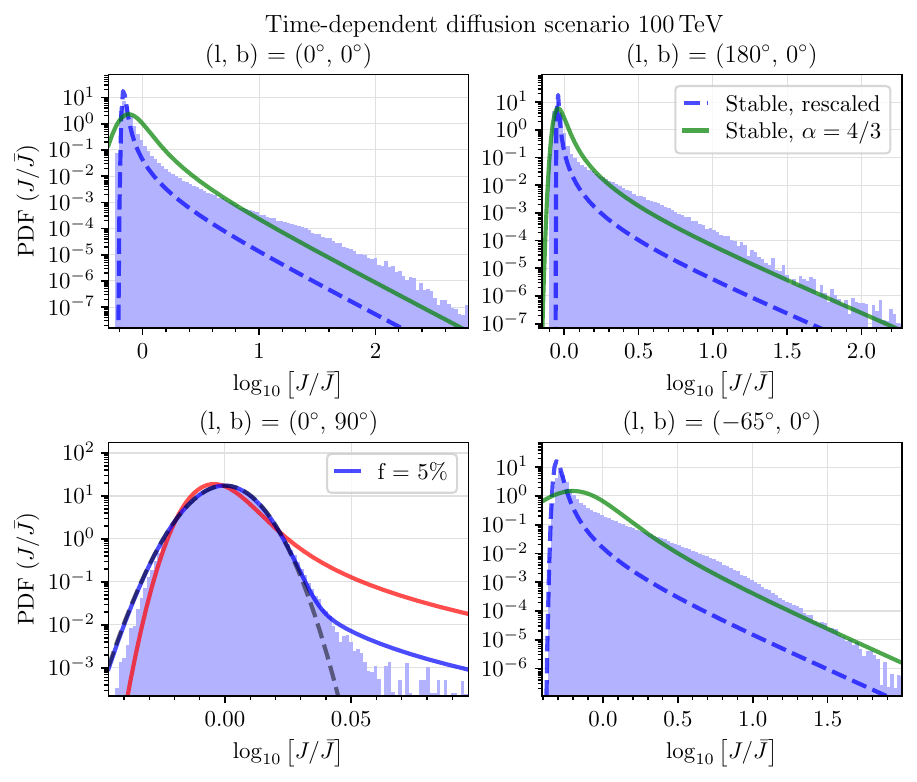}
    \caption{Same as Fig.~\ref{fig:LOS_deviation_log_Burst_high}, but for the \tdd{} scenario.
    For conciseness, the stable laws with $\alpha = 5/3$ are omitted.
    Instead, rescaled versions of the stable laws with $\alpha = 4/3$ in the \tdd{} scenario presented in Fig.~\ref{fig:LOS_deviation_log_TDD_low} are added for the LOS in the Galactic plane (blue dashed). 
    }
    \label{fig:LOS_deviation_log_TDD_high}
\end{figure}
In Figs.~\ref{fig:LOS_deviation_log_TDD_low} and~\ref{fig:LOS_deviation_log_TDD_high}, we show the distributions of intensities in the \tdd{} scenario.
As we have seen before, the effective confinement of CRs around their sources in this scenario leads to enhanced GDE intensity values.
This is reflected in the higher proportion of high intensity deviations in the simulations.
We check, whether the predictions made in Sec.~\ref{sec:Cosmic rays and stable law distributions} can also be used to explain the data in the \tdd{} scenario.

We begin with the low-energy results presented in Fig.~\ref{fig:LOS_deviation_log_TDD_low}.
The histograms for the LOS along the Galactic plane ($b = 0^{\circ}$) are not at all described by normal distributions.
However, the stable laws with $\alpha = 4/3$ describe the histograms nicely, generally much better than the stable laws with $\alpha = 5/3$, which described the distributions in the \burst{} scenario at this energy better.
This could hint to a stronger influence of sources farther away from the LOS of interest in the \tdd{} scenario compared to the \burst{} scenario.
Then, the distribution of high $J$-values would tend to be dominated by the disk-like spatial distribution of CR sources that is reflected in the stable law with $\alpha = 4/3$.
However, it is complicated to directly draw conclusions with our predictions from Sec~\ref{sec:Cosmic rays and stable law distributions} as the \tdd{} scenario has a more complicated transport mechanism that is not considered in the CR intensity distributions in Ref.~\cite{2017GenoliniSalatiA&A}.

The results at \SI{100}{\tera\electronvolt} are shown in Fig.~\ref{fig:LOS_deviation_log_TDD_high}.
We can see that those histograms extend to very extreme values for $J$.
Here, the stable laws with fitted scale parameters are not a good fit to the data.
For the LOS through the Galactic plane, we show the slightly better fitting stable law with $\alpha = 4/3$.
While those somehow explain the tails of the distributions, there is more structure due to the \tdd{}.
One attempt to explain the bulk of the data better is to use the distributions for the low energy case presented in Fig.~\ref{fig:LOS_deviation_log_TDD_low} and rescale their scale parameters by the respective mean values $\bar{J}$.
The resulting PDFs are shown by the blue dashed lines in Fig.~\ref{fig:LOS_deviation_log_TDD_high}.
While they describe the bulk of the data better than the stable laws with a fitted scale parameter, there is much more structure in the distribution, that we cannot capture with our simple model.

The LOS $(0^{\circ}, 90^{\circ})$, however, can be explained well for both energies.
We attribute this to a small influence of close-by CR sources and hence, the exact source injection and near-source transport properties.
\paragraph{Summary}
We simulated GDEs for selected LOS with large sample sizes of $10^6$ realisations to study the GDE intensity distributions in detail.
Only these large sample sizes allow us to resolve the tails of the distributions, which made it possible to test predictions based on the known distributions of CR intensities (see Sec.~\ref{sec:Cosmic rays and stable law distributions}).
We found that stable laws are able to describe main features of the distribution of GDEs for LOS along the Galactic plane.
Normal distributions fail to describe the simulation data.
The explanation with stable laws works best for the \burst{} scenario and reasonably well for the \credit{} scenario.
At a GDE energy of \SI{100}{\tera\electronvolt}, we find deviations from our predictions for high GDE intensity values, that can be attributed to the implementation of a causality criterion in our model.
In the \tdd{} scenario, the simple models are not able to capture the effects of the more complex near-source transport.

For the LOS out of the Galactic plane, a mixture model that combines contributions to the GDEs distributed following a stable law and ones following a normal distribution can describe the histograms in all models well.

We have established the influence of discrete CR sources on the GDEs by successfully testing our predictions from Sec.~\ref{sec:Cosmic rays and stable law distributions}.
This allows us to underpin the link between CR and GDE intensity distributions.
\subsection{Sky windows and profiles}\label{sec:On sky windows and profiles}
In the last sections, we have investigated the influence discrete sources have on the morphology and distribution of GDE intensities.
We can understand the imprints of individual CR sources qualitatively and have a good understanding of the maximum deviations from smooth model predictions that we can expect along individual lines of sight.
However, measurements of GDEs, especially at energies above some teraelectronvolts, are often not spatially resolved like the sky maps we presented before.
Instead, intensity measurements over extended sky windows are averaged and energy spectra for those sky windows are presented~\cite{2023CaoAharonianPhRvL,2025CaoAharonianPhRvL,2021AmenomoriBaoPhRvL}.
Naturally, the question arises, whether discrete sources influence the model predictions on those GDE spectra averaged over sky windows.

We want to study this influence of source stochasticity on the predictions of GDE spectra in the different injection and near-source transport scenarios.
For that, we analyse $1\,000$ realisations over a broad energy range in two selected sky windows, representing the inner and outer Galaxy sky windows as seen by LHAASO~\cite{2023CaoAharonianPhRvL,2025CaoAharonianPhRvL}.
The uncertainties on the GDE intensity stemming from our limited knowledge about the source positions and ages have to be compared with uncertainties of other components that enter the model predictions, like the interaction cross sections, the gas density, the CR source distribution, and the CR model parameters.
For the extent of these uncertainties, we refer to the CRINGE model~\cite{2023SchweferMertschApJ}, where they have been studied carefully.

With increased sensitivities of experiments and more involved analysis methods, it also becomes feasible to resolve the diffuse emissions in longitude.
We can expect that the influence of discrete sources to be larger if the regions that measurements are averaged over becomes smaller and more structure in the GDEs gets resolved.
We investigate this by discussing longitude profiles at a GDE energy of \SI{100}{\tera\electronvolt} in the different source injection and near-source transport models.
Again, we use $1\,000$ realisations each to determine the expected spread of GDEs across the realisations.
\paragraph{\BURST{} scenario}
\begin{figure}  
     \centering
    \includegraphics[width=\textwidth]{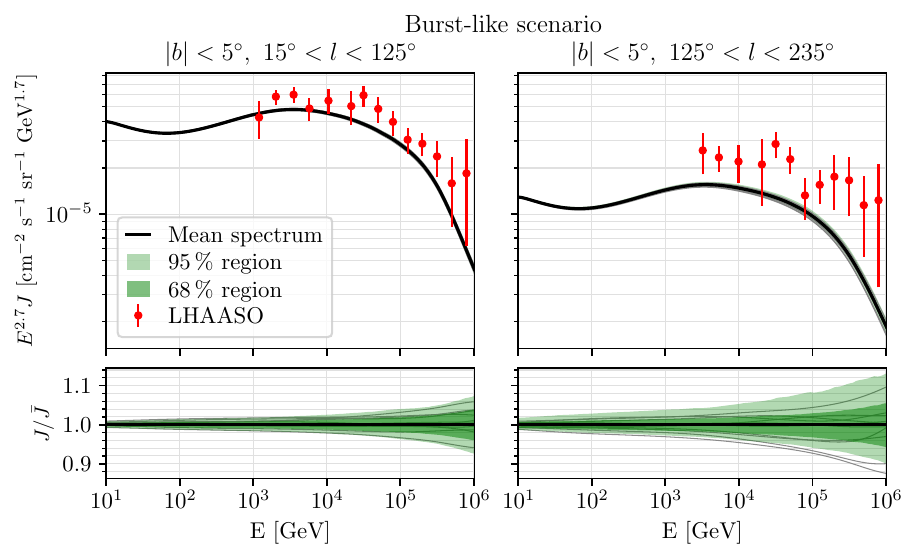}
    \caption{Spectra of the GDE intensity averaged over two sky windows (LHAASO inner and outer Galaxy)~\cite{2023CaoAharonianPhRvL,2025CaoAharonianPhRvL} for the \burst{} scenario.
    The sample mean of the spectra is displayed as a black line and the green shaded bands indicate different uncertainty intervals.
    The lower panel show the same normalised to the respective mean spectra.
    $10$ spectra from individual source realisations are shown in grey.
    Additionally, we include the diffuse emission intensities as measured by LHAASO WCDA-KM2A~\cite{2025CaoAharonianPhRvL}.
    The errors are the $1\sigma$ statistical uncertainties and systematic uncertainties added in quadrature.
    }
    \label{fig:5a}
\end{figure}
\begin{figure}  
     \centering
    \includegraphics[width=\textwidth]{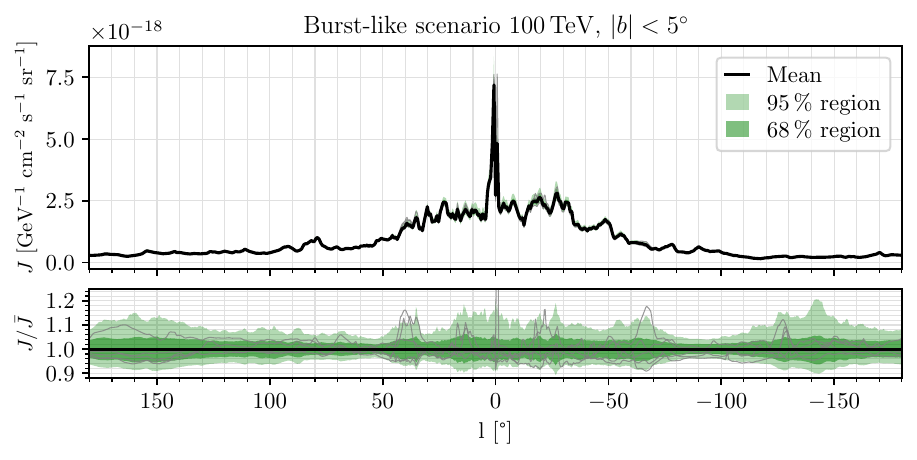}
    \caption{GDE intensity averaged for $|b|<5^\circ$ as a function of Galactic longitude at an energy of \SI{100}{\tera\electronvolt} for the \burst{} scenario.
    The sample mean of the spectra is displayed as a black line and the green shaded bands indicate different uncertainty intervals.
    The lower panel show the same normalised to the respective mean spectra.
    $5$ spectra from individual source realisations are drawn in grey.
    }
    \label{fig:lon_profile_a}
\end{figure}
In Fig.~\ref{fig:5a}, we show GDE intensity spectra in the LHAASO inner and outer Galaxy sky windows~\cite{2023CaoAharonianPhRvL,2025CaoAharonianPhRvL} for the \burst{} scenario.
The mean spectra as well as different uncertainty intervals are shown to illustrate the spread of the spectra in our sample.
We clearly see that the intensity in the inner Galaxy is higher than in the outer Galaxy sky window, but the spectral shape is similar in both cases.
We also show $10$~realisations of the spectrum from individual source realisations in grey to illustrate the spectral shape of GDE realisations with discrete sources.

The lower panels show the same lines normalised by the mean spectrum $\bar{J}$.
This makes it easier to quantify the spread due to discrete sources.
We use the light-green region as a proxy for this uncertainty.
It is on a sub-percent level up to the teraelectronvolt range and reaches a maximum of less than \SI{10}{\percent} for petaelectronvolt energies.
We can compare these to other model uncertainties like CR model parameters or cross sections.
Those were found to be of the order of \SI{10}{\percent} at gigaelectronvolt energies and increase to around \SI{50}{\percent} at about \SI{100}{\tera\electronvolt} (see~\cite{2023SchweferMertschApJ}).
Hence, the uncertainty due to discrete sources on the diffuse emissions averaged over these sky windows is subdominant in this scenario.
Considering the large sky windows that the GDEs are averaged over, it is not surprising that the influence of discrete sources on the GDE intensity spectra is small in the \burst{} scenario.
Averaging GDE intensities over larger regions of the sky diminishes the influence of individual sources which only effect a small region of the sky around their direction in the sky map (see Fig.~\ref{fig:1a} in Sec.~\ref{sec: A glimpse into the diffuse sky}).

We also display the measurements of LHAASO WCDA-KM2A~\cite{2025CaoAharonianPhRvL} with their statistic and systematic uncertainties.
The small uncertainty of the GDEs due to the discreteness of the sources in this \burst{} scenario also means that CR source stochasticity cannot be expected to explain the discrepancy between the GDE model predictions and the LHAASO measurements.
Other uncertainties or unresolved sources need to be invoked to bridge the gap in this scenario.

In Fig.~\ref{fig:lon_profile_a}, we show a longitude profile of the GDE intensities averaged for $|b|<5^\circ$ at an energy of \SI{100}{\tera\electronvolt} for the \burst{} scenario.
Again, we show different uncertainty intervals to illustrate the spread of the profiles in our sample.
This longitude profile clearly illustrates that GDE intensities are much stronger in the Galactic centre region compared to the outer parts of the Galaxy.
The lower panel shows the same lines normalised by the mean profile, which allows us to investigate the influence of source stochasticity.
We find that it leads to an uncertainty of around \SI{10}{\percent} on the predictions of longitude profiles.

We also show the longitude spectra of $5$ individual realisations in grey.
It can be seen that their deviation from the mean profile has a lot of structure in longitude with fluctuations being limited to a couple of degrees in longitude.
The sample mean of the profiles can be viewed as a proxy for the smooth model predictions for the profile, similarly to our considerations in Sec.~\ref{sec:stochastic vs smooth models}.
As the deviations occur due to the influence of individual sources, we can expect them to have a limited range in longitude, just as they are limited to small regions in the sky maps (see Sec.~\ref{sec: A glimpse into the diffuse sky}).
This implies that there is a chance that the source configuration realised in our Galaxy could have similar imprints on the longitude profile.
Thus, discrete sources should be considered as a modelling uncertainty in this case.
\paragraph{\CREDIT{} scenario}
\begin{figure}  
     \centering
    \includegraphics[width=\textwidth]{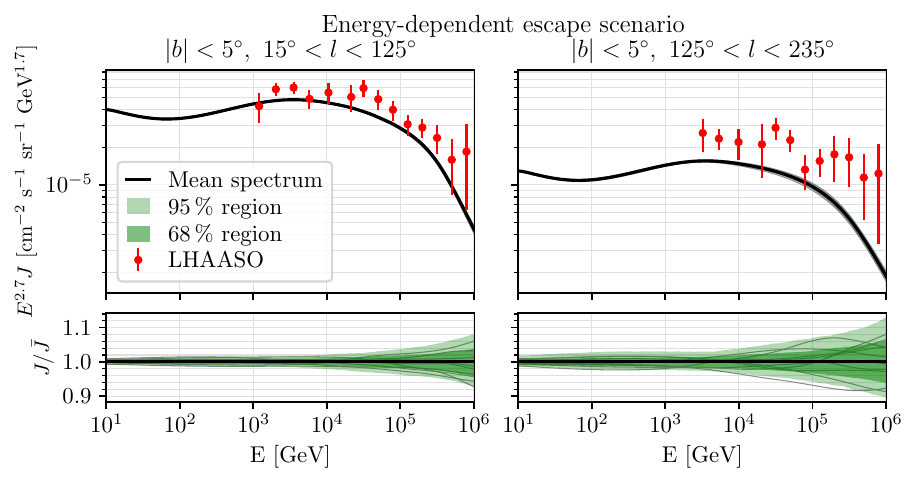}
    \caption{Same as Fig.~\ref{fig:5a}, but for the \credit{} scenario.
    }
    \label{fig:5b}
\end{figure}
\begin{figure}  
     \centering
    \includegraphics[width=\textwidth]{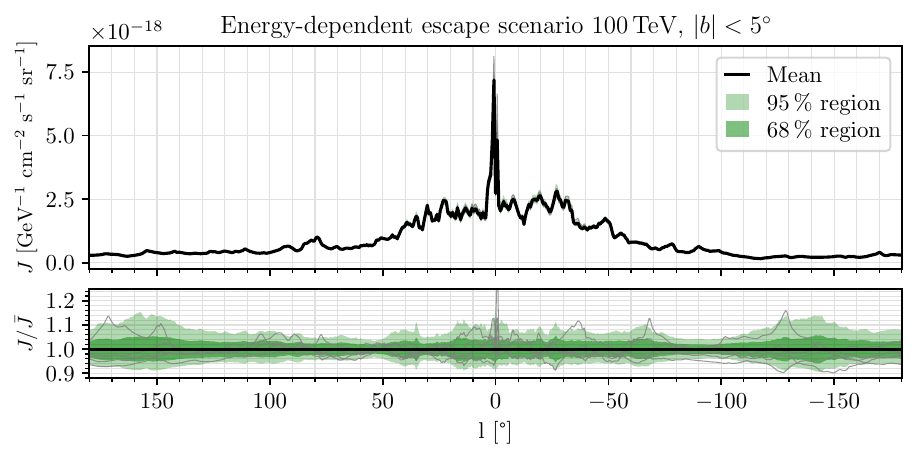}
    \caption{Same as Fig.~\ref{fig:lon_profile_a}, but for the \credit{} scenario.
    }
    \label{fig:lon_profile_b}
\end{figure}
The results for the GDE spectra in the \credit{} scenario can be seen in Fig.~\ref{fig:5b}.
The longitude profiles are shown in Fig.~\ref{fig:lon_profile_b}.
As in the analyses in the chapters before, the extent of the deviations is similar to the \burst{} scenario, while it is slightly less extreme at GDE energies that are influenced by CRs of energies above $E_{\mathrm{b}} \simeq e\mathcal{R}_{\mathrm{b}} = \SI{10}{\tera\electronvolt}$, where the energy-dependent escape sets in (see Eq.~\eqref{eq:CREDIT energy dependent escape} in Sec.~\ref{sec: source injection and near-source transport}).
For the spectra in Fig.~\ref{fig:5b}, this means that we can come to similar conclusions as in the \burst{} scenario considering the uncertainties and the compatibility to LHAASO data.
The $10$ individual realisations also do not show any strong spectral features deviating from the spectral shape of the mean, just as in the \burst{} scenario.
The longitude profile (Fig.~\ref{fig:lon_profile_b}) shows a similar structure to the one in the \burst{} scenario, but with less extended uncertainty bands.
This smaller imprint of CR sources at high energies in the \credit{} scenario is discussed in more detail in App.~\ref{app:Burst vs. CREDIT}.
\paragraph{\TDD{} scenario}
\begin{figure}  
     \centering
    \includegraphics[width=\textwidth]{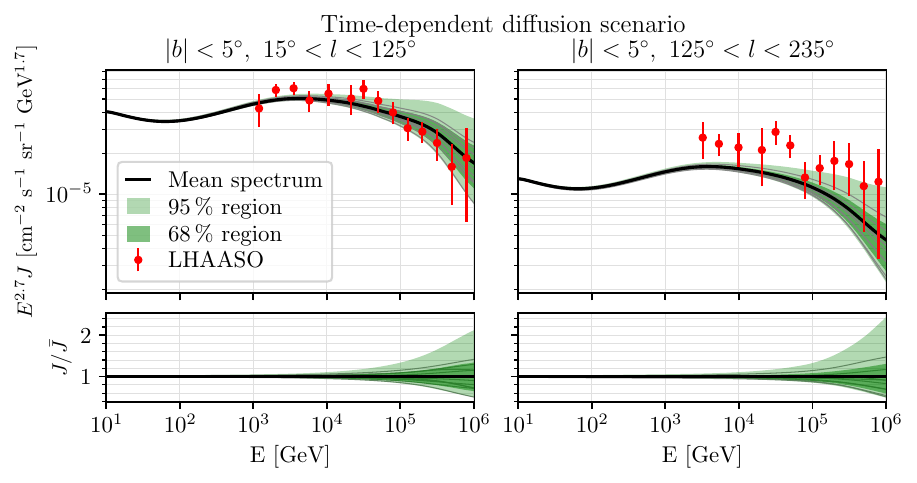}
    \caption{Same as Fig.~\ref{fig:5a}, but for the \tdd{} scenario.
    }
    \label{fig:5c}
\end{figure}
\begin{figure}  
     \centering
    \includegraphics[width=\textwidth]{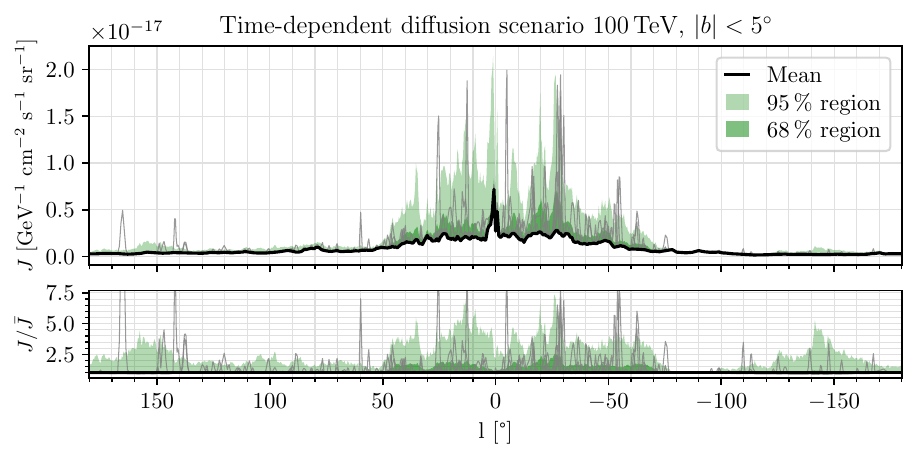}
    \caption{Same as Fig.~\ref{fig:lon_profile_a}, but for the \tdd{} scenario.
    }
    \label{fig:lon_profile_c}
\end{figure}
The spectra in the \tdd{} scenario in Fig.~\ref{fig:5c} show a much more extreme influence of discrete sources on the GDE spectra.
In fact, we find that the uncertainty due to source stochasticity becomes a relevant contribution to the total model uncertainties, as presented in Ref.~\cite{2023SchweferMertschApJ}, above some tens of teraelectronvolts.
Further, we observe that the mean spectrum increases compared to the mean spectra in the \burst{} and \credit{} scenario.
Together with the wider uncertainty bands, this model closes the gap between the smooth model predictions\footnote{These are approximately the mean spectra in the \burst{} scenario, as discussed before and in Sec.~\ref{sec:stochastic vs smooth models}.} and the LHAASO measurements at least in the inner Galaxy sky window.

The longitude profile results in the \tdd{} scenario are shown in Fig.\ref{fig:lon_profile_c}.
We can clearly see the extreme imprints of sources in this scenario that we have already discussed in the context of Fig.~\ref{fig:1c} in Sec.~\ref{sec: A glimpse into the diffuse sky}.
They are large and extend to several hundred percent in this \tdd{} scenario.
The grey lines representing $5$ individual realisations show that the source imprints are very peaked, which fits the spatially concentrated imprint of CR sources on the GDEs in this scenario.

The effects of the \tdd{} scenario on the GDEs offers an opportunity to constrain source injection and near-source transport models.
But as already discussed in Sec.\ref{sec: A glimpse into the diffuse sky}, a non-observation of imprints like in Fig.~\ref{fig:lon_profile_c} could also hint to a lower rate of CR sources accelerating particles up to petaelectronvolt energies rather than excluding the \tdd{} scenario altogether.
\paragraph{Summary}
In this section, we have discussed the stochastic effects of discrete CR sources on GDE spectra in the LHAASO GDE sky windows and on GDE intensity profiles in Galactic longitude.
For the spectra, the influence of discrete sources is limited because contributions from an extended region of the sky are averaged.
We found that GDE intensity spectra of individual realisations mainly follow the spectral shape of the smooth model predictions without strong spectral features.
The uncertainty bands obtained from our samples of $1\,000$ realisations each represent the modelling uncertainties due to source stochasticity.
They have to be compared to the uncertainties due to other modelling ingredients like the CR model parameters or interaction cross sections (see~\cite{2023SchweferMertschApJ}).
The source stochasticity contributions are subdominant for the \burst{} and \credit{} scenario, but are sizeable above some tens of teraelectronvolts in the \tdd{} scenario.

We also displayed the LHAASO WCDA-KM2A measurements~\cite{2025CaoAharonianPhRvL}.
We can see that the spectra in the \burst{} and \credit{} scenario undershoot the data with our choice of CR model parameters, which have been determined by fit to local CR data (see~\cite{2023SchweferMertschApJ}).
In those cases, other GDE model uncertainties or unresolved sources need to be invoked to explain the gap between the model predictions and the data.
In the \tdd{} scenario, however, the measurements are compatible with the model within the uncertainties.

For the longitude profiles, we found a much stronger signature of the discrete sources, especially in the profiles of individual realisations.
This is promising, as the real source configuration realised in our Galaxy might lead to such signatures.
The uncertainties on the longitude profiles are significant modelling uncertainties in all three source injection and near-source transport models, but they are clearly dominant in the \tdd{} scenario.

We conclude that the influence of discrete CR sources on the GDE predictions gets larger, the smaller the probed region.
With experimental advancements and better analysis methods, smaller regions can be probed and measurements of the longitude profiles will become available.
This could even make it possible to constrain the most extreme source injection and near-source transport models like the \tdd{} scenario (see also~\cite{2025KaciGiacintiJCAP}).
\subsection{(Dis-)Connections of diffuse emissions at \SI{10}{\giga\electronvolt} and \SI{100}{\tera\electronvolt}}\label{sec:correlation results}
The example sky maps in Sec.~\ref{sec: A glimpse into the diffuse sky} have shown that the influence of discrete sources can be visibly different at different energies.
In this section, we want to discuss the correlation between GDE intensities as well as GDE deviations from the smooth model predictions at \SI{10}{\giga\electronvolt} and \SI{100}{\tera\electronvolt} in more detail.
First, we argue that smooth source models predict a strong correlation between low and high energy GDEs.
Then, we investigate how this correlation is broken if discrete CR sources are considered.

Popular literature models for GDEs, like the $\pi^0$-model~\cite{2012AckermannAjelloApJ} and the $\mathrm{KRA}_{\gamma}$-models~\cite{2015GaggeroUrbanoPhRvD, 2015GaggeroGrassoApJL}, do not consider discrete sources and rather assume a smooth CR source distribution.
Under this assumption, the predictions for the GDE sky maps at low and high energies turn out to be well-correlated.
By that, we mean that the GDE intensities representing the sky maps at low and high energies can be related approximately as
\begin{equation}\label{eq:correlation assumption smooth}
    J_{\SI{100}{\tera\electronvolt}}\left(l,b\right) \simeq X \cdot J_{\SI{10}{\giga\electronvolt}}\left(l,b\right)
\end{equation}
with a fixed factor $X$.

A measure to test such a linear dependence is the Pearson correlation coefficient $\rho$.
For two random variables $X$ and $Y$, it is defined as
\begin{equation}
    \rho_{\text{X,Y}} = \sigma_{\text{X,Y}} / \sigma_{\text{X}}\sigma_{\text{Y}} \, ,
\end{equation}
where $\sigma_{\text{X,Y}}$ is the covariance of $X$ and $Y$, and $\sigma_{\text{X}}$ is the standard deviation of $X$.
Its value lies between $-1$ and $1$, where a positive (negative) sign indicates (anti-)~correlation.
The closer $|\rho_{\text{X,Y}}|$ is to $1$, the larger the (anti-)~correlation between $X$ and $Y$.

As discussed in Sec.~\ref{sec:stochastic vs smooth models}, the sample mean in the \burst{} scenario approximates the predictions of a smooth CR source distribution well.
To test whether the assumption~\eqref{eq:correlation assumption smooth} holds for the sample means $\langle J_{\SI{10}{\giga\electronvolt}} \rangle$ and $\langle J_{\SI{100}{\tera\electronvolt}} \rangle$, we calculate the correlation of the GDE intensities along the different LOS and find $\rho_{\langle J_{\SI{10}{\giga\electronvolt}} \rangle,\langle J_{\SI{100}{\tera\electronvolt}} \rangle} = 0.9988$.

This almost perfect correlation is not surprising considering the various ingredients of diffuse emissions discussed in Sec.~\ref{sec: diffuse emissions}.
The gas distribution does not change with energy, the cross sections only add an energy dependence, and the CR intensity only has a small spatial dependence in a smooth source model.
The main energy-dependent spatial modulations are added by the influence of absorption, which only effects the high GDE intensities, but this effect is limited to a relatively small region in the Galactic disk, especially in the Galactic centre.

As we have seen in Sec.~\ref{sec: A glimpse into the diffuse sky}, the consideration of discrete sources adds structures to the sky maps which are in general different at low and high energies.
We expect that Eq.~\eqref{eq:correlation assumption smooth} cannot simply be extended to a pair of GDE sky maps at different energies in a discrete source model.
There will be differences from the almost perfectly correlated smooth model predictions at \SI{10}{\giga\electronvolt} and \SI{100}{\tera\electronvolt}, respectively.
We will test whether those deviations ($\Delta J_{\SI{10}{\giga\electronvolt}}\left(l,b\right)$ and $\Delta J_{\SI{100}{\tera\electronvolt}}\left(l,b\right)$) or the total GDE intensities ($J_{\SI{10}{\giga\electronvolt}}\left(l,b\right)$ and $J_{\SI{100}{\tera\electronvolt}}\left(l,b\right)$) are correlated well in the different source injection and near-source transport models.
\paragraph{\BURST{} and \credit{} scenario}
\begin{figure}  
     \centering
    \includegraphics[width=\textwidth]{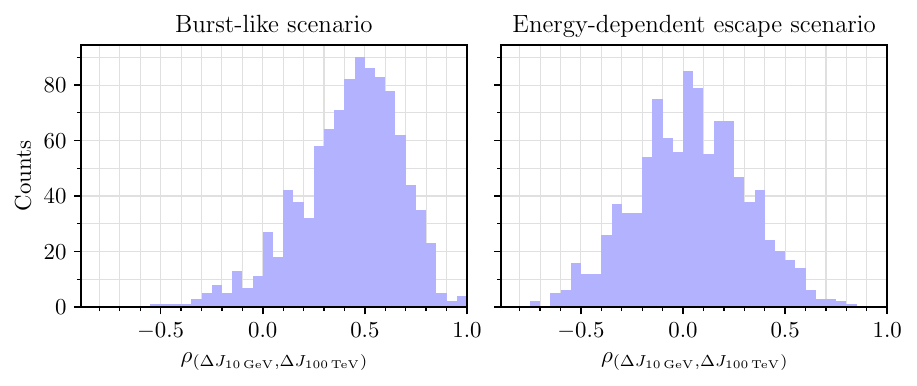}
    \caption{Histograms of the Pearson correlation coefficients, $\rho$, between $\Delta J_{\SI{10}{\giga\electronvolt}}\left(l,b\right)$ and $\Delta J_{\SI{100}{\tera\electronvolt}}\left(l,b\right)$ calculated for each of $1\,000$ source realisations in the \burst{}  and the \credit{} scenario (left and right).
    The histograms show that the deviations of GDE intensities from the mean are not strongly correlated in those two scenarios as most correlation coefficients are clearly smaller than $1$.
    The correlation tends to be stronger in the \burst{} scenario, where the same sources contribute to the GDE intensity at different energies.
    }
    \label{fig:6a_b}
\end{figure}
In Fig.~\ref{fig:6a_b}, we show the Pearson correlation coefficients of $\Delta J_{\SI{10}{\giga\electronvolt}}\left(l,b\right)$ and $\Delta J_{\SI{100}{\tera\electronvolt}}\left(l,b\right)$ for each of the $1\,000$ realisations in histograms for the \burst{} and the \credit{} scenario.
Each individual correlation coefficient encodes the goodness of an affine linear relation between $\Delta J$ at \SI{10}{\giga\electronvolt} and \SI{100}{\tera\electronvolt} for one of the $1\,000$ realisations.
As can be seen in both panels of Fig.~\ref{fig:6a_b}, the correlation coefficient is in general clearly smaller than $1$, i.e., such a relation does not hold.

The deviations from the smooth model predictions, which are described by the $\Delta J$'s, can mainly be attributed to the effect of some most-contributing individual sources (see Sec.~\ref{sec:stochastic vs smooth models}).
This tells us that, due to the different transport properties of CRs at low and high energies, the morphology of their imprints on the GDEs is different.
The correlations are generally worse in the \credit{} scenario shown on the right in Fig.~\ref{fig:6a_b}, because in this scenario, different CR sources at different positions contribute most to the GDE deviations.
This fully breaks the correlation between $\Delta J_{\text{low}}$ and $\Delta J_{\text{high}}$ in this scenario.

However, the differences from the well-correlated smooth model predictions are relatively small compared to the smooth model predictions.
Hence, the correlation between $J_{\SI{10}{\giga\electronvolt}}\left(l,b\right)$ and $J_{\SI{100}{\tera\electronvolt}}\left(l,b\right)$ turns out to be still very good with most values being within a percent of $1$.
This tells us that the influence of sources on the GDEs in this scenario is limited to small regions of the sky, while the influence on the total GDE intensity across the sky is rather small.
This has already been observed in Sec.~\ref{sec:On sky windows and profiles}, where the influence of the \burst{} and the \credit{} scenario on the spectrum in the LHAASO sky windows is very small.
\paragraph{\TDD{} scenario}
\begin{figure}  
     \centering
    \includegraphics[width=\textwidth]{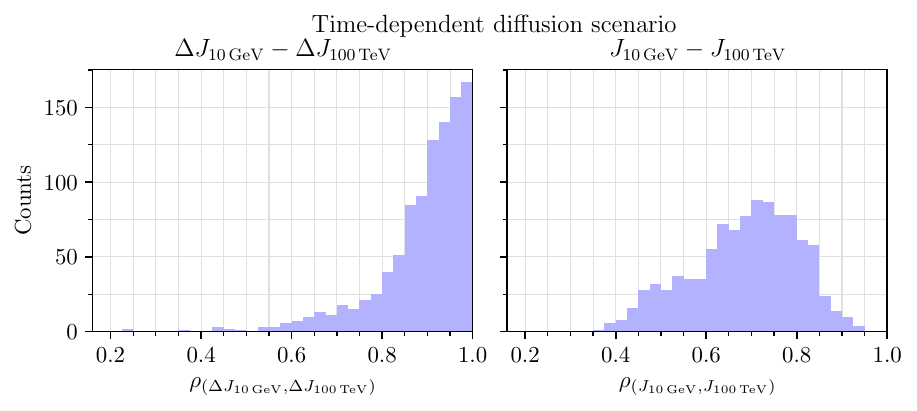}
    \caption{The left panel shows the same kind of histogram as the ones in Fig.~\ref{fig:6a_b}, but for the \tdd{} scenario.
    It can be seen that the correlation is much better than in the \burst{}  and \credit{} scenario.
    On the right, a histogram of the total GDE intensities $J_{\SI{10}{\giga\electronvolt}}\left(l,b\right)$ and $J_{\SI{100}{\tera\electronvolt}}\left(l,b\right)$ is shown, which reveals that those are not very strongly correlated.
    }
    \label{fig:6c}
\end{figure}
The results of the \tdd{} scenario in Fig.~\ref{fig:6c} show a different behaviour.
Here, the deviations from the smooth model predictions at \SI{10}{\giga\electronvolt} and \SI{100}{\tera\electronvolt} are rather well correlated.
The $\Delta J$'s describe the imprint of dominant individual sources.
Due to the near-source transport model in the \tdd{} scenario, the transport of CRs is energy-independent in an initial phase of their propagation and should coincide directly at the different energies.
Besides the sources in the low-diffusion regime, other older sources might contribute strongly which breaks the correlation between $\Delta J_{\SI{10}{\giga\electronvolt}}\left(l,b\right)$ and $\Delta J_{\SI{100}{\tera\electronvolt}}\left(l,b\right)$ for some of the realisations.
It is this effect, that we see in the histogram on the left in Fig.~\ref{fig:6c}.

In contrast to the small influence of sources on the total GDE intensities in the previous scenarios, the \tdd{} scenario can lead to strong GDE intensities.
As can be seen in the right panel in Fig.~\ref{fig:6c}, this breaks the correlation between $J_{\SI{10}{\giga\electronvolt}}\left(l,b\right)$ and $J_{\SI{100}{\tera\electronvolt}}\left(l,b\right)$ that we still found in the \burst{}  and \credit{} scenario.
This underlines the more extreme influence CR sources have on the GDEs in the \tdd{} scenario, which we already discussed in the previous sections.
\paragraph{Summary}
We confirmed that the predictions of GDE sky maps at \SI{10}{\giga\electronvolt} and \SI{100}{\tera\electronvolt} in a smooth source model are well-correlated.
In contrast to that, the differences of the GDE sky maps in the discrete model from those smooth model predictions ($\Delta J_{\SI{10}{\giga\electronvolt}}\left(l,b\right)$ and $\Delta J_{\SI{100}{\tera\electronvolt}}\left(l,b\right)$) do not show such a good correlation.
In the \credit{} scenario, the correlation tends to be the weakest, it is still poor in the \burst{} scenario, but good in the \tdd{} scenario.
We discussed how these observations can be directly linked to the characteristics of the different source injection and near-source transport models.

The correlation between the total GDE intensities ($J_{\SI{10}{\giga\electronvolt}}\left(l,b\right)$ and $J_{\SI{100}{\tera\electronvolt}}\left(l,b\right)$), however, is very good for the moderate \burst{} and \credit{} scenarios, but the correlation gets broken for the extreme \tdd{} scenario.
As we discussed, this shows the extent to which the separate models can influence the total GDE intensities.
While the influence on individual CR sources is only important in small regions around a LOS passing the source in the \burst{} and \credit{} scenario, it can have a more substantial impact on the total GDE intensity in the \tdd{} scenario.
\subsection{Discussion}\label{sec:Discussion}
We showed the general features of discrete CR source models on GDEs and highlighted different aspects in the previous sections.
Particularly, we focused on the specific impact the different source injection and near-source transport models have.
We will give a more general overview of the main findings in the subsequent Sec.~\ref{sec:Summary and Conclusion}.
Here, we will discuss some of the limitations of the study presented in this work.

A main caveat of our study is the assumption of a single source population of SNRs with a common source spectrum that extends up to the CR knee.
This simple assumption facilitates stochastic studies like this one and has been a common choice before (see, e.g.,~\cite{2011MertschJCAP,2017GenoliniSalatiA&A, 2021EvoliAmatoPhRvDa, 2012BlasiAmatoJCAP,2025KaciGiacintiJCAP,2021PhanSchulzePhRvL}).
The restriction to the consideration of a single type of source makes it possible to study the impact of specific source models like we have done by discussing the three source injection and near-source transport models introduced in Sec.~\ref{sec: source injection and near-source transport}.
More complex compositions of source populations with source parameters varying between individual sources are likely closer to reality but currently mostly unknown.
A study of those should only be a next step after a single-population study like this one.
Some of the effects discussed in this work will be bigger, others likely smaller.
Similar considerations hold for the inclusion of continuously injecting CR sources like star clusters, which we exclude from our model for conciseness.

A further limitation concerns the simple isotropic homogeneous diffusion transport model that we implemented.
This is also a popular choice in stochastic source models as it allows to solve the Green's function analytically.
While other approaches including anisotropic diffusion exist (see, e.g.,~\cite{2023GiacintiSemikozarXiv}), the structure of magnetic fields and the diffusion coefficient in the Galaxy are poorly constrained.
We can expect anisotropic transport to alter the morphology of the CR source imprints on the GDEs.
Determining the impact of anisotropic diffusion on the results presented in this work lies beyond our scope.

Besides the impact of the different source injection and near-source transport models, the influence of many more CR model parameters could be studied.
However, it is not computationally feasible to perform an extended study like this one for a large parameter grid.\footnote{This is a general problem in Monte Carlo studies like this one.
Machine learning methods have been identified as a possible remedy for this~\cite{2025FredianiKramerJCAP} and might help facilitate parameter studies in the future.}
We reckon that it could be especially interesting to investigate the influence of the source rate on the stochasticity of the GDEs.
Lowering the source rate would lead to a larger influence of individual CR sources in the CR intensities, thus increasing their individual impact on the GDEs.
A similar effect has been observed in other stochastic CR studies (e.g.,~\cite{2018MertschJCAP}).
For discussions of the impact of source rates in the \tdd{} scenario, we refer to Ref.~\cite{2025KaciGiacintiJCAP}.

A different choice for the halo height $H$ is also expected to have an influence on the results.
Larger values, as suggested by measurements of stable and unstable isotopes at low energies~\cite{2020EvoliMorlinoPhRvD}, require a larger normalisation for the diffusion coefficient to fit local secondary-to-primary ratios.
This effectively decreases the stochasticity and is expected to lead to lower deviations in the GDEs as well.
A different parametrisation of the diffusion coefficient can also be expected to have an effect on the GDE results which can be expected to follow the same rationale:
If the diffusion coefficient at a given energy were lower in a different parametrisation, the stochasticity and thus the expected deviations at that energy would increase, and vice versa.
Further a change of the spectral index in the diffusion coefficient would directly translate into a different expected spectral index for the GDE intensities.

Lastly, we want to comment on leptonic contributions to the total diffuse gamma-ray emissions.
We exclude them in our study as their contribution can be assumed to be subdominant, making up less than \SI{10}{\percent} of the GDEs for small latitudes ($|b| < 10^\circ$)~\cite{2018LipariVernettoPhRvD}.
Local measurements of the CR electron spectrum suggest that the CR electron spectrum has a cut-off at some tens of teraelectronvolts~\cite{2018MertschJCAP}, which limits the leptonic contributions at energies above \SI{1}{\tera\electronvolt} to a maximum of \SI{5}{\percent}~\cite{2018LipariVernettoPhRvD}, which can also be approximated in the Thomson limit.
Thus, we do not expect this to have a significant effect on our results.
Actually, we would only expect the correspondence between the low- and high-energy GDEs to decrease further.
However, there are stochastic GDE studies considering leptonic GDE contributions~\cite{2023ThalerKissmannAPh,2025MarinosPorterApJ}.
\section{Summary and Conclusion}\label{sec:Summary and Conclusion}
Cosmic rays (CRs) are generally assumed to be accelerated by discrete sources like supernova remnants.
Yet, it is not common to include this effect in models of the CR intensity distribution across the Galaxy from which Galactic diffuse emissions (GDEs) can be predicted (see, however,~\cite{2023MarinosRowellMNRAS}).
Due to the limited knowledge of the coordinates of current and past CR sources, such a model has to be done in a stochastic way, i.e., many source realisations have to be considered to cover the space of possible source coordinates appropriately.

In this work, we considered a stochastic model for hadronic GDEs produced by CR protons.
We focused on three different CR source injection and near-source transport models (Sec.~\ref{sec: source injection and near-source transport}).
We described the influence of discrete sources by determining the relative difference of GDE predictions with discrete sources to predictions of the corresponding smooth source model (Sec.~\ref{sec:stochastic vs smooth models}).
Examples of these source imprints in different source realisations can be seen in Sec.~\ref{sec: A glimpse into the diffuse sky}.
Then, we determined the most extreme deviations in thousands of realisations to describe the range of possible excesses caused by individual CR sources in the different models in Sec.~\ref{sec:The most extreme deviations}.

In Sec.~\ref{sec:High-statistics assessment of diffuse emission distributions}, we studied GDEs along selected lines of sight (LOS) with large sample sizes of $10^6$ realisations per model.
We found that the distribution of CR intensities is key for understanding the general features of the GDE intensity distribution along a single LOS.
Stable laws and normal distributions (Sec.~\ref{sec:Cosmic rays and stable law distributions}) can be used to describe the main features of the GDE intensity distributions in the \burst{} and \credit{} scenarios.
This underlines the close connection between CR intensity fluctuations throughout the Galaxy and locally observed GDEs.
For more complicated models like the \tdd{} scenario the Monte Carlo simulation of these GDE distributions is the most reliable way to study the GDE intensity distributions in detail.

In Sec.~\ref{sec:On sky windows and profiles}, we studied the influence of discrete sources on the GDEs averaged over sky windows and in longitude profiles.
Our limited knowledge about the exact source positions and ages adds an additional model uncertainty to others like the ones linked to CR model parameters or interaction cross sections.
For the sky windows, we found that the influence of discrete sources is small in the \burst{} and \credit{} scenarios and cannot explain the gap between model predictions and LHAASO measurements~\cite{2025CaoAharonianPhRvL}.
In the \tdd{} scenario, however, the uncertainties are much larger and source stochasticity is a significant model uncertainty.
For spatial more resolved analyses like longitude profiles, the influence of discrete sources is more important, which is consistent with the morphology of source imprints on the GDEs
With the advent of higher spatial resolutions in analyses of GDEs at high energies, the  discreteness of CR sources becomes an important model uncertainty.

Finally, we investigated the correlation of GDEs at \SI{10}{\giga\electronvolt} and \SI{100}{\tera\electronvolt} in Sec.~\ref{sec:correlation results}.
We found that it depends on the general impact the discrete CR source injection and near-source transport models have on the total GDE intensities.
For the moderate \burst{} and \credit{} scenarios, the correlation of total GDE intensities at different energies is good, while it is poor for the \tdd{} scenario with more extreme effects on GDEs.
However, the correlation of deviations from smooth model predictions at different energies is influenced by the injection and transport model.
Here, it is good for the \tdd{} scenario, where the transport is energy-independent in an initial phase after injection, and poor for the other scenarios.
This marks yet another way of understanding GDEs with CR model assumptions and points out a general flaw in the correspondence between GDEs at different energies.

Whenever we discussed GDEs in this work, we were specifically referring to hadronic gamma-ray emissions.
However, as we have pointed out in Eq.~\eqref{eq:neutrino intensity connection} in Sec.~\ref{sec: diffuse emissions}, those can be related to diffuse neutrino emissions from $pp$-interactions.
This allows us to extend the conclusions from GDEs in gamma rays to those in neutrinos at the respective adjusted energies (see Eq.~\eqref{eq:neutrino intensity connection}).
Especially, this includes our discussion on the model uncertainties on GDE intensity spectra in sky windows in Sec.~\ref{sec:On sky windows and profiles}, where we found that the uncertainty due to CR source stochasticity is subdominant in the \burst{} and \credit{} scenarios, but is relevant in the \tdd{} scenario.

The results of this work improve the understanding of the influence of discrete sources on GDEs.
They clearly show the connection between CR intensities and GDEs and further our understanding of imprints of CR sources on the GDEs.
This helps to use GDE measurements, e.g. from LHAASO, as a probe for CR source injection and near-source transport models.
We showed that such studies can be expected to provide valuable insights, especially once increased spatial resolution of GDE measurements at high energies, exceeding tens of teraelectronvolts, are available.
\acknowledgments{This work has been funded by the Deutsche Forschungsgemeinschaft (DFG, German Research Foundation) — project number 490751943.
The authors also gratefully acknowledge the computing time provided to them at the NHR Center NHR4CES at RWTH Aachen University (project number p0021785).

Further, we would like to thank Leonard Kaiser, who tested the waters for the diffuse emission calculations in his Master's thesis.}
\bibliographystyle{JHEP}
\bibliography{references.bib}

\providecommand{\href}[2]{#2}\begingroup\raggedright\begin{thebibliography}{10}

\bibitem{2021AguilarAliCavasonzaPhR}
M.~{Aguilar}, L.~{Ali Cavasonza}, G.~{Ambrosi}, L.~{Arruda}, N.~{Attig},
  F.~{Barao} et~al., \emph{{The Alpha Magnetic Spectrometer (AMS) on the
  international space station: Part II - Results from the first seven years}},
  \href{https://doi.org/10.1016/j.physrep.2020.09.003}{\emph{\physrep}
  {\bfseries 894} (2021) 1}.

\bibitem{2024AlemannoAltomarePhRvD}
F.~{Alemanno}, C.~{Altomare}, Q.~{An}, P.~{Azzarello}, F.C.T.~{Barbato},
  P.~{Bernardini} et~al., \emph{{Measurement of the cosmic p +He energy
  spectrum from 50 GeV to 0.5 PeV with the DAMPE space mission}},
  \href{https://doi.org/10.1103/PhysRevD.109.L121101}{\emph{\prd} {\bfseries
  109} (2024) L121101} [\href{https://arxiv.org/abs/2304.00137}{{\ttfamily
  2304.00137}}].

\bibitem{2019SchaelAtanasyanNIMPA}
S.~{Schael}, A.~{Atanasyan}, J.~{Berdugo}, T.~{Bretz}, M.~{Czupalla},
  B.~{Dachwald} et~al., \emph{{AMS-100: The next generation magnetic
  spectrometer in space - An international science platform for physics and
  astrophysics at Lagrange point 2}},
  \href{https://doi.org/10.1016/j.nima.2019.162561}{\emph{Nuclear Instruments
  and Methods in Physics Research A} {\bfseries 944} (2019) 162561}
  [\href{https://arxiv.org/abs/1907.04168}{{\ttfamily 1907.04168}}].

\bibitem{2025StallLooApJL}
A.~{Stall}, C.K.~{Loo} and P.~{Mertsch}, \emph{{Investigating the CREDIT
  History of Supernova Remnants as Cosmic-Ray Sources}},
  \href{https://doi.org/10.3847/2041-8213/adaea8}{\emph{\apjl} {\bfseries 980}
  (2025) L21} [\href{https://arxiv.org/abs/2409.11012}{{\ttfamily
  2409.11012}}].

\bibitem{2021TibaldoGaggeroUniv}
L.~{Tibaldo}, D.~{Gaggero} and P.~{Martin}, \emph{{Gamma Rays as Probes of
  Cosmic-Ray Propagation and Interactions in Galaxies}},
  \href{https://doi.org/10.3390/universe7050141}{\emph{Universe} {\bfseries 7}
  (2021) 141} [\href{https://arxiv.org/abs/2103.16423}{{\ttfamily
  2103.16423}}].

\bibitem{2012AckermannAjelloApJ}
M.~{Ackermann}, M.~{Ajello}, W.B.~{Atwood}, L.~{Baldini}, J.~{Ballet},
  G.~{Barbiellini} et~al., \emph{{Fermi-LAT Observations of the Diffuse
  {\ensuremath{\gamma}}-Ray Emission: Implications for Cosmic Rays and the
  Interstellar Medium}},
  \href{https://doi.org/10.1088/0004-637X/750/1/3}{\emph{\apj} {\bfseries 750}
  (2012) 3} [\href{https://arxiv.org/abs/1202.4039}{{\ttfamily 1202.4039}}].

\bibitem{2023CaoAharonianPhRvL}
Z.~{Cao}, F.~{Aharonian}, Q.~{An}, B.~{Axikegu}, Y.~X., Y.W.~{Bao},
  D.~{Bastieri} et~al., \emph{{Measurement of Ultra-High-Energy Diffuse
  Gamma-Ray Emission of the Galactic Plane from 10 TeV to 1 PeV with
  LHAASO-KM2A}},
  \href{https://doi.org/10.1103/PhysRevLett.131.151001}{\emph{\prl} {\bfseries
  131} (2023) 151001} [\href{https://arxiv.org/abs/2305.05372}{{\ttfamily
  2305.05372}}].

\bibitem{2025CaoAharonianPhRvL}
Z.~{Cao}, F.~{Aharonian}, {Axikegu}, Y.X.~{Bai}, Y.W.~{Bao}, D.~{Bastieri}
  et~al., \emph{{Measurement of Very-High-Energy Diffuse Gamma-Ray Emissions
  from the Galactic Plane with LHAASO-WCDA}},
  \href{https://doi.org/10.1103/PhysRevLett.134.081002}{\emph{\prl} {\bfseries
  134} (2025) 081002} [\href{https://arxiv.org/abs/2411.16021}{{\ttfamily
  2411.16021}}].

\bibitem{2023IcecubeCollaborationAbbasiSci}
{Icecube Collaboration}, R.~{Abbasi}, M.~{Ackermann}, J.~{Adams},
  J.A.~{Aguilar}, M.~{Ahlers} et~al., \emph{{Observation of high-energy
  neutrinos from the Galactic plane}},
  \href{https://doi.org/10.1126/science.adc9818}{\emph{Science} {\bfseries 380}
  (2023) 1338} [\href{https://arxiv.org/abs/2307.04427}{{\ttfamily
  2307.04427}}].

\bibitem{2006KelnerAharonianPhRvD}
S.R.~{Kelner}, F.A.~{Aharonian} and V.V.~{Bugayov}, \emph{{Energy spectra of
  gamma rays, electrons, and neutrinos produced at proton-proton interactions
  in the very high energy regime}},
  \href{https://doi.org/10.1103/PhysRevD.74.034018}{\emph{\prd} {\bfseries 74}
  (2006) 034018} [\href{https://arxiv.org/abs/astro-ph/0606058}{{\ttfamily
  astro-ph/0606058}}].

\bibitem{2021KoldobskiyKachelriessPhRvD}
S.~{Koldobskiy}, M.~{Kachelrie{\ss}}, A.~{Lskavyan}, A.~{Neronov},
  S.~{Ostapchenko} and D.V.~{Semikoz}, \emph{{Energy spectra of secondaries in
  proton-proton interactions}},
  \href{https://doi.org/10.1103/PhysRevD.104.123027}{\emph{\prd} {\bfseries
  104} (2021) 123027} [\href{https://arxiv.org/abs/2110.00496}{{\ttfamily
  2110.00496}}].

\bibitem{2023SchweferMertschApJ}
G.~{Schwefer}, P.~{Mertsch} and C.~{Wiebusch}, \emph{{Diffuse Emission of
  Galactic High-energy Neutrinos from a Global Fit of Cosmic Rays}},
  \href{https://doi.org/10.3847/1538-4357/acc1e2}{\emph{\apj} {\bfseries 949}
  (2023) 16} [\href{https://arxiv.org/abs/2211.15607}{{\ttfamily 2211.15607}}].

\bibitem{2025VecchiottiPeronJCAP}
V.~{Vecchiotti}, G.~{Peron}, E.~{Amato}, S.~{Menchiari}, G.~{Morlino},
  G.~{Pagliaroli} et~al., \emph{{Interpreting the LHAASO Galactic diffuse
  emission data}},
  \href{https://doi.org/10.1088/1475-7516/2025/09/041}{\emph{\jcap} {\bfseries
  2025} (2025) 041} [\href{https://arxiv.org/abs/2411.11439}{{\ttfamily
  2411.11439}}].

\bibitem{2019GabiciEvoliIJMPD}
S.~{Gabici}, C.~{Evoli}, D.~{Gaggero}, P.~{Lipari}, P.~{Mertsch}, E.~{Orlando}
  et~al., \emph{{The origin of Galactic cosmic rays: Challenges to the standard
  paradigm}},
  \href{https://doi.org/10.1142/S0218271819300222}{\emph{International Journal
  of Modern Physics D} {\bfseries 28} (2019) 1930022}
  [\href{https://arxiv.org/abs/1903.11584}{{\ttfamily 1903.11584}}].

\bibitem{1977KrymskiiSPhD}
G.F.~{Krymskii}, \emph{{A regular mechanism for the acceleration of charged
  particles on the front of a shock wave}}, {\emph{Soviet Physics Doklady}
  {\bfseries 22} (1977) 327}.

\bibitem{2021MorlinoBlasiMNRAS}
G.~{Morlino}, P.~{Blasi}, E.~{Peretti} and P.~{Cristofari}, \emph{{Particle
  acceleration in winds of star clusters}},
  \href{https://doi.org/10.1093/mnras/stab690}{\emph{\mnras} {\bfseries 504}
  (2021) 6096} [\href{https://arxiv.org/abs/2102.09217}{{\ttfamily
  2102.09217}}].

\bibitem{2017GenoliniSalatiA&A}
Y.~{Genolini}, P.~{Salati}, P.D.~{Serpico} and R.~{Taillet}, \emph{{Stable laws
  and cosmic ray physics}},
  \href{https://doi.org/10.1051/0004-6361/201629903}{\emph{\aap} {\bfseries
  600} (2017) A68} [\href{https://arxiv.org/abs/1610.02010}{{\ttfamily
  1610.02010}}].

\bibitem{2011MertschJCAP}
P.~{Mertsch}, \emph{{Cosmic ray electrons and positrons from discrete
  stochastic sources}},
  \href{https://doi.org/10.1088/1475-7516/2011/02/031}{\emph{\jcap} {\bfseries
  2011} (2011) 031} [\href{https://arxiv.org/abs/1012.0805}{{\ttfamily
  1012.0805}}].

\bibitem{2021EvoliAmatoPhRvDa}
C.~{Evoli}, E.~{Amato}, P.~{Blasi} and R.~{Aloisio}, \emph{{Stochastic nature
  of Galactic cosmic-ray sources}},
  \href{https://doi.org/10.1103/PhysRevD.104.123029}{\emph{\prd} {\bfseries
  104} (2021) 123029} [\href{https://arxiv.org/abs/2111.01171}{{\ttfamily
  2111.01171}}].

\bibitem{2012BlasiAmatoJCAP}
P.~{Blasi} and E.~{Amato}, \emph{{Diffusive propagation of cosmic rays from
  supernova remnants in the Galaxy. I: spectrum and chemical composition}},
  \href{https://doi.org/10.1088/1475-7516/2012/01/010}{\emph{\jcap} {\bfseries
  2012} (2012) 010} [\href{https://arxiv.org/abs/1105.4521}{{\ttfamily
  1105.4521}}].

\bibitem{2021PhanSchulzePhRvL}
V.H.M.~{Phan}, F.~{Schulze}, P.~{Mertsch}, S.~{Recchia} and S.~{Gabici},
  \emph{{Stochastic Fluctuations of Low-Energy Cosmic Rays and the
  Interpretation of Voyager Data}},
  \href{https://doi.org/10.1103/PhysRevLett.127.141101}{\emph{\prl} {\bfseries
  127} (2021) 141101} [\href{https://arxiv.org/abs/2105.00311}{{\ttfamily
  2105.00311}}].

\bibitem{2015KachelriessNeronovPhRvL}
M.~{Kachelrie{\ss}}, A.~{Neronov} and D.V.~{Semikoz}, \emph{{Signatures of a
  Two Million Year Old Supernova in the Spectra of Cosmic Ray Protons,
  Antiprotons, and Positrons}},
  \href{https://doi.org/10.1103/PhysRevLett.115.181103}{\emph{\prl} {\bfseries
  115} (2015) 181103} [\href{https://arxiv.org/abs/1504.06472}{{\ttfamily
  1504.06472}}].

\bibitem{2023ThalerKissmannAPh}
J.~{Thaler}, R.~{Kissmann} and O.~{Reimer}, \emph{{Cosmic-ray propagation under
  consideration of a spatially resolved source distribution}},
  \href{https://doi.org/10.1016/j.astropartphys.2022.102776}{\emph{Astroparticle
  Physics} {\bfseries 144} (2023) 102776}
  [\href{https://arxiv.org/abs/2209.02295}{{\ttfamily 2209.02295}}].

\bibitem{2023MarinosRowellMNRAS}
P.D.~{Marinos}, G.P.~{Rowell}, T.A.~{Porter} and G.~{J{\'o}hannesson},
  \emph{{The steady-state multi-TeV diffuse {\ensuremath{\gamma}}-ray emission
  predicted with GALPROP and prospects for the Cherenkov Telescope Array}},
  \href{https://doi.org/10.1093/mnras/stac3222}{\emph{\mnras} {\bfseries 518}
  (2023) 5036} [\href{https://arxiv.org/abs/2211.01619}{{\ttfamily
  2211.01619}}].

\bibitem{2025MarinosPorterApJ}
P.D.~{Marinos}, T.A.~{Porter}, G.P.~{Rowell}, G.~{J{\'o}hannesson} and
  I.V.~{Moskalenko}, \emph{{On the Temporal Variability of the Galactic
  Multi-TeV Interstellar Emissions}},
  \href{https://doi.org/10.3847/1538-4357/adadfb}{\emph{\apj} {\bfseries 981}
  (2025) 93} [\href{https://arxiv.org/abs/2411.03553}{{\ttfamily 2411.03553}}].

\bibitem{2025KaciGiacintiJCAP}
S.~{Kaci} and G.~{Giacinti}, \emph{{Imprints of PeV cosmic-ray sources on the
  diffuse gamma-ray emission}},
  \href{https://doi.org/10.1088/1475-7516/2025/01/049}{\emph{\jcap} {\bfseries
  2025} (2025) 049} [\href{https://arxiv.org/abs/2406.11015}{{\ttfamily
  2406.11015}}].

\bibitem{2018LipariVernettoPhRvD}
P.~{Lipari} and S.~{Vernetto}, \emph{{Diffuse Galactic gamma-ray flux at very
  high energy}}, \href{https://doi.org/10.1103/PhysRevD.98.043003}{\emph{\prd}
  {\bfseries 98} (2018) 043003}
  [\href{https://arxiv.org/abs/1804.10116}{{\ttfamily 1804.10116}}].

\bibitem{2018MertschJCAP}
P.~{Mertsch}, \emph{{Stochastic cosmic ray sources and the TeV break in the
  all-electron spectrum}},
  \href{https://doi.org/10.1088/1475-7516/2018/11/045}{\emph{\jcap} {\bfseries
  2018} (2018) 045} [\href{https://arxiv.org/abs/1809.05104}{{\ttfamily
  1809.05104}}].

\bibitem{2025SodingEdenhoferA&A}
L.~{S{\"o}ding}, G.~{Edenhofer}, T.A.~{En{\ss}lin}, P.~{Frank}, R.~{Kissmann},
  V.H.M.~{Phan} et~al., \emph{{Spatially coherent 3D distributions of HI and CO
  in the Milky Way}},
  \href{https://doi.org/10.1051/0004-6361/202451361}{\emph{\aap} {\bfseries
  693} (2025) A139} [\href{https://arxiv.org/abs/2407.02859}{{\ttfamily
  2407.02859}}].

\bibitem{1991vandenBerghTammannARA&A}
S.~{van den Bergh} and G.A.~{Tammann}, \emph{{Galactic and extragalactic
  supernova rates.}},
  \href{https://doi.org/10.1146/annurev.aa.29.090191.002051}{\emph{\araa}
  {\bfseries 29} (1991) 363}.

\bibitem{1994TammannLoefflerApJS}
G.A.~{Tammann}, W.~{Loeffler} and A.~{Schroeder}, \emph{{The Galactic Supernova
  Rate}}, \href{https://doi.org/10.1086/192002}{\emph{\apjs} {\bfseries 92}
  (1994) 487}.

\bibitem{2021EvoliAmatoPhRvDb}
C.~{Evoli}, E.~{Amato}, P.~{Blasi} and R.~{Aloisio}, \emph{{Galactic factories
  of cosmic-ray electrons and positrons}},
  \href{https://doi.org/10.1103/PhysRevD.103.083010}{\emph{\prd} {\bfseries
  103} (2021) 083010} [\href{https://arxiv.org/abs/2010.11955}{{\ttfamily
  2010.11955}}].

\bibitem{2010Steiman-CameronWolfireApJ}
T.Y.~{Steiman-Cameron}, M.~{Wolfire} and D.~{Hollenbach}, \emph{{COBE and the
  Galactic Interstellar Medium: Geometry of the Spiral Arms from FIR Cooling
  Lines}}, \href{https://doi.org/10.1088/0004-637X/722/2/1460}{\emph{\apj}
  {\bfseries 722} (2010) 1460}.

\bibitem{2001FerriereRvMP}
K.M.~{Ferri{\`e}re}, \emph{{The interstellar environment of our galaxy}},
  \href{https://doi.org/10.1103/RevModPhys.73.1031}{\emph{Reviews of Modern
  Physics} {\bfseries 73} (2001) 1031}
  [\href{https://arxiv.org/abs/astro-ph/0106359}{{\ttfamily
  astro-ph/0106359}}].

\bibitem{1990BerezinskiiBulanovacr}
V.S.~{Berezinskii}, S.V.~{Bulanov}, V.A.~{Dogiel} and V.S.~{Ptuskin},
  \emph{{Astrophysics of cosmic rays}}, North-Holland (1990).

\bibitem{2012BlasiAmatoPhRvL}
P.~{Blasi}, E.~{Amato} and P.D.~{Serpico}, \emph{{Spectral Breaks as a
  Signature of Cosmic Ray Induced Turbulence in the Galaxy}},
  \href{https://doi.org/10.1103/PhysRevLett.109.061101}{\emph{\prl} {\bfseries
  109} (2012) 061101} [\href{https://arxiv.org/abs/1207.3706}{{\ttfamily
  1207.3706}}].

\bibitem{2018EvoliBlasiPhRvL}
C.~{Evoli}, P.~{Blasi}, G.~{Morlino} and R.~{Aloisio}, \emph{{Origin of the
  Cosmic Ray Galactic Halo Driven by Advected Turbulence and Self-Generated
  Waves}}, \href{https://doi.org/10.1103/PhysRevLett.121.021102}{\emph{\prl}
  {\bfseries 121} (2018) 021102}
  [\href{https://arxiv.org/abs/1806.04153}{{\ttfamily 1806.04153}}].

\bibitem{2021MalkovMoskalenkoApJ}
M.A.~{Malkov} and I.V.~{Moskalenko}, \emph{{The TeV Cosmic-Ray Bump: A Message
  from the Epsilon Indi or Epsilon Eridani Star?}},
  \href{https://doi.org/10.3847/1538-4357/abe855}{\emph{\apj} {\bfseries 911}
  (2021) 151} [\href{https://arxiv.org/abs/2010.02826}{{\ttfamily
  2010.02826}}].

\bibitem{2021FornieriGaggeroPhRvD}
O.~{Fornieri}, D.~{Gaggero}, D.~{Guberman}, L.~{Brahimi}, P.D.L.T.~{Luque} and
  A.~{Marcowith}, \emph{{Diffusive origin for the cosmic-ray spectral hardening
  reveals signatures of a nearby source in the leptons and protons data}},
  \href{https://doi.org/10.1103/PhysRevD.104.103013}{\emph{\prd} {\bfseries
  104} (2021) 103013} [\href{https://arxiv.org/abs/2007.15321}{{\ttfamily
  2007.15321}}].

\bibitem{2001MalkovDruryRPPh}
M.A.~{Malkov} and L.O.~{Drury}, \emph{{Nonlinear theory of diffusive
  acceleration of particles by shock waves}},
  \href{https://doi.org/10.1088/0034-4885/64/4/201}{\emph{Reports on Progress
  in Physics} {\bfseries 64} (2001) 429}.

\bibitem{1978BellMNRAS}
A.R.~{Bell}, \emph{{The acceleration of cosmic rays in shock fronts - I.}},
  \href{https://doi.org/10.1093/mnras/182.2.147}{\emph{\mnras} {\bfseries 182}
  (1978) 147}.

\bibitem{2020CaprioliHaggertyApJ}
D.~{Caprioli}, C.C.~{Haggerty} and P.~{Blasi}, \emph{{Kinetic Simulations of
  Cosmic-Ray-modified Shocks. II. Particle Spectra}},
  \href{https://doi.org/10.3847/1538-4357/abbe05}{\emph{\apj} {\bfseries 905}
  (2020) 2} [\href{https://arxiv.org/abs/2009.00007}{{\ttfamily 2009.00007}}].

\bibitem{2011BellSchureMNRAS}
A.R.~{Bell}, K.M.~{Schure} and B.~{Reville}, \emph{{Cosmic ray acceleration at
  oblique shocks}},
  \href{https://doi.org/10.1111/j.1365-2966.2011.19571.x}{\emph{\mnras}
  {\bfseries 418} (2011) 1208}
  [\href{https://arxiv.org/abs/1108.0582}{{\ttfamily 1108.0582}}].

\bibitem{1959SyrovatskiiSvA}
S.I.~{Syrovatskii}, \emph{{The Distribution of Relativistic Electrons in the
  Galaxy and the Spectrum of Synchrotron Radio Emission.}}, {\emph{\sovast}
  {\bfseries 3} (1959) 22}.

\bibitem{2022RecchiaGalliA&A}
S.~{Recchia}, D.~{Galli}, L.~{Nava}, M.~{Padovani}, S.~{Gabici}, A.~{Marcowith}
  et~al., \emph{{Grammage of cosmic rays in the proximity of supernova remnants
  embedded in a partially ionized medium}},
  \href{https://doi.org/10.1051/0004-6361/202142558}{\emph{\aap} {\bfseries
  660} (2022) A57} [\href{https://arxiv.org/abs/2106.04948}{{\ttfamily
  2106.04948}}].

\bibitem{2012ThoudamHorandelMNRAS}
S.~{Thoudam} and J.R.~{H{\"o}randel}, \emph{{Nearby supernova remnants and the
  cosmic ray spectral hardening at high energies}},
  \href{https://doi.org/10.1111/j.1365-2966.2011.20385.x}{\emph{\mnras}
  {\bfseries 421} (2012) 1209}
  [\href{https://arxiv.org/abs/1112.3020}{{\ttfamily 1112.3020}}].

\bibitem{2009CaprioliBlasiMNRAS}
D.~{Caprioli}, P.~{Blasi} and E.~{Amato}, \emph{{On the escape of particles
  from cosmic ray modified shocks}},
  \href{https://doi.org/10.1111/j.1365-2966.2008.14298.x}{\emph{\mnras}
  {\bfseries 396} (2009) 2065}
  [\href{https://arxiv.org/abs/0807.4259}{{\ttfamily 0807.4259}}].

\bibitem{2009GabiciAharonianMNRAS}
S.~{Gabici}, F.A.~{Aharonian} and S.~{Casanova}, \emph{{Broad-band non-thermal
  emission from molecular clouds illuminated by cosmic rays from nearby
  supernova remnants}},
  \href{https://doi.org/10.1111/j.1365-2966.2009.14832.x}{\emph{\mnras}
  {\bfseries 396} (2009) 1629}
  [\href{https://arxiv.org/abs/0901.4549}{{\ttfamily 0901.4549}}].

\bibitem{2019CelliMorlinoMNRAS}
S.~{Celli}, G.~{Morlino}, S.~{Gabici} and F.A.~{Aharonian}, \emph{{Exploring
  particle escape in supernova remnants through gamma rays}},
  \href{https://doi.org/10.1093/mnras/stz2897}{\emph{\mnras} {\bfseries 490}
  (2019) 4317} [\href{https://arxiv.org/abs/1906.09454}{{\ttfamily
  1906.09454}}].

\bibitem{2022JacobsMertschJCAP}
H.~{Jacobs}, P.~{Mertsch} and V.H.M.~{Phan}, \emph{{Self-confinement of
  low-energy cosmic rays around supernova remnants}},
  \href{https://doi.org/10.1088/1475-7516/2022/05/024}{\emph{\jcap} {\bfseries
  2022} (2022) 024} [\href{https://arxiv.org/abs/2112.09708}{{\ttfamily
  2112.09708}}].

\bibitem{2022MukhopadhyayLindenPhRvD}
P.~{Mukhopadhyay} and T.~{Linden}, \emph{{Self-generated cosmic-ray turbulence
  can explain the morphology of TeV halos}},
  \href{https://doi.org/10.1103/PhysRevD.105.123008}{\emph{\prd} {\bfseries
  105} (2022) 123008} [\href{https://arxiv.org/abs/2111.01143}{{\ttfamily
  2111.01143}}].

\bibitem{2018FangBiApJ}
K.~{Fang}, X.-J.~{Bi}, P.-F.~{Yin} and Q.~{Yuan}, \emph{{Two-zone Diffusion of
  Electrons and Positrons from Geminga Explains the Positron Anomaly}},
  \href{https://doi.org/10.3847/1538-4357/aad092}{\emph{\apj} {\bfseries 863}
  (2018) 30} [\href{https://arxiv.org/abs/1803.02640}{{\ttfamily 1803.02640}}].

\bibitem{2016GaisserEngelcrpp}
T.K.~{Gaisser}, R.~{Engel} and E.~{Resconi}, \emph{{Cosmic Rays and Particle
  Physics}}, Cambridge University Press (2016).

\bibitem{2014AhlersMurasePhRvD}
M.~{Ahlers} and K.~{Murase}, \emph{{Probing the Galactic origin of the IceCube
  excess with gamma rays}},
  \href{https://doi.org/10.1103/PhysRevD.90.023010}{\emph{\prd} {\bfseries 90}
  (2014) 023010} [\href{https://arxiv.org/abs/1309.4077}{{\ttfamily
  1309.4077}}].

\bibitem{2023KaiserMaster}
L.~Kaiser, \emph{The diffuse emission of the galaxy at high energies},
  {M}aster's thesis, RWTH Aachen University, TTK, Aachen, Germany, October,
  2023.

\bibitem{2005GorskiHivonApJ}
K.M.~{G{\'o}rski}, E.~{Hivon}, A.J.~{Banday}, B.D.~{Wandelt}, F.K.~{Hansen},
  M.~{Reinecke} et~al., \emph{{HEALPix: A Framework for High-Resolution
  Discretization and Fast Analysis of Data Distributed on the Sphere}},
  \href{https://doi.org/10.1086/427976}{\emph{\apj} {\bfseries 622} (2005) 759}
  [\href{https://arxiv.org/abs/astro-ph/0409513}{{\ttfamily
  astro-ph/0409513}}].

\bibitem{2015CasandjianApJ}
J.-M.~{Casandjian}, \emph{{Local H i Emissivity Measured with Fermi-LAT and
  Implications for Cosmic-Ray Spectra}},
  \href{https://doi.org/10.1088/0004-637X/806/2/240}{\emph{\apj} {\bfseries
  806} (2015) 240} [\href{https://arxiv.org/abs/1506.00047}{{\ttfamily
  1506.00047}}].

\bibitem{2004HondaKajitaPhRvD}
M.~{Honda}, T.~{Kajita}, K.~{Kasahara} and S.~{Midorikawa}, \emph{{New
  calculation of the atmospheric neutrino flux in a three-dimensional scheme}},
  \href{https://doi.org/10.1103/PhysRevD.70.043008}{\emph{\prd} {\bfseries 70}
  (2004) 043008} [\href{https://arxiv.org/abs/astro-ph/0404457}{{\ttfamily
  astro-ph/0404457}}].

\bibitem{2019KachelriessMoskalenkoCoPhC}
M.~{Kachelrie{\ss}}, I.V.~{Moskalenko} and S.~{Ostapchenko}, \emph{{AAfrag:
  Interpolation routines for Monte Carlo results on secondary production in
  proton-proton, proton-nucleus and nucleus-nucleus interactions}},
  \href{https://doi.org/10.1016/j.cpc.2019.08.001}{\emph{Computer Physics
  Communications} {\bfseries 245} (2019) 106846}
  [\href{https://arxiv.org/abs/1904.05129}{{\ttfamily 1904.05129}}].

\bibitem{jax2018github}
J.~Bradbury, R.~Frostig, P.~Hawkins, M.J.~Johnson, C.~Leary, D.~Maclaurin
  et~al., \emph{{JAX}: composable transformations of {P}ython+{N}um{P}y
  programs},  2018.

\bibitem{2020HarrisMillmanNatur}
C.R.~{Harris}, K.J.~{Millman}, S.J.~{van der Walt}, R.~{Gommers},
  P.~{Virtanen}, D.~{Cournapeau} et~al., \emph{{Array programming with NumPy}},
  \href{https://doi.org/10.1038/s41586-020-2649-2}{\emph{\nat} {\bfseries 585}
  (2020) 357} [\href{https://arxiv.org/abs/2006.10256}{{\ttfamily
  2006.10256}}].

\bibitem{2021DundovicEvoliA&A}
A.~{Dundovic}, C.~{Evoli}, D.~{Gaggero} and D.~{Grasso}, \emph{{Simulating the
  Galactic multi-messenger emissions with HERMES}},
  \href{https://doi.org/10.1051/0004-6361/202140801}{\emph{\aap} {\bfseries
  653} (2021) A18} [\href{https://arxiv.org/abs/2105.13165}{{\ttfamily
  2105.13165}}].

\bibitem{2022PorterJohannessonApJS}
T.A.~{Porter}, G.~{J{\'o}hannesson} and I.V.~{Moskalenko}, \emph{{The GALPROP
  Cosmic-ray Propagation and Nonthermal Emissions Framework: Release v57}},
  \href{https://doi.org/10.3847/1538-4365/ac80f6}{\emph{\apjs} {\bfseries 262}
  (2022) 30} [\href{https://arxiv.org/abs/2112.12745}{{\ttfamily 2112.12745}}].

\bibitem{2018PorterRowellPhRvD}
T.A.~{Porter}, G.P.~{Rowell}, G.~{J{\'o}hannesson} and I.V.~{Moskalenko},
  \emph{{Galactic PeVatrons and helping to find them: Effects of galactic
  absorption on the observed spectra of very high energy {\ensuremath{\gamma}}
  -ray sources}}, \href{https://doi.org/10.1103/PhysRevD.98.041302}{\emph{\prd}
  {\bfseries 98} (2018) 041302}
  [\href{https://arxiv.org/abs/1808.07596}{{\ttfamily 1808.07596}}].

\bibitem{2012RobitailleChurchwellA&A}
T.P.~{Robitaille}, E.~{Churchwell}, R.A.~{Benjamin}, B.A.~{Whitney}, K.~{Wood},
  B.L.~{Babler} et~al., \emph{{A self-consistent model of Galactic stellar and
  dust infrared emission and the abundance of polycyclic aromatic
  hydrocarbons}},
  \href{https://doi.org/10.1051/0004-6361/201219073}{\emph{\aap} {\bfseries
  545} (2012) A39} [\href{https://arxiv.org/abs/1208.4606}{{\ttfamily
  1208.4606}}].

\bibitem{2016VernettoLipariPhRvD}
S.~{Vernetto} and P.~{Lipari}, \emph{{Absorption of very high energy gamma rays
  in the Milky Way}},
  \href{https://doi.org/10.1103/PhysRevD.94.063009}{\emph{\prd} {\bfseries 94}
  (2016) 063009} [\href{https://arxiv.org/abs/1608.01587}{{\ttfamily
  1608.01587}}].

\bibitem{1925Levy}
P.~Lévy, \emph{Calcul des probabilités}, Gauthier-Villars, Paris (1925).

\bibitem{1999UchaikinZolotarev}
V.V.~Uchaikin and V.M.~Zolotarev, \emph{Chance and Stability}, De Gruyter,
  Berlin, Boston (1999),
  \href{https://doi.org/10.1515/9783110935974}{10.1515/9783110935974}.

\bibitem{2020Nolan}
J.P.~Nolan, \emph{Univariate stable distributions}, Springer Series in
  Operations Research and Financial Engineering, Springer Nature, Cham,
  Switzerland, 2020~ed. (Sept., 2020),
  \href{https://doi.org/10.1007/978-3-030-52915-4}{10.1007/978-3-030-52915-4}.

\bibitem{1986McCulloch}
J.H.~McCulloch, \emph{Simple consistent estimators of stable distribution
  parameters},
  \href{https://doi.org/10.1080/03610918608812563}{\emph{Communications in
  Statistics - Simulation and Computation} {\bfseries 15} (1986) 1109–1136}.

\bibitem{2020VirtanenGommersNatMe}
P.~{Virtanen}, R.~{Gommers}, T.E.~{Oliphant}, M.~{Haberland}, T.~{Reddy},
  D.~{Cournapeau} et~al., \emph{{SciPy 1.0: fundamental algorithms for
  scientific computing in Python}},
  \href{https://doi.org/10.1038/s41592-019-0686-2}{\emph{Nature Methods}
  {\bfseries 17} (2020) 261}
  [\href{https://arxiv.org/abs/1907.10121}{{\ttfamily 1907.10121}}].

\bibitem{2021AmenomoriBaoPhRvL}
M.~{Amenomori}, Y.W.~{Bao}, X.J.~{Bi}, D.~{Chen}, T.L.~{Chen}, W.Y.~{Chen}
  et~al., \emph{{First Detection of sub-PeV Diffuse Gamma Rays from the
  Galactic Disk: Evidence for Ubiquitous Galactic Cosmic Rays beyond PeV
  Energies}}, \href{https://doi.org/10.1103/PhysRevLett.126.141101}{\emph{\prl}
  {\bfseries 126} (2021) 141101}
  [\href{https://arxiv.org/abs/2104.05181}{{\ttfamily 2104.05181}}].

\bibitem{2015GaggeroUrbanoPhRvD}
D.~{Gaggero}, A.~{Urbano}, M.~{Valli} and P.~{Ullio}, \emph{{Gamma-ray sky
  points to radial gradients in cosmic-ray transport}},
  \href{https://doi.org/10.1103/PhysRevD.91.083012}{\emph{\prd} {\bfseries 91}
  (2015) 083012} [\href{https://arxiv.org/abs/1411.7623}{{\ttfamily
  1411.7623}}].

\bibitem{2015GaggeroGrassoApJL}
D.~{Gaggero}, D.~{Grasso}, A.~{Marinelli}, A.~{Urbano} and M.~{Valli},
  \emph{{The Gamma-Ray and Neutrino Sky: A Consistent Picture of Fermi-LAT,
  Milagro, and IceCube Results}},
  \href{https://doi.org/10.1088/2041-8205/815/2/L25}{\emph{\apjl} {\bfseries
  815} (2015) L25} [\href{https://arxiv.org/abs/1504.00227}{{\ttfamily
  1504.00227}}].

\bibitem{2023GiacintiSemikozarXiv}
G.~{Giacinti} and D.~{Semikoz}, \emph{{Model of Cosmic Ray Propagation in the
  Milky Way at the Knee}},
  \href{https://doi.org/10.48550/arXiv.2305.10251}{\emph{arXiv e-prints} (2023)
  arXiv:2305.10251} [\href{https://arxiv.org/abs/2305.10251}{{\ttfamily
  2305.10251}}].

\bibitem{2025FredianiKramerJCAP}
N.~{Frediani}, M.~{Kr{\"a}mer}, P.~{Mertsch} and K.~{Nippel}, \emph{{SECRET:
  Stochasticity Emulator for Cosmic Ray Electrons}},
  \href{https://doi.org/10.1088/1475-7516/2025/08/007}{\emph{\jcap} {\bfseries
  2025} (2025) 007} [\href{https://arxiv.org/abs/2501.06011}{{\ttfamily
  2501.06011}}].

\bibitem{2020EvoliMorlinoPhRvD}
C.~{Evoli}, G.~{Morlino}, P.~{Blasi} and R.~{Aloisio}, \emph{{AMS-02 beryllium
  data and its implication for cosmic ray transport}},
  \href{https://doi.org/10.1103/PhysRevD.101.023013}{\emph{\prd} {\bfseries
  101} (2020) 023013} [\href{https://arxiv.org/abs/1910.04113}{{\ttfamily
  1910.04113}}].

\bibitem{1996Whittaker_Watson}
E.T.~Whittaker and G.N.~Watson, \emph{A Course of Modern Analysis}, Cambridge
  Mathematical Library, Cambridge University Press, 4~ed. (1996),
  \href{https://doi.org/10.1017/CBO9780511608759}{10.1017/CBO9780511608759}.

\bibitem{2009FixsenApJ}
D.J.~{Fixsen}, \emph{{The Temperature of the Cosmic Microwave Background}},
  \href{https://doi.org/10.1088/0004-637X/707/2/916}{\emph{\apj} {\bfseries
  707} (2009) 916} [\href{https://arxiv.org/abs/0911.1955}{{\ttfamily
  0911.1955}}].

\end{thebibliography}\endgroup
\appendix
\section{Details of the analytical solution}\label{app: Jacobi}
In our transport model, we impose free-escape boundary conditions at the edge of the Galactic halo ($z=\pm H$).
The Green's function $G\left(t, \mathbf{x}; t_i, \mathbf{x}_i\right)$ (see Eq.~\eqref{eq: analytic Greens function}) solving the transport Eq.~\eqref{eq:transport_equation} for discrete sources gets a correction that we capture by the function $\vartheta\left(z, z_i, \sigma^2, H\right)$.
Here, we are going to describe how this function can be derived by the method of \textit{mirror-charges}, commonly known in electrostatics problems.
This extends the discussion in Ref.~\cite{2011MertschJCAP} as we are searching for the implementation of a solution for the case of the evaluation of the Green's function at arbitrary $z$-values and for source positions out of the plane ($z_i \neq 0$).

The solution of the transport Eq.~\eqref{eq:transport_equation} without a boundary condition (free) is given by
\begin{equation}
    G_{\text{free}}\left(t, \mathbf{x}; t_i, \mathbf{x}_i\right) =  \frac{Q_{\mathcal{R}}\left(\mathcal{R}\right)}{\left(2 \pi \sigma^2\right)^{\frac{3}{2}}} e^{-\frac{\left(\mathbf{x}_i-\mathbf{x}\right)^2}{2 \sigma^2}} \, ,
\end{equation}
where the variance $\sigma^2$ does neither depend on $\mathbf{x}$ nor $\mathbf{x}_i$ as described in Eq.~\eqref{eq: variance Green's function}.

The boundary condition can now be satisfied by placing sources with alternating signs (\textit{mirror-charges}) on the opposite site of the free-escape boundaries.
As there are two such free-escape boundaries, an infinite amount of such \textit{mirror-charges} have to be placed to satisfy the boundary conditions.
The resulting Green's function with the consideration of the boundary conditions can be written as
\begin{equation}\label{eq: mirror charges}
    G\left(t, \mathbf{x}; t_i, \mathbf{x}_i\right) = \sum_{n=-\infty}^{\infty} \left(-1\right)^n G_{\text{free}}\left(t, \mathbf{x}; t_i, \mathbf{x}_{i,n}\right) \, \text{where } \mathbf{x}_{i,n} = \begin{pmatrix}
        x_i \\ y_i \\ 2 H n + \left(-1\right)^n z_i
    \end{pmatrix} \, .
\end{equation}
It can easily be seen that $\mathbf{x}_{i,0} = \mathbf{x}_i$ denotes the position of the original source.
We define the source distance as $d_n$ and find
\begin{align}
    d_n^2 &= \left(\mathbf{x}_{i,n}-\mathbf{x}\right)^2 \nonumber \\
    &= \left(x_i - x\right)^2 + \left(y_i - y\right)^2 + \left(2 H n + \left(-1\right)^n z_i - z\right)^2 \nonumber \\
    &= \left(x_i - x\right)^2 + \left(y_i - y\right)^2 + \left(z_i - z\right)^2 + \begin{cases}
        4 H^2 n^2 - 4 H n (z-z_i) \quad &n \text{ even}\\
        4 H^2 n^2 - 4 H n (z+z_i) + 4 z z_i \quad &n \text{ odd}
    \end{cases} \, ,
\end{align}
which allows us to rewrite Eq.~\eqref{eq: mirror charges} as
\begin{equation}
    G\left(t, \mathbf{x}; t_i, \mathbf{x}_i\right) = G_{\text{free}}\left(t, \mathbf{x}; t_i, \mathbf{x}_i\right) \cdot \left(A - B\right) \, ,
\end{equation}
where
\begin{equation}
    A = \sum_{n \text{ even}} e^{-\frac{2 H^2 n^2 - 2 H n (z-z_i)}{\sigma^2}} \quad \text{and} \quad B = \sum_{n \text{ odd}} e^{-\frac{2 H^2 n^2 - 2 H n (z+z_i) + 2 z z_i}{\sigma^2}} \, .
\end{equation}

We can rewrite $A$ by setting $n = 2k$
\begin{equation}\label{eq: A}
    A = \sum_{k = -\infty}^{\infty} e^{-\frac{8 H^2 k^2 - 4 H (z-z_i) k}{\sigma^2}}
\end{equation}
and $B$ by setting $n = 2k+1$, respectively,
\begin{equation}\label{eq: B}
    B = e^{-\frac{2 H^2 + 2 z z_i - 2 H (z+z_i)}{\sigma^2}} \sum_{k = -\infty}^{\infty} e^{-\frac{8 H^2 k^2 - 4 H (z+z_i) k + 8 H^2 k}{\sigma^2}} \, .
\end{equation}

It is possible to evaluate these sums up to a finite $n$ to approximate their value.
However, we found that it is faster and numerically more stable to evaluate these sums in a different way by reformulating them as products.
We expect the function $\vartheta\left(z, z_i, \sigma^2, H\right) = A - B$ to strongly suppress the Green's function's value close to the boundary at $z = \pm H$ and if the diffusion length is similar in value to the halo height $H$.
The addition and subtraction of numbers of different magnitudes can lead to rounding errors which can lead to wrong, potentially even negative, suppression factors.
The evaluation of a product is numerically more stable and less terms have to be calculated to achieve a satisfactory result.

The infinite series in $A$ and $B$ can be written in terms of Jacobi theta functions.
The third type Jacobi theta function $\vartheta_3$, can be rewritten as a product~\cite{1996Whittaker_Watson}
\begin{equation}\label{eq: jacobi theta 3 product}
    \vartheta_3\left(l, \xi\right) = \sum_{k=-\infty}^{\infty} e^{-l k^2} e^{\xi k} = K\left(l\right) \cdot \prod_{k=1}^{\infty} \left[1 + 2 e^{-l \left(2 k - 1\right)} \cosh\left(\xi\right) + e^{-l \left(4 k - 2\right)} \right] \, ,
\end{equation}
where $K\left(l\right) = \prod_{k=1}^{\infty} \left(1 - e^{-2 l k}\right)$.

By comparing the functions $A$ and $B$ in~\eqref{eq: A} and~\eqref{eq: B} with the functional form in Eq.~\eqref{eq: jacobi theta 3 product}, we find that
\begin{align}
    \vartheta\left(z, z_i, \sigma^2, H\right) &= K\left(\frac{8 H^2}{\sigma^2}\right) \cdot \Biggl[\vartheta_3\left(\frac{8 H^2}{\sigma^2}, \frac{4 H \left(z-z_i\right)}{\sigma^2}\right)\nonumber \\
    &\quad - e^{-\frac{2 H^2 + 2 z z_i - 2 H (z+z_i)}{\sigma^2}} \vartheta_3\left(\frac{8 H^2}{\sigma^2}, \frac{4 H \left(z+z_i\right) - 8 H^2}{\sigma^2}\right)\Biggr] \, .
\end{align}

The evaluation of the emerging third type theta functions via the product formula~\eqref{eq: jacobi theta 3 product} is very fast as we find that the evaluation of $8$ factors is sufficient to achieve a satisfactory accuracy for the suppression.
\section{Absorption coefficients}\label{app:Absorption coefficients}
In this section, we want to give details about the absorption coefficients affecting gamma-ray emissions at energies above some tens of teraelectronvolts as discussed in Sec.~\ref{sec : absorption}.
The absorption mechanism is the creation of electron-positron pairs in photon-photon interactions $\gamma \gamma \rightarrow e^+ e^-$~\cite{2021DundovicEvoliA&A}.
The interaction cross section of this process is given by~\cite{2016VernettoLipariPhRvD}
\begin{equation}
    \sigma_{\gamma\gamma} = \sigma_{\mathrm{T}} \ \frac{3}{16} \left(1-\beta^2\right) \left[2 \ \beta \left(\beta^2-2\right) + \left(3 - \beta^4\right) \ln \left(\frac{1+\beta}{1-\beta}\right)\right],
\end{equation}
where $\sigma_{\mathrm{T}}$ is the Thomson cross section and $\beta = \sqrt{1-1/x}$.
The variable $x$ depends on the square of the centre of mass energy $s = 2 E_{\gamma} E_{\mathrm{ph}} \left(1- \cos \theta\right)$ via $ x = s/(4 m_e^2)$, where $m_e$ is the electron mass, $E_{\gamma}$ is the gamma-ray energy, $E_{\mathrm{ph}}$ is the energy of the target photon, and $\theta$ is the angle between the momenta of the interacting photons.
This cross section vanishes below the threshold $\sqrt{s} = 2m_e$.
It falls off quickly around its maximum which is reached for a gamma-ray energy of roughly $E_{\gamma} \approx m_e^2 / E_{\mathrm{ph}}$~\cite{2016VernettoLipariPhRvD}.

To include the absorption effects in the calculation of the GDEs according to Eq.~\eqref{eq: diffuse emissions general}, we need the optical depth $\tau\left(E, \boldsymbol{x}(l, b, s)\right)$.
It can be obtained by integrating the absorption coefficient $K$, which we will define in Eq.~\eqref{eq: absorption coefficient}, along the line of sight
\begin{equation}
    \tau\left(E, \boldsymbol{x}(l, b, s)\right) = \int_0^s \mathrm{d} t \ K\left(E_{\gamma}, \boldsymbol{\hat{u}}, t \cdot \boldsymbol{\hat{u}}\right),
\end{equation}
where $\boldsymbol{\hat{u}}$ is the unit vector pointing from the source to the observer's position.
In general, the absorption coefficient depends on the gamma ray's energy and direction, as well as its position.
It can be calculated from the differential number density of the target photon field $\frac{\mathrm{d} n_{\mathrm{ph}}}{\mathrm{d} E_{\mathrm{ph}}}$ and the interaction cross section $\sigma_{\gamma\gamma}$ via
\begin{equation}\label{eq: absorption coefficient}
    K\left(E_{\gamma}, \boldsymbol{\hat{u}}, \boldsymbol{x}\right) = \int_{E_{\mathrm{ph}}^{\mathrm{min}}}^{\infty} \mathrm{d} E_{\mathrm{ph}} \int_{4\pi} \mathrm{d} \Omega \ \frac{1-\cos \theta}{2} \ \frac{\mathrm{d} n_{\mathrm{ph}}}{\mathrm{d} E_{\mathrm{ph}}} \left(E_{\mathrm{ph}}, \boldsymbol{\hat{u}}, \boldsymbol{x}\right) \ \sigma_{\gamma\gamma} \left(E_{\gamma}, E_{\mathrm{ph}}, \theta\right).
\end{equation}
The factor $\frac{1-\cos \theta}{2}$ accounts for the dependence on the relative velocity on the interaction angle $\theta$.
The factor $1/2$ normalises it as a probability.

Eq.~\eqref{eq: absorption coefficient} shows that, for an isotropic homogeneous target radiation field, the absorption coefficient does not depend on the position or direction.
Thus, the optical depth is just a function of the gamma-ray energy and the distance along the line of sight.
One example for this is the cosmic microwave background (CMB), relevant above some tens of teraelectronvolts.
The differential number density of the CMB is
\begin{equation}
    \frac{\mathrm{d} n_{\mathrm{ph}}}{\mathrm{d} E_{\mathrm{ph}}} \left(E_{\mathrm{ph}}\right) = \frac{8\pi}{\left(hc\right)^3} \ \frac{E_{\mathrm{ph}}^2}{\exp\left(\frac{E_{\mathrm{ph}}}{k_{\mathrm{B}} T_{\mathrm{CMB}}}\right) - 1}
\end{equation}
where we use a CMB temperature of $T_{\mathrm{CMB}} = \SI{2.7255}{\kelvin}$~\cite{2009FixsenApJ}.
This component can easily be incorporated in the calculations.
Other components, like the infrared emissions of dust, are more complicated.
As mentioned in Sec.~\ref{sec : absorption}, we use precomputed absorption coefficients provided in \textsc{GALPROP v.57}~\cite{2022PorterJohannessonApJS} for those.
\section{\BURST{} vs. \credit{}}\label{app:Burst vs. CREDIT}
\begin{figure}  
     \centering
    \includegraphics[width=\textwidth]{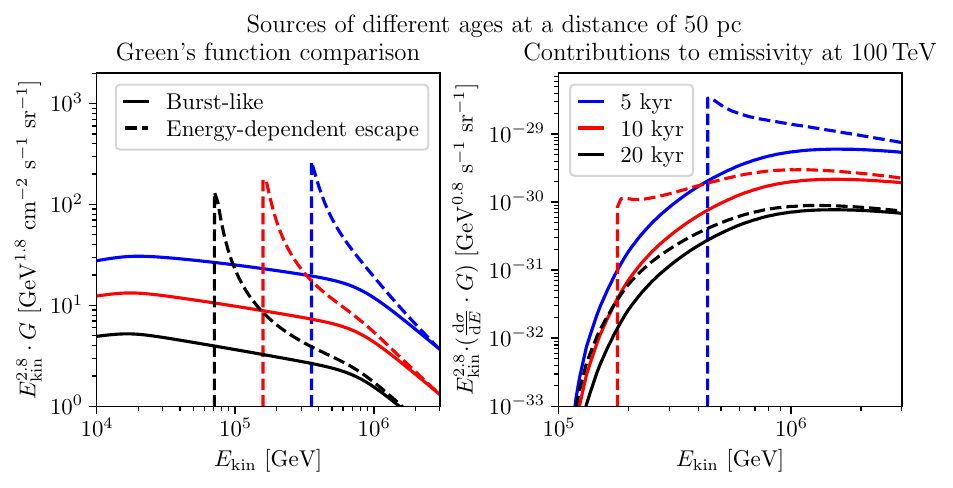}
    \caption{\textit{Left:} The CR spectra produced by single sources (Green's functions) at a distance of \SI{50}{\parsec} are shown.
    Spectra in the \burst{} model are shown with solid lines and in the \credit{} model with dashed lines.
    Different ages (time since beginning of the injection) are shown with different colours.
    The spectra are multiplied by $E_{\mathrm{kin}}^{2.8}$.
    One can clearly see the low-energy cut-off in the \credit{} scenario and the ensuing peak (see also~\cite{2025StallLooApJL}).
    \textit{Right:} The contributions of these sources to the emissivity at a gamma-ray energy of \SI{100}{\tera\electronvolt} over a range of CR energies is shown.
    It is obtained by multiplying the Green's functions with the differential cross sections.
    }
    \label{fig:App1}
\end{figure}
The statistical analysis of our Monte Carlo samples shows that the deviations in the \burst{} scenario are more extreme than the ones in the \credit{} scenario (see especially Secs.~\ref{sec:The most extreme deviations} and~\ref{sec:High-statistics assessment of diffuse emission distributions}).
In this section, we want to illustrate how this effect can be understood.
The left panel of Fig.~\ref{fig:App1} shows the Green's functions (see Eq.~\eqref{eq: analytic Greens function}) for the \burst{} and the \credit{} scenario for different source ages and a fixed distance of \SI{50}{\parsec}.
One can clearly see the peaks produced in the \credit{} scenario that emerge due to the energy-dependence of the escape in that model and have been discussed in Ref.~\cite{2025StallLooApJL}.
On the one hand, the peaks are strong enhancements of the CR intensity in a specific energy range and could thus potentially lead to extreme contributions to the GDEs.
On the other hand, the low-energy cut-off due to the energy-dependent escape times limits the energy range in which a single source contributes.

The right panel of Fig.~\ref{fig:App1} shows the contributions to the emissivity (Eq.~\eqref{eq: emissivity}) at a gamma-ray energy of \SI{100}{\tera\electronvolt} as a function of the CR kinetic energies.
Only CR energies above \SI{100}{\tera\electronvolt} contribute.
Further, we can see the influence of the differential cross section by comparing to the left panel.
Depending on the parameters of the source, the contributions to the emissivity at a specific CR energy are higher in the \credit{} case or in the \burst{} case.
To obtain the total emissivity, these functions have to be integrated in CR energy according to Eq.~\eqref{eq: emissivity}.

\begin{figure}  
     \centering
    \includegraphics[width=\textwidth]{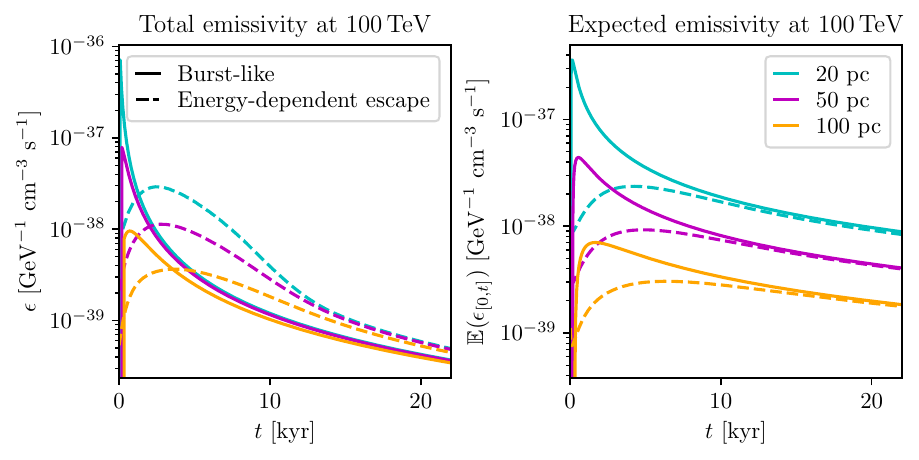}
    \caption{Contributions of discrete sources to the emissivity at a gamma-ray energy of \SI{100}{\tera\electronvolt}.
    As in Fig.~\ref{fig:App1}, the line styles show the source model, whereas the different colours display different source distances.
    The emissivity is shown as a function of the source age.
    \textit{Left:} We show the single-source-emissivity $\epsilon$ calculated according to Eq.~\eqref{eq: emissivity} as a function of the source age for different distances in the \burst{} and \credit{} model with an assumed gas density of \SI{1}{\per\centi\meter\cubed}.
    The emissivity peaks for older sources in the \credit{} scenario as the peak in the CR spectrum (see Fig~\ref{fig:App1}) moves towards a higher-contributing energy range later.
    \textit{Right:} With random source ages, the expected emissivity can be obtained by calculating the expectation value $\mathbb{{E}}(\epsilon_{{[0,t]}})$ of the emissivity for source ages from $0$ to some age $t$ (see Eq.~\eqref{eq:expectation value emissivity}).
    This time average is always higher for the \burst{} compared to the \credit{} model, which explains why we can expect higher GDE deviations in this scenario.
    }
    \label{fig:App2}
\end{figure}
The total emissivity at \SI{100}{\tera\electronvolt} is shown in the left panel of Fig.~\ref{fig:App2}.
The curves show the total emissivities as a function of the age of the sources.
This means that each value of the curve represents the CR energy integral of a curve like the ones shown in the right panel of Fig.~\ref{fig:App1}.
The different colours represent different source distances.
For all source distances, the emissivities of the \burst{} sources peak for earlier times than in the \credit{} scenario and show much higher values.
This dependence on the age of the source can be understood by considering that the CR intensity in the \burst{} scenario is highest across all energies for very early times, while the escape of particles is still ongoing in the \credit{} scenario.
The peak in the CR spectrum in the \credit{} scenario moves to lower CR energies with increasing age of the source.
First, this increases the total emissivity. Then, it decreases again with increasing age before finally approaching the emissivity in the \burst{} scenario for older ages.

We want to determine, which scenario can be expected to produce more extreme GDE intensities.
In our discrete model, deviations are produced by individual sources.
The most extreme ones are produces by especially young sources close to a grid point.
Thus, we know that the most extreme deviations are produced by very young sources.
To compare the \burst{} and the \credit{} scenario, we calculate the expectation value
\begin{equation}\label{eq:expectation value emissivity}
    \mathbb{{E}}(\epsilon_{{[0,t]}}) = \frac{1}{t} \int_0^t \mathrm{d}t' \ \epsilon \left(t', d_0; \mathbf{x}, \SI{100}{\tera\electronvolt}\right) \, ,
\end{equation}
where $\epsilon \left(t', d_0; \mathbf{x}, \SI{100}{\tera\electronvolt}\right)$ is the emissivity for GDEs at \SI{100}{\tera\electronvolt} for a source of age $t'$ at a distance of $d_0$ at some grid point $\mathbf{x}$.
The factor $1/t$ normalises the uniform distribution of ages.

This quantity is shown in the right panel of Fig.~\ref{fig:App2} and allows us to compare the two scenarios.
We always find a larger expectation value for the emissivity in the \burst{} scenario.
Thus, we can expect that realisations that include very young sources, which lead to the most extreme GDE intensities, are more extreme in the \burst{} scenario.
The effect of a strong enhancement of CR intensity in a narrow CR energy energy range around the peaks in the \credit{} scenario (see Fig~\ref{fig:App1}) does not outweigh the strong contribution across the whole CR energy range early-on in the \burst{} scenario.
\end{document}